\documentclass[fleqn,usenatbib]{mnras}

\usepackage{newtxtext,newtxmath}
\usepackage{booktabs}

\usepackage[T1]{fontenc}

\usepackage[normalem]{ulem}

\DeclareRobustCommand{\VAN}[3]{#2}
\let\VANthebibliography\thebibliography
\def\thebibliography{\DeclareRobustCommand{\VAN}[3]{##3}\VANthebibliography}

\usepackage{graphicx}	
\usepackage{amsmath}	
\usepackage{subfig}      

\title[Insights from the brightest blazar flare]{Insights from leptohadronic modelling of the brightest blazar flare}

\author[E. Podlesnyi \& F. Oikonomou]{
Egor Podlesnyi,$^{1}$\thanks{E-mail: egor.podlesnyi@ntnu.no}
Foteini Oikonomou$^{1}$
\\
$^{1}$Department of Physics, Norwegian University of Science and Technology, Høgskoleringen 5, Trondheim 7491, Norway\\
}

\date{Accepted XXX. Received YYY; in original form ZZZ}

\pubyear{\the\year{}}

\begin{document}
\label{firstpage}
\pagerange{\pageref{firstpage}--\pageref{lastpage}}
\maketitle

\begin{abstract}
The blazar 3C 454.3 experienced a major flare in November 2010, making it the brightest $\gamma$-ray source in the sky of the \textit{Fermi} Large Area Telescope (LAT). We obtain seven daily consecutive spectral-energy distributions (SEDs) of the flare in the infrared, optical, ultraviolet, X-ray and $\gamma$-ray bands with publicly available data. We simulate the physical conditions in the blazar and show that the observed SEDs are well reproduced in the framework of a ``standing feature'' where the position of the emitting region is almost stationary, located beyond the outer radius of the broad-line region and into which fresh blobs of relativistically moving magnetised plasma are continuously injected. Meanwhile, a model with a single ``moving blob'' does not describe the data well. We obtain a robust upper limit to the amount of high-energy protons in the jet of 3C 454.3 from the electromagnetic SED. We construct a neutrino light curve of 3C 454.3 and estimate the expected neutrino yield at energies $\geq 100$~TeV for 3C 454.3 to be up to $6 \times 10^{-3}$ $\nu_{\mu}$ per year. Finally, we extrapolate our model findings to the light curves of all \textit{Fermi}-LAT flat-spectrum radio quasars. We find that next-generation neutrino telescopes are expected to detect approximately one multimessenger ($\gamma + \nu_{\mu}$) flare per year from bright blazars with neutrino peak energy in the hundreds TeV -- hundreds PeV energy range and show that the electromagnetic flare peak can precede the neutrino arrival by months to years.
\end{abstract}




\section{Introduction}

Blazars -- bright active galactic nuclei (AGNs) with their relativistic jets pointing at small angles towards Earth -- dominate the $\gamma$-ray sky visible by the \textit{Fermi} Large Area Telescope (\textit{Fermi}-LAT) \citep{2009ApJ...697.1071A} in the region far from the galactic plane \citep{2020ApJS..247...33A}. Since the evidence of the association of the \textit{IceCube} neutrino event IceCube-170922A with the electromagnetic flare of blazar TXS 0506+056 in 2017 \citep{2018Sci...361.1378I}, blazars became objects of intensive theoretical studies as potential sources of the astrophysical neutrinos \citep[e.g.,][]{2018ApJ...864...84K,2018ApJ...863L..10A,2018ApJ...865..124M,2018ApJ...866..109S,2019NatAs...3...88G,2019MNRAS.489.4347O,2019MNRAS.483L..12C,2019ApJ...881...46R,2019ApJ...874L..29R,2019PhRvD..99f3008L,2021MNRAS.500.3613J,2020ApJ...891..115P,Das:2020hev,Das:2022nyp,Kalashev:2022scs,2024arXiv241007905V,2024ApJ...977...42R}. Multiple searches for correlations between the \textit{IceCube} neutrinos and different electromagnetic counterparts among blazars resulted in the conclusion that blazars contribute to the total \textit{IceCube} astrophysical neutrino flux at the level of around $\lesssim 1-30$~\% (for a review, see \citealp{2021PhyU...64.1261T}, table 4 and references therein; \citealp{2023ApJ...954...75A,2024PhyU...67..349T}).

Although some individual blazars were suggested as candidates for the sources of astrophysical neutrinos -- besides TXS 0506+056, there are PKS 0735+178 by \citet{2023MNRAS.519.1396S,2024MNRAS.527.8746P,2023ApJ...954...70A}; PKS 1424+240 by \citet{2022MNRAS.511.4697P}; PKS 1502+106 by \citet{Rodrigues:2020fbu,2021JCAP...10..082O} -- there is at the moment no firmly established paradigm on what particular properties of blazars make them the most efficient neutrino emitters, or, in other words, what subclass of the blazars should contribute the most to the astrophysical neutrino flux. Traditionally blazars are divided into the subclasses of flat-spectrum radio quasars (FSRQs) and BL Lacertae-like objects (BL Lacs), where the former are characterised by the presence of strong line emission from the broad-line region (BLR) and higher accretion rate, while the latter lack the presence of the BLR and have a lower accretion rate \citep{2019ARA&A..57..467B}. FSRQs hosting strong photon fields serving as a target for photopion production may be promising candidates for the sources of high-energy neutrinos \citep{2014PhRvD..90b3007M,2018ApJ...854...54R}.

FSRQs are sources with high flaring activity \citep{2011ApJ...743..171A,2017ApJ...846...34A,2023ApJ...954..194Y} with enhanced observable electromagnetic intensity due to the increase of the luminosity of charged particles in the source and/or the apparent Doppler factor, which leads to the amplified neutrino production. Depending on the details of a particular model, flares may dominantly contribute to the diffuse neutrino flux \citep{2018ApJ...865..124M}. Besides, the atmospheric background is limited by the time window of a flare.

Among FSRQs in the \textit{Fermi}-LAT catalogue 4FGL, 3C 454.3 is the brightest one \citep{2020ApJS..247...33A}. In November 2010, it flared in the broad electromagnetic spectrum (from radio waves to $\gamma$ rays) and became, for some time, the brightest object in the \textit{Fermi}-LAT sky \citep{2011ApJ...733L..26A}. \citet{2013MNRAS.430.1324N} ranked that flare with the peak around MJD 55520 as the brightest blazar flare (defined by the flux at the peak). Observations with different instruments covered it very well. Hence, we choose this source as the study object.

In this work, we develop a model which describes the observational data during the flare and calculate the expected neutrino yield observable by \textit{IceCube} from 3C 454.3. After that, we extrapolate the obtained results on the whole population of FSRQs observed by the \textit{Fermi}-LAT to estimate their contribution to the \textit{IceCube} neutrino flux at energies $\geq 100$~TeV within the framework of our single-zone leptohadronic model.

Using publicly available data from \textit{SMARTS}, \textit{Swift}-UVOT, \textit{Swift}-XRT, and \textit{Fermi}-LAT, we obtain seven multi-wavelength quasi-simultaneous one-day averaged SEDs observed during the flare period.
We describe the multi-wavelength data analysis in Sect. \ref{sec:data}. We model (Sect. \ref{sec:modelling}) the observable flare SEDs with the publicly-available \texttt{AM$^{3}$} program \citep{2023ascl.soft12031K,2024ApJS..275....4K} within three different frameworks: 1)~assuming a standard leptonic model where the observed electromagnetic emission comes from relativistic electrons losing energy via synchrotron and inverse-Compton (IC) radiation in compact blobs passing through a quasi-stationary feature in the jet into which relativistic plasma is injected with large bulk Lorentz factor (``standing-feature'' model, Sects. \ref{sec:standing_blob_model_description}, \ref{sec:standing_blob_model_results}); 2)~utilizing the \texttt{AM$^{3}$}'s capabilities of the time-dependent modelling, we attempt to model the whole flare in framework of a ``moving-blob'' model where the whole radiation comes from a single blob co-moving with the bulk motion of the relativistic jet plasma (``moving-blob'' model, Sects. \ref{sec:moving_blob_model_description}, \ref{sec:moving_blob_model_results}); 3)~assuming that in addition to electrons, protons might be co-accelerated in the jet with the same efficiency, we calculate the maximum-allowed by the data proton energy density $u^{\prime \mathrm{UL}}_{p}$ within a leptohadronic ``standing-feature'' model (Sects. \ref{sec:leptohadronic}, \ref{sec:results_leptohadronic}), providing our method to estimate the expected \textit{IceCube} neutrino yield in Sect. \ref{sec:neutrino_yield_estimates} with corresponding results in Sect.~\ref{sec:neutrino_light_curve}. Our method to extrapolate the results for 3C~454.3 on the whole population of \textit{Fermi}-LAT FSRQs is presented in Sect.~\ref{sec:fsrqs_contribution_estimates} and the results of this are presented in Sect.~\ref{sec:IceCat1}. We discuss our findings in Sect.~\ref{sec:discussion}. In Sect. \ref{sec:conclusions}, we conclude that it is not possible to describe the seven days of the flare with a single ``moving-blob'' model, while a ``standing-feature'' model does it successfully. Moreover, we find that multi-wavelength flares of \textit{Fermi}-LAT FSRQs similar to the studied 3C 454.3 flare contribute to the neutrino flux at energies $\geq 100$~TeV at the level~of~$\sim 0.5$\% under the assumption that their emission can be described within a single-zone ``standing-feature'' leptohadronic model with the scaling between the observed integral neutrino and $\gamma$-ray fluxes inferred from the 3C 454.3 modelling for MJD 55517--55524.

\section{Data analysis}
\label{sec:data}

\subsection{\textit{Fermi}-LAT -- $\gamma$ rays}
\label{sec:fermi}

For our object of interest, 3C 454.3, we select the \textit{Fermi}-LAT data in the energy range $100$~MeV $\leq E \leq$ $1$~TeV and in time intervals corresponding to one day within the period MJD 55517--55524. The analysis is performed with the standard \texttt{fermipy} \citep{Wood2017} package for the binned likelihood analysis. During the \textit{Fermi}-LAT data analysis of the $15^{\circ}$ ROI (region of interest) centred at the position of 3C 454.3 we apply a cut on the zenith angle value $\theta_z \leq 90^{\circ}$ in order to remove contamination from the Earth's limb. We use the \texttt{{P8R3\_SOURCE\_V3}} instrument response function (\texttt{evclass}~$=128$) with detected $\gamma$ rays both in \texttt{FRONT} and \texttt{BACK} of the \textit{Fermi}-LAT (\texttt{evtype}~$=3$)\footnote{\url{https://fermi.gsfc.nasa.gov/ssc/data/analysis/documentation/Cicerone/Cicerone_LAT_IRFs/IRF_overview.html}}. We divide the data into six bins per decade in energy, and the angular binning is $0.1^\circ$ per pixel. To describe the observed $\gamma$-ray emission we construct a model containing all sources from the 4$^{\mathrm{th}}$ \textit{Fermi}-LAT source catalogue 4FGL \citep{2020ApJS..247...33A} located within $15^{\circ}$ from the centre of the ROI\footnote{3C 454.3 is present itself in the 4FGL with the name 4FGL J2253.9+1609.}, the model of the diffuse galactic background \texttt{gll\_iem\_v07}, and the model of the isotropic $\gamma$-ray background \texttt{iso\_P8R3\_SOURCE\_V3\_v1}\footnote{\url{https://fermi.gsfc.nasa.gov/ssc/data/access/lat/BackgroundModels.html}}. We model the spectral shape of 3C 454.3 as a power law with a subexponential cutoff\footnote{The \textit{Fermi}-LAT Collaboration uses the term ``superexponential cutoff'' \url{https://fermi.gsfc.nasa.gov/ssc/data/analysis/scitools/source\_models.html}, but in our case ${b} < 1$, so we use the term ``subexponential'' instead.} as it was modelled in the 4FGL catalogue:
\begin{equation}
        \frac{dN}{dE} = N_{0} \left(\frac{E}{E_0} \right)^{-\gamma_{1}} \exp{\left(- aE^{b}\right)}.
    \label{eq:superexp_dn_de}
\end{equation}
The normalizations of the diffuse backgrounds, 3C 454.3 and other sources within $7.5^{\circ}$ from the centre of the ROI are left free during the initial fitting of the ROI with all spectral shapes fixed according to the catalogue; both normalizations and spectral shapes of other sources outside $7.5^{\circ}$ from the ROI centre are fixed to their catalogue values. After the initial fit, the spectral shape of 3C 454.3 is freed, and the fit is repeated. The results of the global \textit{Fermi}-LAT SED fits are presented in Table~\ref{table:fermi_lat_data_analysis}. The $\gamma$-ray SED of 3C 454.3 is then obtained using the \texttt{fermipy.GTAnalysis.sed} method with the default parameter values\footnote{\url{https://fermipy.readthedocs.io/en/latest/advanced/sed.html\#id3}}.

\begin{table*}
    \centering
    \begin{tabular}{|l|c||c|c|c|c|c|c|c}
    \toprule
    Parameter & 55517 & 55518 & 55519 & 55520 & 55521 & 55522 & 55523 & 7-d-averaged \\
    \midrule
    $N_{0}$ & $3.22 \pm 0.27$ & $3.53 \pm 0.07$ & $3.60 \pm 0.08$ & $3.55 \pm 0.08$ & $1.67 \pm 0.25$ & $1.42 \pm 0.05$ & $1.22 \pm 0.05$ & $2.63 \pm 0.26$ \\
    $\gamma_{1}$ & $1.78 \pm 0.05$ & $1.952 \pm 0.021$ & $1.959 \pm 0.022$ & $1.806 \pm 0.021$ & $1.88 \pm 0.08$ & $1.96 \pm 0.03$ & $1.95 \pm 0.04$ & $1.889 \pm 0.009$\\
    $E_{0}$ &\multicolumn{8}{c}{$517.5$} \\
    $a$ & \, $1.99$ & \, $1.991$ & \, $1.9908$ & \, $1.9908$ & \, $1.8$ & \, $1.99076$ & \, $1.9908$ & \, $1.990755$ \\
        & $\pm 0.27$ & $\pm 0.003$ & $\pm 0.0009$ & $\pm 0.0018$ & $\pm 0.5$ & $\pm 0.00019$ & $\pm 0.0016$ & $\pm 0.000012$ \\
    $b$ & \multicolumn{8}{c}{$0.5184$} \\
    \bottomrule
    \end{tabular}
    \caption{Results of the \textit{Fermi}-LAT $\gamma$-ray global SED fit in the energy range $100$~MeV $\leq E \leq$ $1$~TeV obtained with \texttt{fermipy} for each of the seven days and the seven-day-averaged period. The columns indicate the start time of the one-day averaged observations in MJD. $N_{0}$ is expressed in [$10^{-8}$ MeV$^{-1}$ s$^{-1}$ cm$^{-2}$]; $E_{0}$ is in [MeV]; and $a$ is in [$10^{-2}$ MeV$^{-1}$]. \label{table:fermi_lat_data_analysis}}
\end{table*}
\begin{table*}
    \centering
    \begin{tabular}{|l|c||c|c|c|c|c|c|c}
    \toprule
    Parameter & 55517 & 55518 & 55519 & 55520 & 55521 & 55522 & 55523 & 7-d-averaged \\
    \midrule
    obsID & 00035030133 & 00035030134 & 00035030136 & 00035030138 & 00035030140 & 00035030142 & 00035030144 & all shown obsIDs \\
     &  & 00035030135 & 00035030137 & 00035030139 & 00035030141 & 00035030143 & 00035030145 & \\
    $\kappa$ & $5.3^{+4.2}_{-0.0}$ & $5.3^{+4.2}_{-0.0}$ & $5.3^{+2.9}_{-0.0}$ & $5.3^{+3.0}_{-0.0}$ & $9.5^{+0.0}_{-4.2}$ & $9.5^{+0.0}_{-4.2}$ & $8.3^{+1.2}_{-3.0}$ & $7.7^{+1.8}_{-2.4}$\\
    $N_{X}$ & $19.4^{+5.1}_{-1.2}$ & $23.3^{+2.9}_{-0.6}$ & $22.0^{+2.7}_{-0.5}$ & $20.3^{+2.5}_{-0.5}$ & $16.6^{+0.5}_{-1.9}$ & $15.4^{+1.4}_{-1.8}$ & $12.8^{+0.8}_{-1.3}$ & $18.5^{+1.0}_{-1.2}$\\
    $\Gamma_{1}$ & $0.97^{+0.40}_{-0.28}$ & $1.00^{+0.26}_{-0.09}$ & $1.06^{+0.22}_{-0.07}$ & $1.05^{+0.22}_{-0.08}$ & $1.25^{+0.08}_{-0.35}$ & $1.30^{+0.10}_{-0.42}$ & $1.10^{+0.15}_{-0.28}$ & $1.16^{+0.10}_{-0.16}$\\
    $E_{\mathrm{br}}$ & $1.6^{+0.9}_{-0.6}$ & $1.74^{+0.44}_{-0.20}$ & $2.09^{+0.79}_{-0.31}$ & $2.0^{+0.5}_{-0.3}$ & $2.0^{+0.7}_{-0.5}$ & $1.6^{+1.5}_{-0.6}$ & $1.9^{+0.5}_{-0.4}$ & $1.96^{+0.25}_{-0.25}$\\
    $\Gamma_{2}$ & $1.69^{+0.25}_{-0.16}$ & $1.58^{+0.08}_{-0.06}$ & $1.60^{+0.11}_{-0.08}$ & $1.54^{+0.10}_{-0.08}$ & $1.69^{+0.15}_{-0.12}$ & $1.56^{+0.16}_{-0.08}$ & $1.64^{+0.14}_{-0.11}$ & $1.61^{+0.04}_{-0.04}$\\
    $\chi^{2}_{\mathrm{red}}$ & $0.88$ & $1.75$ & $0.79$ & $0.80$ & $1.29$ & $1.09$ & $0.39$ & $1.06$\\
    \bottomrule
    \end{tabular}
    \caption{Results of the \textit{Swift}-XRT X-ray SED fit obtained with \texttt{ISIS} for each of the seven days and the whole seven-day-averaged period. The asymmetric errors display the $68$\% confidence interval estimated using \texttt{conf\_loop} \texttt{ISIS} function. Parameter $\kappa$ for each single day is forced to be within $68$\% confidence interval of the seven-day-averaged fit for the stability of the fits. The columns indicate the start time of the one-day-averaged observations in MJD. Parameter $\kappa$ is expressed in [$10^{20}$ cm$^{-2}$], $N_{X}$ is in [$10^{-3}$ cm$^{-2}$ s$^{-1}$ keV$^{-1}$], and $E_{\mathrm{br}}$ is in [keV]. \label{table:swift_xrt_data_analysis}}
\end{table*}
\begin{table*}
    \centering
    \begin{tabular}{|l|c||c|c|c|c|c|c|c}
    \toprule
    Filter & 55517 & 55518 & 55519 & 55520 & 55521 & 55522 & 55523 & 7-d-averaged \\
    \midrule
    $v$ & $-$ & $-$ & $12.7 \pm 0.3$ & $-$ & $-$ & $-$ & $-$ & $-$ \\
    $b$ & $-$ & $-$ & $8.03 \pm 0.19$ & $-$ & $-$ & $-$ & $-$ & $-$ \\
    $u$ & $-$ & $-$ & $5.10 \pm 0.13$ & $6.64 \pm 0.14$ & $-$ & $-$ & $-$ & $-$ \\
    $uw1$ & $-$ & $-$ & $2.31 \pm 0.05$ & $-$ & $-$ & $-$ & $1.72 \pm 0.04$ & $-$ \\
    $um2$ & $-$ & $1.60 \pm 0.04$ & $1.48 \pm 0.04$ & $-$ & $-$ & $1.241 \pm 0.029$ & $-$ & $1.44 \pm 0.15$ \\
    $uw2$ & $1.030 \pm 0.023$ & $-$ & $1.189 \pm 0.029$ & $-$ & $0.940 \pm 0.020$ & $-$ & $-$ & $1.05 \pm 0.10$ \\
    \bottomrule
    \end{tabular}
    \caption{Spectral flux densities and their uncertainties in mJy observed (before deredenning) with each filter by the \textit{Swift}-UVOT for each of the seven days and the whole seven-day-averaged period. The columns indicate the start time of the one-day averaged observations in MJD. The \textit{Swift}-UVOT uses the same obsIDs as the \textit{Swift}-XRT (see Table~\ref{table:swift_xrt_data_analysis}). Filter $v$ corresponds to the wavelength of $5468$~Å, $b$ --- $4392$~Å, $u$ --- $3465$~Å, $uw1$ --- $2600$~Å, $um2$ --- $2246$~Å, $uw2$ --- $1928$~Å. \label{table:swift_uvot_data_analysis}}
\end{table*}

\subsection{\textit{Swift}-XRT -- X rays}
\label{sec:xrt}

The \textit{Swift}-XRT data analysis is performed with \texttt{ISIS} \citep{2000ASPC..216..591H} and \texttt{swifttools.ukssdc/xrt\_prods}\footnote{\url{https://www.swift.ac.uk/user_objects/API/}} \citep{2009MNRAS.397.1177E} which rely on \texttt{HEASOFT}, version $6.32.1$ \citep{2014ascl.soft08004N,1995ASPC...77..367B}. To retrieve the \textit{Swift}-XRT SED quasi-simultaneous to the \textit{Fermi}-LAT one-day-averaged SED, we download the data from all the \textit{Swift}-XRT pointings to 3C 454.3 within the given one-day-long period. We analyse the photon counts within the energy range $\left[ 0.3; 10.0 \right]$ keV, requiring the number of channels per energy bin to be $\geq 64$ and the signal-to-noise ratio in each bin $\geq 5.0$.

The observed spectrum is modelled as a broken power law with the normalization $N_{X}$, break energy $E_{\mathrm{br}}$, and the two spectral indices ($\Gamma_1$ before the break and $\Gamma_2$ after the break) accounting for the photoelectric absorption along the LoS (line of sight) with the hydrogen column density $\kappa$. The details of the deabsorption procedure are presented in Appendix~\ref{appendix:X_ray_deabsorption}. The results of the \textit{Swift}-XRT SED fits are presented in Table~\ref{table:swift_xrt_data_analysis}.

\subsection{\textit{Swift}-UVOT, \textit{SMARTS}, and \textit{Steward} -- UV, optical, IR}
\label{sec:optical}

\subsubsection{\textit{Swift}-UVOT}

The \textit{Swift}-UVOT SEDs are obtained with the \texttt{swift-uvot-analysis-tools}\footnote{\url{https://github.com/KarlenS/swift-uvot-analysis-tools}} program, which provides the values of the observed SED averaged over a given day within the time interval MJD 55517--55524. For the seven-day-averaged period, we calculate the arithmetic average optical/UV SED value in each filter, which was used at least three times within the period and incorporate the standard deviation of the flux values as an additional uncertainty, adding it quadratically to the statistical uncertainties (the same is done for the optical data from {\textit{SMARTS} and \textit{Steward} Observatory).

\subsubsection{\textit{SMARTS} and \textit{Steward} Observatory}

For each considered period, when available, we use the data from the \textit{SMARTS} program from optical/IR observations of \textit{Fermi} blazars \citep{2012ApJ...756...13B} provided as photometric measurements in B, V, R, J, and K bands at the corresponding web page\footnote{\url{http://www.astro.yale.edu/smarts/glast/tables/3C454.tab}} and the data from the \textit{Steward} observatory \citep{2009arXiv0912.3621S} provided as photometric measurements in R and V bands\footnote{\url{https://james.as.arizona.edu/~psmith/Fermi/DATA/Objects/3c454.3.html}}. The conversion of apparent Vega magnitudes into the observed SEDs was done using the zero-flux points from table A2 by \citet{1998A&A...333..231B}.

\subsubsection{Deredenning}
The deredenning for IR, optical and UV data was performed using the \texttt{extinction} program \citep{2016zndo....804967B} assuming the extinction law of \citet{1999PASP..111...63F}, with the mean value of $E(B-V)$ extracted from \cite{https://doi.org/10.26131/irsa537} using \texttt{astroquery.ipac.irsa.irsa\_dust} module \citep{astroquery_10799414,2019AJ....157...98G} with the assumed value of $R_V = 3.1$.

\subsection{Radio}
\label{sec:radio}

The one-zone model used in the current study invokes compact blobs. In such compact regions, the radio waves get absorbed due to the synchrotron self-absorption (SSA). Thus, the observed radio signal must originate from larger-scale parts of the jet. We use the archival data from NASA/IPAC Extragalactic Database \citep{https://doi.org/10.26132/ned1} as upper limits when comparing our model SEDs to the observations of 3C\,454.3. 

\subsection{Assumed systematic uncertainties}
\label{sec:systematics}
From the comparison of measured magnitudes by \textit{SMARTS} and \textit{Steward} for the same object for the same night, we found a typical value of $\approx 0.08$ for the relative uncertainty of the SED (see Appendix \ref{appendix:SMARTS_VS_Steward}). We add in quadrature 8\% relative systematic uncertainty to the statistical uncertainty of all IR, optical and UV data points. Moreover, both \textit{Fermi}-LAT and \textit{Swift}-XRT have some uncertainty in the estimations of their effective areas, which results in a systematic uncertainty in the measured SEDs. We assume $3\%$ relative systematic uncertainty, which can be expected for both \textit{Fermi}-LAT\footnote{\url{https://fermi.gsfc.nasa.gov/ssc/data/analysis/scitools/Aeff\_Systematics.html}} and \textit{Swift}-XRT \citep[][sect. 5]{2006A&A...451..777M}, adding it in quadrature to all the uncertainties of all the analysed SEDs (including IR, optical, and UV data points; Tables~\ref{table:fermi_lat_data_analysis}, \ref{table:swift_xrt_data_analysis}, and \ref{table:swift_uvot_data_analysis} show values before adding any systematic uncertainty).

The obtained SEDs have been compared to the SEDs available in the Markarian Multiwavelength Data Center (MMDC)\footnote{\url{https://mmdc.am/}} \citep{sahakyan2024markarianmultiwavelengthdatacenter}. The comparison showed a good agreement except for the low-energy part of the \textit{Swift}-XRT spectrum where our data points lie somewhat higher than those from the MMDC. The disagreement is at a level consistent with the expected discrepancy due to different treatments of the photoelectric absorption \citep{2021MNRAS.507.5690G}. The MMDC was not available when the data analysis for this work was performed.

\section{Leptonic models}
\label{sec:modelling}

\begin{figure}
     \includegraphics[width=1\columnwidth]{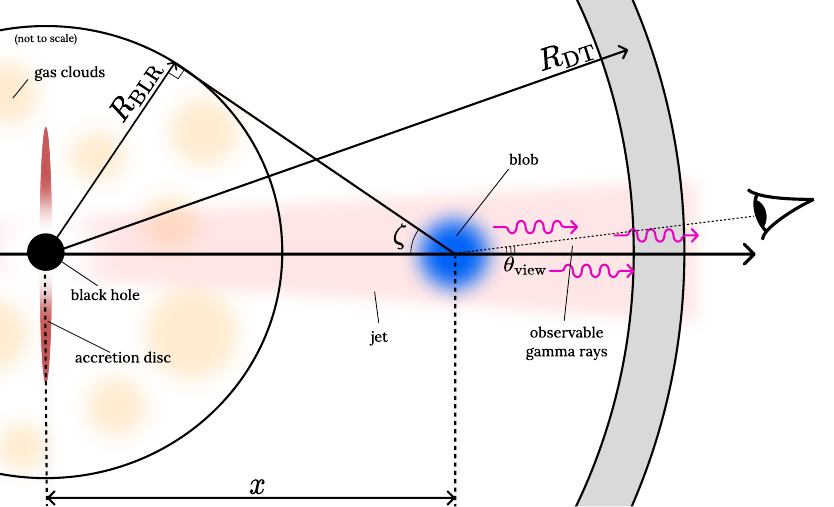}
     \caption{Geometrical scheme of the emitting region in the jet of 3C 454.3 (figure adapted from \citealp{2022MNRAS.515.5242D}).}
     \label{fig:geometry}
\end{figure}

\begin{table*}
    \centering
    \begin{tabular}{llc}
    \toprule
    Variable & Explanation & Value\\
    \midrule
    \multicolumn{2}{c}{Geometry and conditions in the blob} \\
    \midrule
    $B^{\prime} \, [\mathrm{G}]$ & Root-mean-square magnetic field strength in the blob & $f$\\
    $R_b^{\prime} \, [\mathrm{cm}]$ & Blob radius & $f$\\
    $x_s$ [cm] & Blob launching distance from the SMBH$^m$ & $f$$^m$\\
    $x_f$ [cm] & Blob final distance from the SMBH$^m$ & $f$$^m$\\
    $x$ [cm] & Effective (current$^m$) blob distance from the SMBH & $f(d^m$, see Eq. \ref{eq:x_moving_blob}$^m$)\\
    $\theta_{\mathrm{view}} \, [^{\circ}]$ & Viewing angle w.r.t. the LoS & $0.4$\\
    $\beta$ & Blob speed divided by the speed of light $c$ & $d$\\
    $\Gamma$ & Blob Lorentz factor & $f(d^m)$\\
    $\delta$ & Blob Doppler factor & $d$, see Eq. (\ref{eq:observed_doppler})\\
    $z$ & Cosmological redshift & $0.859$\\
    $d_L \, [\mathrm{cm}]$ & Luminosity distance & $d$, $1.74 \times 10^{28}$\\
    \midrule
    \multicolumn{2}{c}{Injected electrons} \\
    \midrule
    $E^{\prime{\mathrm{min}}}_{e}$ [eV] & Minimum electron injection energy & $f$\\
    $E^{\prime{\mathrm{max}}}_{e}$ [eV] & Maximum electron injection energy & $f$\\
    $s_{{e}}$ & Electron injection power-law spectral index & $f$\\
    $L_e^{\prime} \, [\mathrm{erg/s}]$ & Injected electron luminosity & $f$\\
    $u_e^{\prime} \, [\mathrm{erg/cm}^3]$ & Injected electron energy density & $d$\\
    $\eta^{\prime \mathrm{esc}}_{e}$ & Escape factor for electrons w.r.t. photons & $5 \times 10^3$\\
    $t^{\prime \mathrm{esc} }_{e}$ [s] & Electron escape timescale & $d$, see Eq. (\ref{eq:electron_escape})\\
    \midrule
    \multicolumn{2}{c}{Injected protons (for the leptohadronic model)} \\
    \midrule
    $E^{\prime{\mathrm{max}}}_{p}$ [eV] & Minimum proton injection energy & $2 \times 10^{9}$\\
    $E^{\prime{\mathrm{max}}}_{p}$ [eV] & Energy of the exponential cutoff of the proton injection spectrum & $d,$ see Eq. (\ref{eq:proton_max_energy}) \\
    $s_{{p}}$ & Proton injection power-law spectral index & $2.0$\\
    $u_p^{\prime \mathrm{UL}} \, [\mathrm{erg/cm}^3]$ & Upper limit on the proton energy density & $d$, see Eq. (\ref{eq:bayes_factor})\\
    $\eta_{p}^{\prime \mathrm{esc}}$ & Escape factor for protons w.r.t. photons & $5 \times 10^3$\\
    $t^{\prime \mathrm{esc} }_{p}$ [s] & Proton escape timescale & $d$, same as $t^{\prime \mathrm{esc}}_{e}$\\
    \midrule
    \multicolumn{2}{c}{Accretion disc (AD)} \\
    \midrule
    $M_{\mathrm{SMBH}} \, [M_{\odot}]$ & Assumed SMBH mass$^{h}$ & $1.7 \times 10^{9}$\\
    $\eta_{\mathrm{acc}}$ & Accretion efficiency & $1/12$\\
    $L_{\mathrm{AD}}\, [\mathrm{erg/s}]$ & Disc luminosity & $3 \times 10^{46}$, $\approx$ median value of those from Table \ref{table:disc}\\
    $R_g \, [\mathrm{cm}]$ & SMBH gravitational radius & $d$, $5 \times 10^{14}$\\
    $R_{\mathrm{AD,\,in}} \, [R_{g}]$ & Inner radius of the accretion disc & $3$\\
    $R_{\mathrm{AD,\,out}} \, [R_{g}]$ & Outer radius of the accretion disc & $10^3$\\
    \midrule
    \multicolumn{2}{c}{Broad-line region (BLR)} \\
    \midrule
    $R_{\mathrm{BLR}} \, [\mathrm{cm}]$ & BLR radius & $d$, $5.48 \times 10^{17}$, see Eq. (\ref{eq:blr_radius})\\
    $C_{\mathrm{BLR}}$ & BLR continuum reprocessing factor & $0.01$ \\
    $\xi_{\mathrm{BLR}}$ & BLR covering factor for the lines & $0.10$\\
    \midrule
    \multicolumn{2}{c}{Dust torus (DT)} \\
    \midrule
    $R_{\mathrm{DT}} \, [\mathrm{cm}]$ & Dust torus radius & $d$, $1.37 \times 10^{19}$, see Eq. (\ref{eq:dt_radius})\\
    $T_{\mathrm{DT}} \, [\mathrm{K}]$ & Dust torus temperature & $370$\\
    $C_{\mathrm{DT}}$ & Dust torus covering factor & $0.30$\\
    \midrule
    \multicolumn{2}{c}{Time-related variables} \\
    \midrule
    $t^{\prime \mathrm{esc}}_{\gamma}$ [s] & Photon escape timescale & $d$, see Eq. (\ref{eq:gamma_escape}) \\
    $t^{\prime \mathrm{gyro}}_{e}$ [s] & Gyroperiod of an elementary charge in the blob & $d$, see Eq. (\ref{eq:gyroperiod})\\
    $t^{\prime \mathrm{Bohm}}_{e}$ [s] & Bohm diffusion timescale & $d$, see Eq. (\ref{eq:bohm}) \\
    $t^{o}_{\mathrm{obs}}$ [s] & Total time of observations in the observer's rest frame$^m$ & $d$$^m$\\
    $t_{\mathrm{tot}}$ [s] & Total time of the blob motion in the host galaxy rest frame$^m$ & $d$$^m$, see Eq. (\ref{eq:total_time_of_blob_motion})\\
    $t^{o}_{\mathrm{shift}}$ [s] & Time delay of the observations w.r.t. the blob launch$^m$ & $d$$^m$\\
    $t^{\prime}_{\mathrm{sim}}$ [s] & Total simulation time in the blob rest frame & $d$, see Eq. (\ref{eq:simulation_time}) or Eq. (\ref{eq:simulation_time_moving_blob})$^m$\\
    $\Delta t^\prime / t^{\prime}_{\mathrm{sim}}$ & \texttt{AM$^{3}$} relative time step & $0.003$\\
    \bottomrule
    \end{tabular}
    \caption{Definitions of variables mentioned throughout the text. Primed quantities~$[^{\prime}]$ denote variables measured in the blob rest frame. Unprimed variables are either invariant or in the host galaxy rest frame. A superscript~$^{o}$ denotes variables measured in the Earth rest frame at the present cosmological epoch. A letter~$d$ denotes variables whose values are derived from the other variables, and a letter $f$ denotes variables that are free during the optimisation. A superscript~$^m$ denotes variables used in the model with a single ``moving blob''. A superscript~$^{h}$ denotes the average value of those reported by \citet{2010MNRAS.402..497G,2019A&A...631A...4N,2020A&A...633A..73T}.
    \label{table:variables}}
\end{table*}

We assume that the observed multi-wavelength spectral-energy distribution (SED $= E^2 dN/dE = \nu F_{\nu} = E F(E)$) can be modelled as the radiation emitted by a population of relativistic electrons (including positrons which we do not distinguish from electrons unless explicitly stated otherwise) trapped in the turbulent magnetic field of a spherical plasmoid (blob) located in the jet at a distance $x$ from the supermassive black hole (SMBH) where the plasma flows with bulk Lorentz factor $\Gamma$ with a small angle $\theta_{\mathrm{view}}$ to the LoS (see Fig. \ref{fig:geometry}).

For modelling the observable SED, we use the time-dependent program for the simulation of radiative processes \texttt{AM$^{3}$} \citep{2023ascl.soft12031K,2024ApJS..275....4K}, although other solutions are also available (see, e.g., the review of programs simulating radiative processes by \citet{2022Galax..10...85N}, and numerical codes by \citet{2013ApJ...768...54B,2015MNRAS.448..910C,2022MNRAS.509.2102G,2024A&A...683A.225S} compared by \citealp{cerruti2024comprehensivehadroniccodecomparison}). \texttt{AM$^{3}$} is a versatile program for time-dependent modelling of various radiative processes (both leptonic and hadronic) including the electron synchrotron and inverse-Compton (IC) processes (for leptonic models), and the photopion, Bethe-Heitler (BH), proton synchrotron and proton IC processes (for leptohadronic models) with the subsequent cooling of the charged particles. Throughout the paper, we use the variables with their notations explained in Table \ref{table:variables}. We assume the fiducial model of \citet{2012MNRAS.422.3189G} for the calculation of the absorption of $\gamma$ rays on the extragalactic background light (EBL) on their way from 3C 454.3 to Earth and the flat $\Lambda$CDM Universe with the Hubble parameter $H_{0} = 67.74$~km/(s Mpc) and the normalized matter density $\Omega_{m} = 0.3075$ \citep{2016A&A...594A..13P} resulting in the luminosity distance to 3C~454.3 $d_{L} = 1.74 \times 10^{28}$~cm given its redshift $z = 0.859$ \citep{1991MNRAS.250..414J}.

Using the time-dependent solver \texttt{AM$^{3}$}, we model the observed SEDs of the MJD period 55517--55524 in two approaches: 1) assuming that the radiation comes from a region located at a particular distance $x$ from the SMBH which does not change throughout the modelled period, i.e. the emission region is attributed to a stationary feature in the jet -- hereafter the ``standing-feature'' model; 2) alternatively, assuming that the radiation comes from a single blob which moves along with the bulk motion of the jet plasma -- hereafter the ``moving-blob'' model. We perform parameter optimisation utilizing \texttt{iMINUIT} \citep{iminuit, James:1975dr} and \texttt{emcee} \citep{emcee} as well as develop routines to estimate values of some model parameters as described in the following section.

\subsection{Description of the ``standing-feature'' model}
\label{sec:standing_blob_model_description}

    By the ``standing feature'' we mean a quasi-stationary region in the jet in which injection of charged particles and their subsequent cooling are responsible for most of the observed radiation during the flare. The distance $x$ between the ``standing feature'' and the SMBH is kept constant throughout each day of observations. For each of the seven one-day averaged datasets, we obtain a fit with a ``standing-feature'' model, with eight free parameters:
    \begin{enumerate}
        \item injected electron luminosity $L_{e}^{\prime}$;
        \item spectral index $s_{e}$ of the electron power-law injection intensity;
        \item minimum electron injection energy $E^{\prime{\mathrm{min}}}_{e}$;
        \item maximum electron injection energy $E^{\prime{\mathrm{max}}}_{e}$;
        \item blob radius $R_{b}^{\prime}$;
        \item root-mean-square strength of the magnetic field $B^\prime$;
        \item Lorentz factor of the blob $\Gamma$;
        \item dissipation radius $x$.
    \end{enumerate}
    The characteristic timescale over which we average the observed SED is $t^{o}_{\mathrm{av}} = 1$\,d. Therefore, the corresponding time interval in the blob rest frame,
    \begin{equation}
        t^{\prime}_{\mathrm{sim}} = \frac{\delta}{1 + z} t^{o}_{\mathrm{av}},
    \label{eq:simulation_time}
    \end{equation}
    serves as the total time of the calculation of the particle yields after which we stop the \texttt{AM$^{3}$} simulation and extract the particle yields from its last time step, where the apparent Doppler factor is defined as
    \begin{equation}
        \delta = \left[\Gamma (1 - \beta \cos \theta_{\mathrm{view}}) \right]^{-1},
    \label{eq:observed_doppler}
    \end{equation}
    $\beta = \sqrt{1 - \Gamma^{-2}}$, and $\theta_{\mathrm{view}} = 0.4^{\circ}$, is the angle between the jet direction and the LoS \citep[the assumption is based on the radio data by, e.g.,][]{2020MNRAS.495.3576K,2017MNRAS.468.4992P,2013ApJ...773..147J}. We assume that the injection of electrons stops after $t^{\prime}_{\mathrm{sim}}$ and --- since their energy-loss timescale is for most relevant energies much less than $t^{\prime}_{\mathrm{sim}}$ --- their emission stops shortly after that. We verify that in the optical/UV and $\gamma$-ray range, the steady-state SED is reached after around half of $t^{\prime}_{\mathrm{sim}}$ (or, in other words, after around three light-crossing (dynamical) timescales), i.e., after that time, the optical/UV and $\gamma$-ray SED do not change. However, the X-ray SED somewhat depends on $t^{\prime}_{\mathrm{sim}}$ because the timescale for the SSC process is comparable to the simulation timescale. We find that the photon flux in the energy range between $0.2$~keV and $10$~keV varies by $\pm 15$\% after changing the simulation time by a factor of $2$, which is less than or comparable to the relative uncertainty of the model parameters. Thus, we assume that by the end of the \texttt{AM$^{3}$} simulation, we obtain a steady-state model SED, which is fitted to the corresponding daily-averaged quasi-simultaneous observed SED.

    Following \citet{2024A&A...681A.119R}, we model all radiative processes in the blob rest frame having transformed external photon fields into this frame\footnote{Though, an alternative sequence of frame transformations is possible \citep[e.g.,][]{2018MNRAS.481.1455K}.}. The obtained number densities $n^{\prime}(E^{\prime})$, differential per volume and per energy, are transformed into the observer frame SED $\nu F_{\nu}$ as (see, e.g., \citet{Dermer:2009zz}, sect. 5.5.1)
    \begin{equation}
        \nu F_{\nu} = E^{\prime 2} n^{\prime}(E^{\prime}) \frac{\delta^{4}}{4 \pi (1 + z) d_L^2} \frac{V^{\prime}}{t^{\prime \mathrm{esc}}_{\gamma}},
    \label{eq:sed_transformation}
    \end{equation}
    where $d_{L}$ is the cosmological luminosity distance, $V^{\prime} = 4 \pi R_{b}^{\prime 3} / 3$ is the blob volume, $t^{\prime \mathrm{esc}}_{\gamma}$ is the photon escape timescale.

    The emission enhancement factor of $\delta^4$ is applicable in the case of the ``standing feature''\footnote{\citet{1997ApJ...484..108S} use the term ``steady pattern''.} as long as the ratio $N_{b}$ of the duration of emission of each blob $t_{\mathrm{tot}}$ and the time interval between two blobs passing through the ``standing feature'' $t_{\mathrm{btw \, b}}$ is \citep[][appendix B]{1997ApJ...484..108S}
    \begin{equation}
        N_{b} = \frac{t_{\mathrm{tot}}}{t_{\mathrm{btw \, b}}} = \Gamma \delta.
    \label{eq:number_of_blobs}
    \end{equation}
    If, in addition, $t_{\mathrm{btw \, b}} = t^{o}_{\mathrm{av}}$, the observer throughout the time period $t^{o}_{\mathrm{av}}$ will detect photons only from a single ``active'' blob passing through the ``standing feature''. If $t_{\mathrm{btw \, b}} < t^{o}_{\mathrm{av}}$ (and $N_{b}$ still equals $\Gamma \delta$) the observed emission throughout $t^{o}_{\mathrm{av}}$ is the emission averaged over the superposition of emissions of several identical blobs. If $t_{\mathrm{btw \, b}} \gg t^{o}_{\mathrm{av}}$ then all the observed emission comes from a single moving blob. The latter case is considered in Sects. \ref{sec:moving_blob_model_description}, \ref{sec:moving_blob_model_results}. As explained by \citet[][appendix B]{1997ApJ...484..108S}, Eq. (\ref{eq:sed_transformation}) works both for the ``moving blob'' and the ``standing feature'' (for the latter, if Eq. (\ref{eq:number_of_blobs}) is satisfied). 

    In the following subsections, we describe the model parameters in greater detail and present some (semi-)analytical estimates for initial guesses for some of the parameters.

    \subsubsection{External photon fields}
    \label{sec:external_photons}
\begin{table*}
    \centering
    \begin{tabular}{|c|c|c|}
    \toprule
    $L_{\mathrm{AD}}$, erg/s & Reference & Note \\
    \midrule
    $1.00 \times 10^{46}$ & \citet{2010ApJ...714L.303F} & From their assumed ratio $l_{\mathrm{Edd}} = L_{\mathrm{AD}}/L_{\mathrm{Eddington}}= 0.04$ and SMBH mass $M_{\mathrm{SMBH}} = 2 \times10^9 M_{\odot}$\\
    $2.51 \times 10^{46}$ & \citet{2024MNRAS.tmp.2529R} & From power-law + black-body fitting of the IR and optical photometric SED \\
    $3.0 \times 10^{46}$ & This work & Assumed value \\
    $3.0 \times 10^{46}$ & \citet{2010MNRAS.402..497G} & From multi-wavelength SED fitting with a single-zone leptonic model \\
    $3.31 \times 10^{46}$ & \citet{2012MNRAS.421.1764S, 2014MNRAS.441.3375X}& Multiplying their BLR luminosity by 10 \\
    $3.33 \times 10^{46}$ & \citet{2005MNRAS.361..919P} & Multiplying their BLR luminosity by 10 \\
    $4.47 \times 10^{46}$ & \citet{1999MNRAS.307..802C, 2014MNRAS.441.3375X}& Multiplying their BLR luminosity by 10 \\
    $4.79 \times 10^{46}$ & \citet{2006ApJ...637..669L} & Multiplying their BLR luminosity by 10\\
    $6.75 \times 10^{46}$ & \citet{2011MNRAS.410..368B} & Their assumed value \\
    \bottomrule
    \end{tabular}
    \caption{Various values of the 3C 454.3 AD luminosity $L_{\mathrm{AD}}$ present in the literature. \label{table:disc}}
\end{table*}

    Our model includes the following sources of the photon fields external to the moving blob (see also Fig. \ref{fig:geometry}):
    \begin{enumerate}
        \item[\textbf{Accretion disc:}] Direct radiation of the accretion disc (AD) in the model of \citet{1976MNRAS.175..613S} is calculated according to \citet[][sect. 3.1]{2009MNRAS.397..985G} and \citet[][appendix C1]{2024arXiv241105667E}.
        On various estimates of the 3C 454.3 AD luminosity in literature, see Table \ref{table:disc}. We adopt $L_{\mathrm{AD}} = 3 \times 10^{46}$~erg/s, which is approximately the median value of those given in Table \ref{table:disc}.
        \item[\textbf{Broad line region:}] Radiation intercepted and reprocessed by the BLR modelled as the superposition of the \textit{H} I Ly $\alpha$ (10.2~eV), \textit{He} II Ly $\alpha$ (40.8~eV) lines, and thermal continuum, according to the examples in the \texttt{AM}$^3$ documentation \citep{2023ascl.soft12031K, 2024ApJS..275....4K, 2024A&A...681A.119R}, with the continuum re-emitting $C_{\mathrm{BLR}} = 0.01$ \citep[e.g.,][]{1995ApJ...441...79B} of the AD luminosity, where $C_{\mathrm{BLR}}$ is the BLR continuum reprocessing factor. In addition, the BLR \textit{H} I Ly lines reprocess $\xi_{\mathrm{BLR}} = 0.10$ of the AD luminosity, and \textit{He} II Ly --- $0.5 \xi_{\mathrm{BLR}}$ of the AD luminosity, where $\xi_{\mathrm{BLR}}$ is the BLR covering factor \citep{1978ApJ...226....1B,DElia:2002ujp}. The lines are assumed to be emitted at the same $R_{\mathrm{BLR}}$ distance from the SMBH and to have the relative Gaussian widths of $5$\% of their central values. It is the \textit{H} I Ly line which is the most luminous while the continuum and other lines have significantly lower contribution to the external photon field density and have much higher uncertainty of their relative luminosity \citep[][appendix, and table 5]{2016ApJ...830...94F}. For simplicity, our BLR model follows \citet{2024A&A...681A.119R}, who followed \citet{2014PhRvD..90b3007M}, where they claim that their results depend weakly on the ratio of helium-over-hydrogen line luminosities, which is also observationally uncertain \citep[e.g.,][]{Zheng:1998mj}. We check that a factor-of-a-few change of the relative contribution of \textit{He} II Ly or BLR continuum to the external photon field density does not change the observable SED substantially, which confirms the claims of \citet{2014PhRvD..90b3007M}. We calculate the BLR outer radius using the standard relation \citep{2007ApJ...659..997K,2009MNRAS.397..985G}
        \begin{equation}
            R_{\mathrm{BLR}} = 1.0 \times 10^{17} \mathrm{\,cm\,} \left( \frac{L_{\mathrm{AD}}}{10^{45} \mathrm{\,erg/s}} \right)^{\frac{1}{2}},
        \label{eq:blr_radius}
        \end{equation}
        resulting in $R_{\mathrm{BLR}} = 5.48 \times 10^{17}$~cm in our case.
        \item[\textbf{Dust Torus:}] Radiation intercepted and reprocessed by the dust torus (DT) region is modelled as a thermal continuum with temperature $T_{\mathrm{DT}} = 370$~K\footnote{Typical values of $T_{\mathrm{DT}}$ are in between $100-2000$~K \citep{Cleary:2006pe,Malmrose:2011ne}. In our best-fitting ``standing-feature'' model the DT photon field is subdominant with respect to the BLR photon field, so a factor-of-a-few change of $T_{\mathrm{DT}}$ does not cause any substantial change in the observable SED.} and covering factor $C_{\mathrm{DT}} = 0.30$ according to the examples in the \texttt{AM$^{3}$} documentation \citep{2023ascl.soft12031K, 2024ApJS..275....4K, 2024A&A...681A.119R}. The DT outer radius is assumed to follow \citep{2000ApJ...545..107B,2002ApJ...577...78S,2009MNRAS.397..985G}
        \begin{equation}
            R_{\mathrm{DT}} = 2.5 \times 10^{18} \mathrm{\,cm\,} \left( \frac{L_{\mathrm{AD}}}{10^{45} \mathrm{\,erg/s}} \right)^{\frac{1}{2}}
        \label{eq:dt_radius}
        \end{equation}
        resulting in $R_{\mathrm{DT}} = 1.37 \times 10^{19}$~cm.
    \end{enumerate}
    The dependence of the SED and energy density of the BLR and DT photon fields in the blob rest frame on parameters $x$ and $\Gamma$ is modelled according to \citet[][sect. 3.3, 3.4]{2009MNRAS.397..985G} as done by \citet{2024ApJS..275....4K, 2024A&A...681A.119R}.

    \subsubsection{Escape timescales}
    \label{sec:escape}
    For a leptonic setup, the only particles involved in the simulation are photons and electrons. The timescale for photons escaping from the blob is
    \begin{equation}
        t^{\prime \mathrm{esc}}_{\gamma} = \frac{3}{4} \frac{R_b^\prime}{c} = 3 \times 10^5 \mathrm{\, s \,} \times \left( \frac{R_b^\prime}{10^{16} \mathrm{\, cm}} \right).
    \label{eq:gamma_escape}
    \end{equation}
    In principle, electrons can also be assumed to escape with the same timescale as photons (e.g., \citealp{2024A&A...681A.119R}, so-called stream-like escape). However, the period of gyration of the electrons equals
    \begin{equation}
        t^{\prime \mathrm{gyro}}_{e} = 7 \times 10^{-4} \mathrm{\, s\,} \times \left(\frac{E^\prime}{1 \mathrm{\, GeV}} \right) \left(\frac{B^\prime}{1 \mathrm{\,G}} \right)^{-1},
    \label{eq:gyroperiod}
    \end{equation}
   thus, $t^{\prime \mathrm{gyro}}_{e} \ll t^{\prime \mathrm{esc}}_{\gamma}$, i.e. electrons are confined in the magnetic field and are in the diffusion regime which can be characterised, e.g., by Bohm diffusion:
    \begin{equation}
        t^{\prime \mathrm{Bohm}}_{e} = 1 \times 10^{15} \mathrm{\, s} \times \left(\frac{R_b^\prime}{10^{16} \mathrm{\, cm}} \right)^2 \left( \frac{B^\prime}{1 \mathrm{\,G}} \right) \left( \frac{E^\prime}{1 \mathrm{\, GeV}} \right)^{-1}.
    \label{eq:bohm}
    \end{equation}
    As \citet{2018ApJ...854...54R} pointed out, the true electron escape timescale $t^{\prime \mathrm{esc}}_{e}$ should be somewhere in between $t^{\prime \mathrm{esc}}_{\gamma}$ and $t^{\prime \mathrm{Bohm}}_{e}$. However, when modelling the flaring behaviour, it is important rather to compare $t^{\prime \mathrm{esc}}_{e}$ and the total time of the simulation $t^{\prime}_{\mathrm{sim}}$. We assume $t^{\prime \mathrm{esc}}_{e} \gg t^{\prime}_{\mathrm{sim}}$, and we can neglect the dependence of the electron escape timescale on their energy assuming just a large enough factor $\eta_{e} \sim 5 \times 10^3$ by which the electron escape timescale is greater than the photon one:
    \begin{equation}
        t^{\prime \mathrm{esc}}_{e} = \eta_{e} t^{\prime \mathrm{esc}}_{\gamma}.
    \label{eq:electron_escape}
    \end{equation}
    We verify that the confined electrons after $t^{\prime}_{\mathrm{sim}}$, when the injection is stopped, do not produce any substantial signal which could affect the following days. This is expected because, in the energy range of electrons responsible for observable electromagnetic emission, they have an energy-loss timescale much shorter than the simulation timescale.

    \subsubsection{Lorentz and Doppler factors}
    \label{sec:lorentz}

    \citet{2013ApJ...773..147J} observed the flare of 3C 454.3 in radio waves in the period of MJD 55518--55568 using the very-long baseline interferometry (VLBI) and obtained the estimation of the Lorentz factor $\Gamma = 26 \pm 3$. Based on these observations, we use $\Gamma = 25$ as our initial guess and the corresponding Doppler factor $\delta$ calculated with Eq. (\ref{eq:observed_doppler}). 
    
    \subsubsection{Blob radius}
    \label{sec:blob_radius}

    \citet{2019ApJ...877...39M} investigated the same flare of 3C 454.3 analysing the \textit{Fermi}-LAT data. In their fit of the flare light curve, they estimated the observed flare decay time $\tau^{o}_{\mathrm{decay}} = 2.6 \pm 1.0$~h, see their table 5. Assuming that this time is related to the size of the emitting region, we can estimate the blob radius as
    \begin{equation}
        R^{\prime}_{b} = \frac{c \tau^{o}_{\mathrm{decay}} \delta}{1 + z}.
    \label{eq:decay_timescale}
    \end{equation}
    For $\Gamma=25$, $\delta = 50$, $z = 0.859$, we obtain $R^{\prime}_{b} = 7.5 \times 10^{15}$~cm, which we can use as an initial guess for the blob radius.

    \subsubsection{Location of the $\gamma$-ray emitting region}
    \label{sec:dissipation_radius}

    In this subsection, we obtain an analytical estimate of $x$ to use as an initial guess before the \texttt{iMINUIT} optimisation. The distance $x$ of the region emitting the observed radiation from the SMBH establishes the energy at which the high-energy $\gamma$ rays will be absorbed by the photons of the BLR photon field.

    First, for a blob with the Lorentz factor of $\Gamma = 25$ and redshift $z = 0.859$, we calculate the $\gamma\gamma$ interaction rate $r^{\prime}_{\gamma\gamma}(E^{\prime}_{\gamma}, x)$ on the photon field presented in Sect.~\ref{sec:external_photons} for 100 positions of the blob in between $x = R_{\mathrm{BLR}}$ and $x = 2R_{\mathrm{BLR}}$ following the prescriptions of \citet{2012CoPhC.183.1036K} and their equations (5), (6). Then for each value of $x$, we integrate the obtained interaction rate $r^{\prime}_{\gamma\gamma}(E^{\prime}_{\gamma}, x)$ along the LoS to obtain the value of the photon absorption optical depth $\tau_{\gamma\gamma}$ for $\gamma$ rays with the observable energy $E^{o}_{\gamma}$ emitted at distance $x$ from the SMBH:
    \begin{equation}
        \tau_{\gamma\gamma}(E^{o}_{\gamma}, x) = \int_{x}^{+\infty} r^{\prime}_{\gamma\gamma}(E^{\prime}_{\gamma}, y) dy,
    \end{equation}
    \begin{figure}
         \includegraphics[width=1\columnwidth]{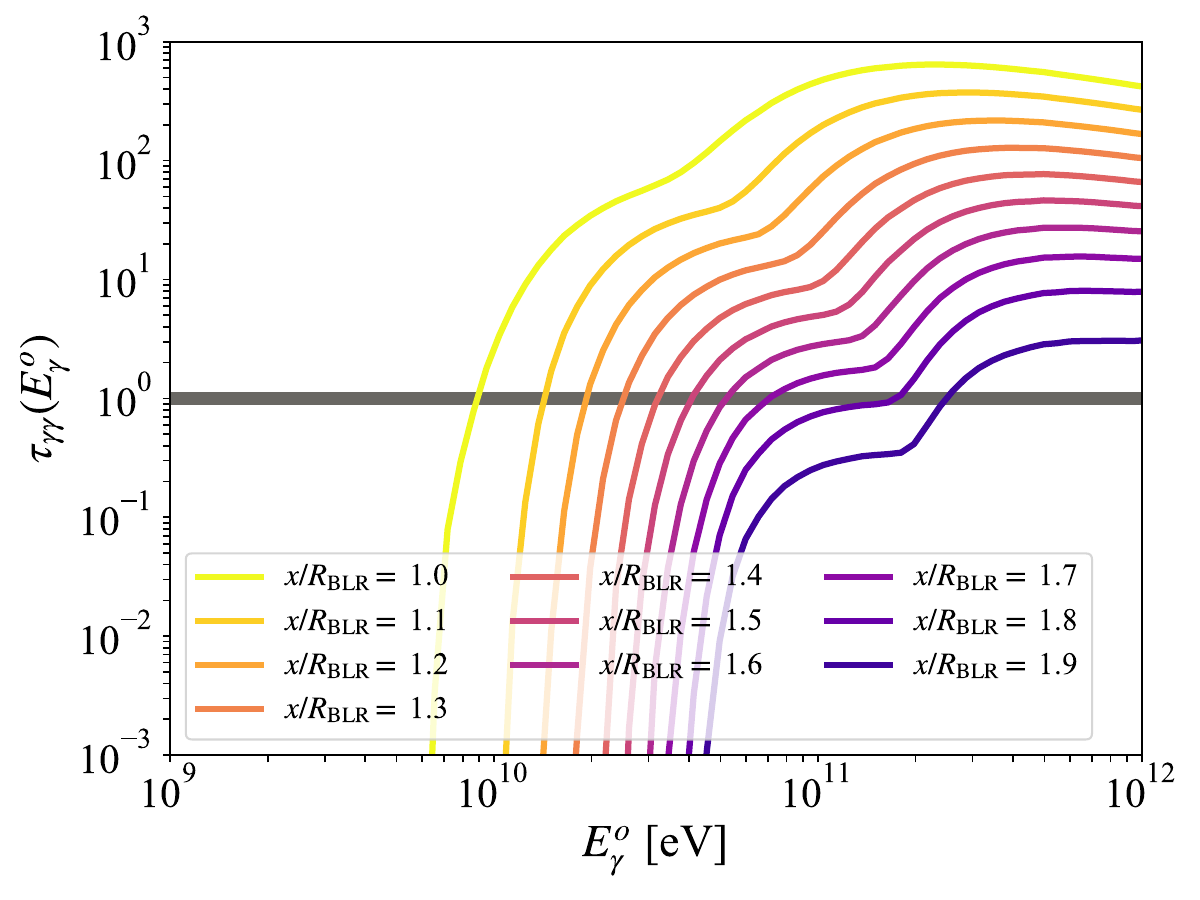}
         \caption{The optical depth of $\gamma\gamma$ pair production for $\gamma$ rays with the observed (at Earth) energy $E^{o}_{\gamma}$ emitted in a blob with $\Gamma = 25$ and $z = 0.859$ at a particular distance $x$ from the SMBH as indicated in the legend.}
         \label{fig:gamma_gamma_optical_depth}
    \end{figure}
    \begin{figure}
         \includegraphics[width=1\columnwidth]{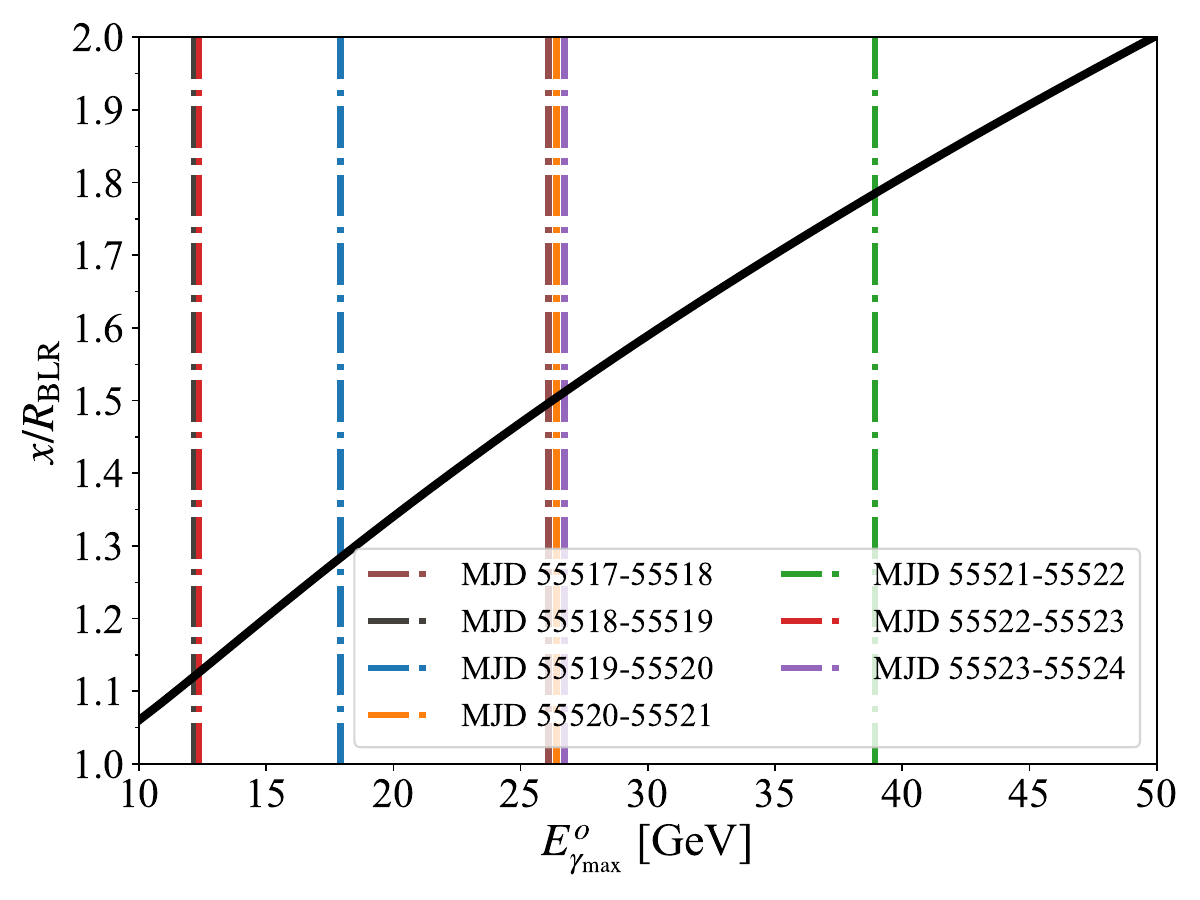}
         \caption{The dependence of the dissipation radius $x$ on the maximum observed (on Earth) $\gamma$-ray energy assuming it corresponds to pair production at the threshold on photons from the \textit{He} II Ly $\alpha$ line in the blob with Lorentz factor $\Gamma = 25$ at redshift $z = 0.859$. The dashed-dotted vertical lines show the central energies of the last observed \textit{Fermi}-LAT bins during one-day periods as indicated in the legend (closely plotted lines correspond to the same energy but slightly shifted for better visibility).}
         \label{fig:maximum_observed_energy}
    \end{figure}
    \begin{figure}
         \includegraphics[width=1\columnwidth]{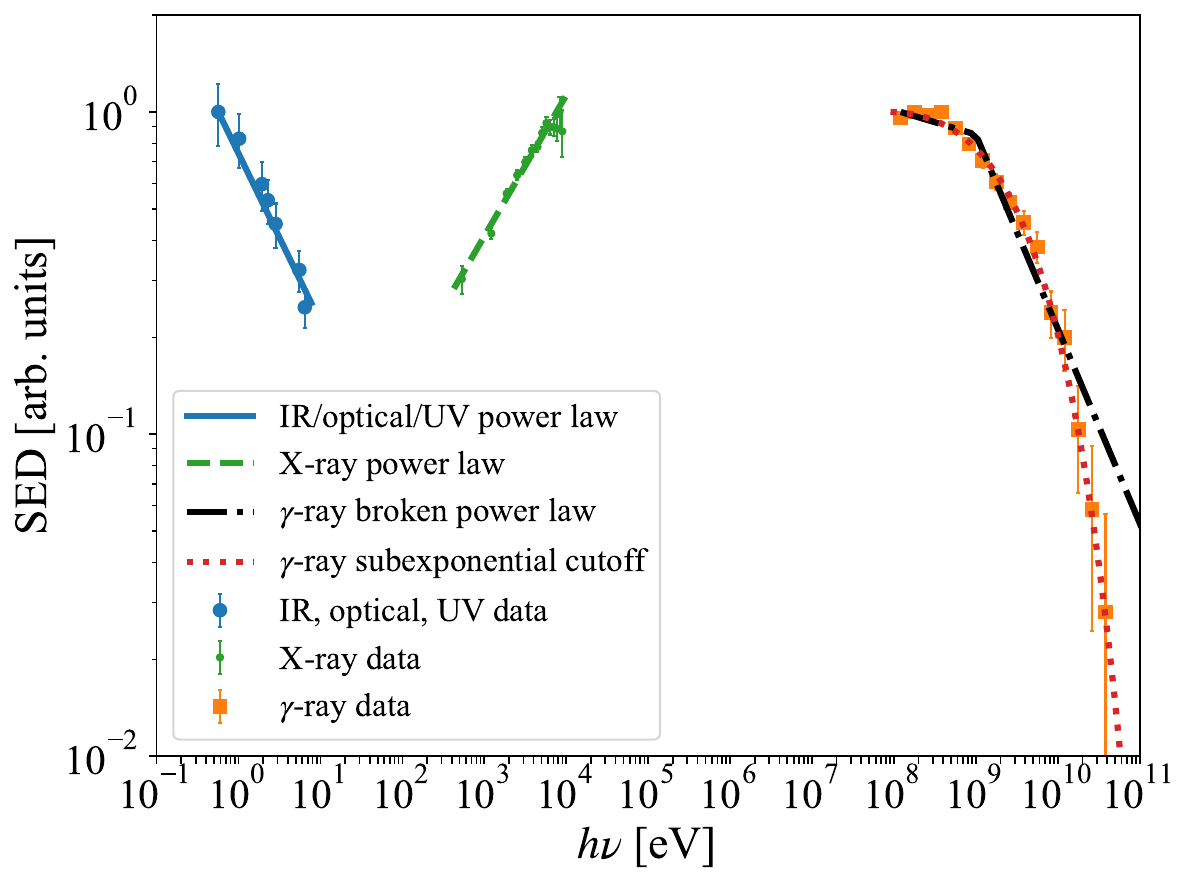}
         \caption{Phenomenological fits to the synchrotron, X-ray, and $\gamma$-ray parts of the SED averaged over 55517--55524. Data for each energy segment are normalized by the maximum value of the SED in the energy range. See Table~\ref{table:game_of_slopes} for the details of the fits.}
         \label{fig:game_of_slopes}
    \end{figure}
    \begin{table}
        \centering
        \begin{tabular}{|l|c|}
        \toprule
        Parameter & Value \\
        \midrule
        \multicolumn{2}{c}{IR/optical/UV power law} \\
        $\Gamma_{\gamma}$ & $2.52 \pm 0.03$\\
        $\chi^{2}_{\mathrm{red}}$ & $0.39$\\
        \midrule
        \multicolumn{2}{c}{X-ray power law} \\
        $\Gamma_{\gamma}$ & $1.562 \pm 0.020$\\
        $\chi^{2}_{\mathrm{red}}$ & $1.17$\\
        \midrule
        \multicolumn{2}{c}{$\gamma$-ray broken power law} \\
        $\Gamma_{\gamma}$ (before the break) & $2.076 \pm 0.014$\\
        $\Gamma_{\gamma}$ (after the break) & $2.60 \pm 0.04$\\
        $E_{\mathrm{break}}$ [GeV] & $1.00 \pm 0.09$\\
        $\chi^{2}_{\mathrm{red}}$ & $2.63$\\
        \midrule
        \multicolumn{2}{c}{$\gamma$-ray subexponential cutoff}\\
        $\gamma_{1}$ & $1.896 \pm 0.026$\\
        $\gamma_{b}$ & $0.49 \pm 0.04$\\
        $E_{\mathrm{cut}}$ [GeV] & $1.85 \pm 0.16$\\
        $\chi^{2}_{\mathrm{red}}$ & $0.80$\\
        \bottomrule
        \end{tabular}
        \caption{Details of the fits to the normalised SEDs shown in Fig. \ref{fig:game_of_slopes}. $\Gamma_{\gamma}$ is the power-law slope of the photon intensity $dN_{\gamma} / dE_{\gamma} \propto E_{\gamma}^{-\Gamma_{\gamma}},$ $E_{\gamma} \equiv h \nu$. Parameters $\gamma_{1}$ and $\gamma_{b}$ are from the parameters of the subexponential function described in Eq. (\ref{eq:superexp}) \label{table:game_of_slopes}}
    \end{table}
    where $E^{o}_{\gamma} = \delta E^{\prime}_{\gamma} / (1 + z)$.
    The obtained optical depth for ten values of $x$ is plotted in Fig.~\ref{fig:gamma_gamma_optical_depth}. The sharp rise of the optical depth happens due to the absorption of $\gamma$ rays on the \textit{He} II Ly $\alpha$ line (40.8~eV) and the second rise -- due to the beginning of pair production on the \textit{H} I Ly $\alpha$ line (10.2~eV). From Fig.~\ref{fig:gamma_gamma_optical_depth}, one can see that for $x \lesssim 1.7 R_{\mathrm{BLR}}$ the optical depth on the \textit{He} II Ly $\alpha$ line is greater than unity.
    
    To obtain an analytical estimate of $x$, we derive analytically the equation linking $x$ with the maximum energy $E^{o}_{\gamma_{\mathrm{max}}}$ among the observed $\gamma$-ray energies assuming it corresponds to the threshold of pair production on \textit{He} II Ly $\alpha$ line. In the blob frame a $\gamma$-ray with energy $E^{\prime}_{\gamma}$ collides with a photon with energy $E^{\prime}_{\gamma_{2}}$ and creates an electron-positron pair at the threshold \citep{1934PhRv...46.1087B, Nikishov1962,1967PhRv..155.1404G}
    \begin{equation}
        E^{\prime}_{\gamma} E^{\prime}_{\gamma_{2}} = \frac{2 m^2_e c^4}{1 - \cos \zeta^{\prime}},
    \label{eq:pair_creation_threshold}
    \end{equation}
    where $\zeta^{\prime}$ is the angle between the momenta of the photons. The photon energy $\epsilon$ is set in the host galaxy rest frame and related to the energy in the blob frame $E^{\prime}_{\gamma_{2}}$ as \citep{1993ApJ...416..458D,1995ApJ...441...79B,2002ApJ...575..667D}
    \begin{equation}
        E^{\prime}_{\gamma_{2}} = \epsilon \left[ \Gamma (1 - \beta \cos \zeta) \right] = \frac{\epsilon}{\Gamma (1 + \beta \cos \zeta^{\prime})},
    \label{eq:line_transformation}
    \end{equation}
    where the cosines in the two frames are related as
    \begin{equation}
        \cos \zeta^{\prime} = \frac{\cos \zeta - \beta}{1 - \beta \cos \zeta}.
    \label{eq:cosine_transformation}
    \end{equation}
    Assuming the blob is located at $x > R_{\mathrm{BLR}}$ and, following \citet{2023ascl.soft12031K, 2024A&A...681A.119R},\footnote{\url{https://am3.readthedocs.io/en/latest/examples/blazar_detailed_example.html}} that the photons coming from the largest angle between the jet axis and the edge of the BLR sphere contribute the most to the BLR photon field observed in the blob (see Fig. \ref{fig:geometry}), and the $\gamma$-ray colliding with the photon propagates along the jet axis at zero angle,
    \begin{equation}
        \sin \zeta = \frac{R_{\mathrm{BLR}}}{x}.
    \label{eq:sin_xi}
    \end{equation}
    Combining Eqs. (\ref{eq:pair_creation_threshold}),
    (\ref{eq:line_transformation}), (\ref{eq:cosine_transformation}), and (\ref{eq:sin_xi}) and taking into account the redshift $z$ and the Doppler factor $\delta$, we obtain the estimate of $x$ given the energy $E^{o}_{\gamma_{\mathrm{max}}}$ of the $\gamma$-ray observed on Earth corresponding to the threshold of pair-production on the line with energy $\epsilon$:
    \begin{equation}
        x = R_{\mathrm{BLR}} \left[1 - \left( 1 -\frac{2 m^2_e c^4}{\epsilon E^{o}_{\gamma_{\mathrm{max}}}} \frac{\delta}{\Gamma (1 + z) (1 + \beta)} \right)^{2} \right]^{-\frac{1}{2}}.
    \label{eq:maximum_observed_energy}
    \end{equation}
    We show as the black solid curve the graph of function (\ref{eq:maximum_observed_energy}) in Fig.~\ref{fig:maximum_observed_energy} for $\Gamma = 25$, and $\epsilon = 40.8$~eV corresponding to the \textit{He} II Ly $\alpha$ line. In Fig.~\ref{fig:maximum_observed_energy}, we superimpose vertical lines showing the central energies of the last bins observed with \textit{Fermi}-LAT in each of the seven days of MJD 55517--55524. From Fig.~\ref{fig:maximum_observed_energy}, we take $x = 1.5 R_{\mathrm{BLR}}$ as our initial guess for the dissipation radius.
    
    \subsubsection{Magnetic field and parameters of the electron injection spectrum}\label{sec:injection}

    In this subsection, we infer initial guesses for the root-mean-square magnetic field strength $B^{\prime}$, the power-law slope of the injected electron spectrum $s_e$ and the maximum electron injection energy $E^{\prime \mathrm{max}}_e$. To do so, we use the SED averaged over the seven-day period of MJD 55517--55524. To focus on the shapes of the parts of the observed SED, we present it normalized by the maximum values as data points with error bars in each energy band (IR/optical/UV photons, X rays, and $\gamma$ rays) in Fig.~\ref{fig:game_of_slopes}.

    \begin{figure}
         \includegraphics[width=1\columnwidth]{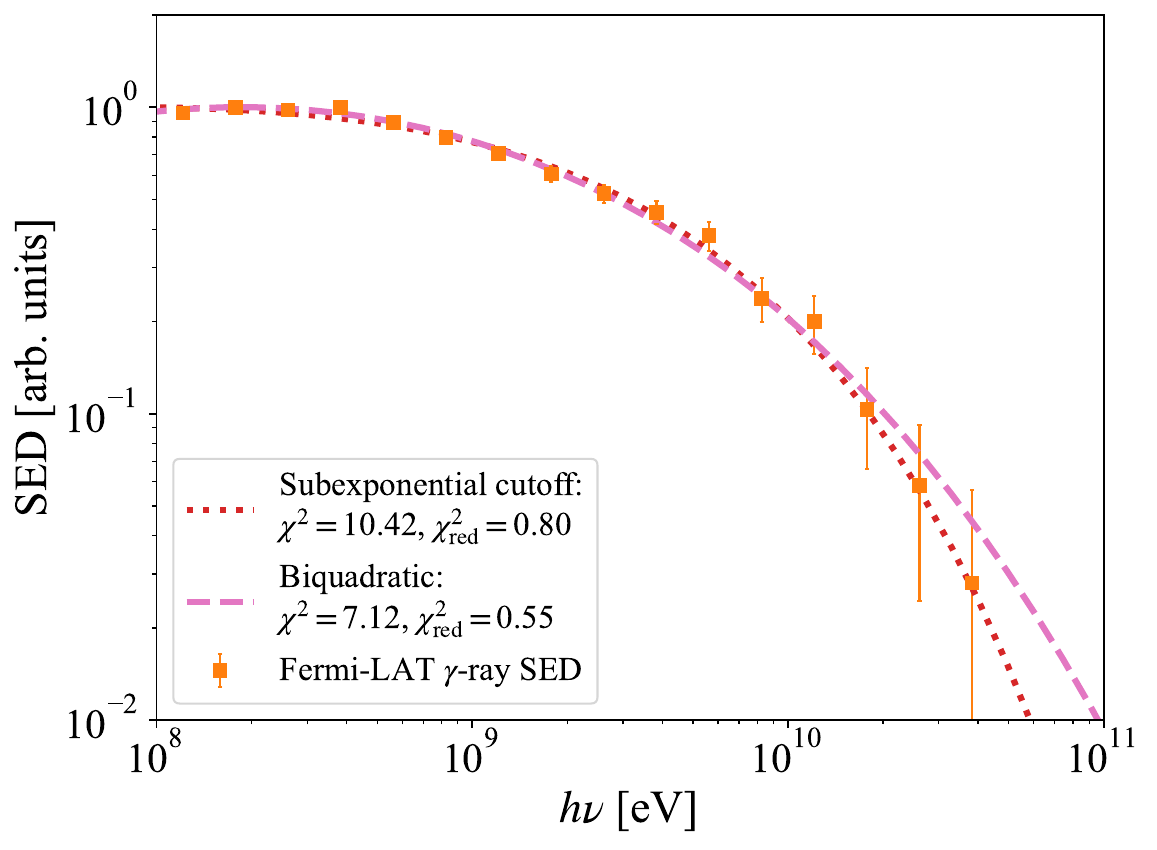}
         \caption{The results of the fits to the normalised \textit{Fermi}-LAT SED averaged over MJD 55517--55524 by subexponential cutoff (Eq. \ref{eq:superexp}) and biquadratic function (Eq. \ref{eq:quartic}).}
         \label{fig:superexp_VS_quartic}
    \end{figure}
    \begin{figure}
         \includegraphics[width=1\columnwidth]{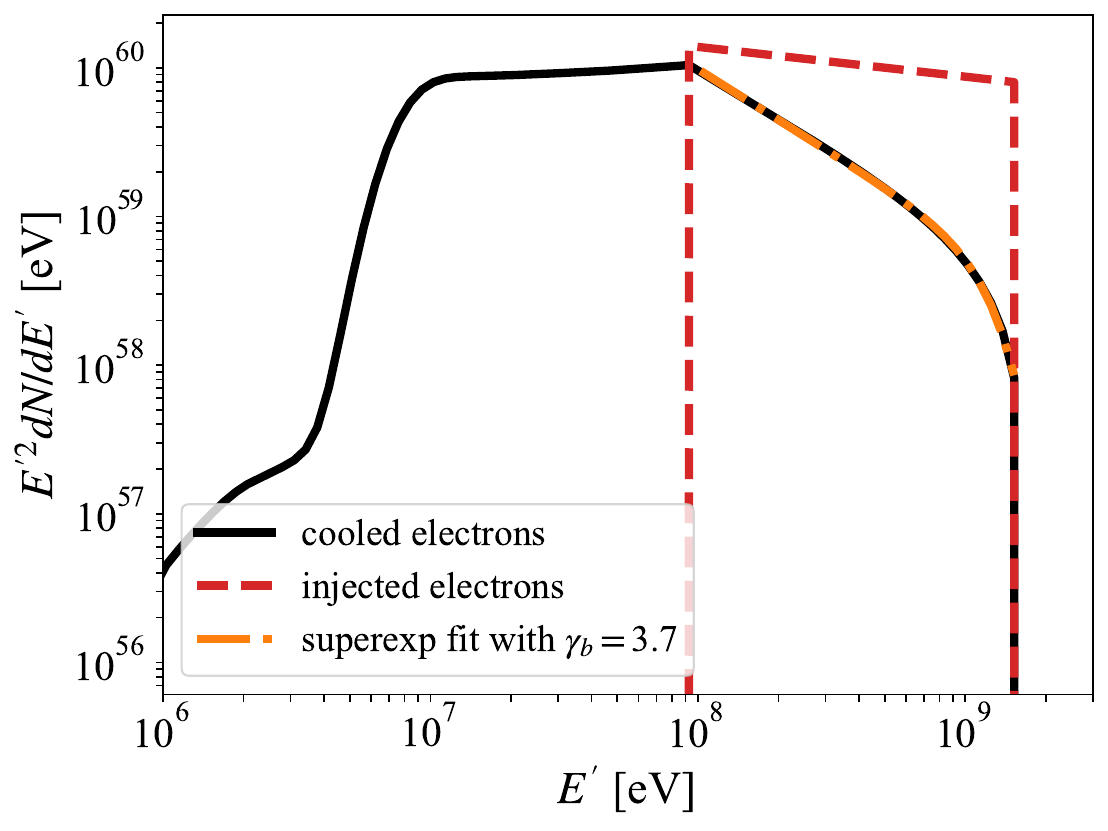}
         \caption{The SEDs of the injected and cooled electrons and the fit with Eq. (\ref{eq:superexp}) to the cooled electron SED.}
         \label{fig:electron_sed}
    \end{figure}
    
    Using \texttt{iMINUIT}, we obtain functional forms fitting the observed bands of the SED presented in Fig.~\ref{fig:game_of_slopes} as curves. For the first two bands, we obtain good fits with simple power laws with the parameter values and their errors presented in Table~\ref{table:game_of_slopes} along with the values of $\chi^2_{\mathrm{red}} = \chi^2 / \mathrm{ndf}$, where ndf stands for the number of degrees of freedom in a particular fit.

    For the \textit{Fermi}-LAT data, it is clear that they cannot be described by a single power law \citep[see also, e.g.,][]{2019MNRAS.486.1781R,2021MNRAS.504.5074S}. We failed to obtain a good fit with a broken power law (BPL) with $\chi^2_{\mathrm{red}} = 2.63$. On the contrary, a fit with a subexponential cutoff (compare with eq. \ref{eq:superexp_dn_de}), 
    \begin{equation}
        \mathrm{SED}(E) = E^{2} \frac{dN}{dE} = E^2 N_{0} \left(\frac{E}{E_0} \right)^{-\gamma_{1}} \exp{\left(-\left(\frac{E}{E_{\mathrm{cut}}} \right)^{\gamma_{b}}\right)},
    \label{eq:superexp}
    \end{equation}
        describes the \textit{Fermi}-LAT range quite well ($\chi^2_{\mathrm{red}} = 0.80$). We interpret this as evidence for the curvature of the spectrum reported previously by \citet{2011ApJ...733L..26A, 2014ApJ...790...45P,2021MNRAS.504.5074S}. Indeed, in the 4FGL catalogue itself, the spectral model for 3C 454.3 is chosen to have a subexponential cutoff \citep{2020ApJS..247...33A} providing the best description of the eight-year data. Besides, \citet{2014ApJ...794....8S} from the analysis of almost five years of \textit{Fermi}-LAT data, found that the SED of 3C 454.3 is best described by a biquadratic function
    \begin{equation}
        \mathrm{SED}(E) = F_{0} \exp{\left(-A \ln^2 \left(E / E_{\mathrm{peak}} \right) -B \ln^4 \left(E / E_{\mathrm{peak}} \right) \right)}
    \label{eq:quartic}
    \end{equation}
    with $A \approx 0$. We argue that:
    \begin{enumerate}
        \item SEDs described by Eqs. (\ref{eq:superexp}) and (\ref{eq:quartic}) can be practically indistinguishable from each other under certain values of the parameters (see Fig. \ref{fig:superexp_VS_quartic});
        \item SEDs described by Eqs. (\ref{eq:superexp}) or (\ref{eq:quartic}) can be a result of the IC scattering of cooled electrons with small maximum injection energies \citep{2012ApJ...753..176L,2018MNRAS.477.4749C} -- see below and Figs. \ref{fig:superexp_VS_quartic} and \ref{fig:electron_sed}.
    \end{enumerate}

    If we assume that photons with maximum observed energy $E^{o}_{\gamma_{\mathrm{max}}}$ are emitted as a result of the IC scattering of electrons at the end of their spectrum, then we can estimate the maximum electron injection energy as
    \begin{equation}
        E^{\prime \mathrm{max}}_e = \frac{1 + z}{\delta} E^{o}_{\gamma_{\mathrm{max}}}
    \label{eq:maximum_electron_injection_energy}
    \end{equation}
    in the Klein-Nishina regime, or, in the Thomson regime \citep[e.g.,][eq. 6.52]{Dermer:2009zz},
    \begin{equation}
        E^{\prime \mathrm{max}}_e = m_e c^{2} \sqrt{\frac{3 E^{o}_{\gamma_{\mathrm{max}}} (1 + z)}{4 \delta \langle \epsilon_{\mathrm{BLR}}^{\prime} \rangle}},
    \label{eq:maximum_electron_injection_energy_thomson}
    \end{equation}
    where $\langle \epsilon_{\mathrm{BLR}}^{\prime} \rangle$ is the average energy of BLR photons in the blob rest frame, which can be estimated with Eqs. (\ref{eq:line_transformation}), and (\ref{eq:cosine_transformation}) for $\epsilon \approx 20$~eV corresponding to the average between \textit{H} I Ly $\alpha$ and \textit{He} II Ly $\alpha$ lines. For the observed maximum photon energy $E^{o}_{\gamma_{\mathrm{max}}} \approx 4~\times~10^{10}$~eV, we obtain $E^{\prime \mathrm{max}}_e \approx 1.5~\times~10^9$~eV according to Eq. (\ref{eq:maximum_electron_injection_energy}) and a similar value of $E^{\prime \mathrm{max}}_e \approx 1.7~\times~10^9$~eV according to Eq. (\ref{eq:maximum_electron_injection_energy_thomson}). To be more conservative, we choose the smaller value as our initial guess\footnote{We also check that the electron inverse Compton scattering off the \textit{He} II Ly $\alpha$ photons enters the Klein-Nishina regime while the scattering off the \textit{H} I Ly $\alpha$ photons is still in the Thomson regime.}.

    From the broken power-law fit to the \textit{Fermi}-LAT data in Fig. \ref{fig:game_of_slopes}, we can assume the power-law slope of the \textit{cooled} electron spectrum $dN_{e}/dE_{e} \propto E^{-p_{e}}$ to be \citep{1970RvMP...42..237B,Rybicki1979rpa..book}
    \begin{equation}
        p_e = 2\Gamma_{\gamma} - 1,
    \label{eq:electron_and_photon_spectral_index_relation}
    \end{equation}
    so the slope of the \textit{injected} electron spectrum $dN_{e}/dE_{e} \propto E^{-s_{e}}$ will be
    \begin{equation}
        s_e = p_e - 1 = 2\Gamma_{\gamma} - 2,
    \end{equation}
    i.e. $s_e \approx 2.2$ for $\Gamma_{\gamma} \approx 2.1$ (see Table~\ref{table:game_of_slopes}).

    Assuming that the UV photons at the energy $E^{\mathrm{o}}_{\mathrm{UV}} \approx 4$~eV are exclusively produced by the cooled electrons at the cutoff $E^{\prime \mathrm{max}}_e$, we can estimate the root-mean-square magnetic field strength as \citep[e.g.,][eq. 6.17c]{Rybicki1979rpa..book}
    \begin{equation}
        B^{\prime} \approx \frac{4 \pi m_e^3 c^5 E^{\mathrm{o}}_{\mathrm{UV}} (1 + z)}{3 h q_e \delta \left( E^{\prime \mathrm{max}}_e \right)^{2}},
    \label{eq:magnetic_field}
    \end{equation}
    which gives $B^{\prime} \approx 1$~G in our case.

    For the initial guess of the minimum electron injection energy $E^{\prime \mathrm{min}}_e$, we assume that the observed by \textit{Fermi}-LAT peak of the $\gamma$-ray SED at around $E^{o}_{\gamma_{\mathrm{peak}}} \approx 2 \times 10^{8}$~eV is due to the break in the cooled electron spectrum happening at $E^{\prime \mathrm{min}}_e$, which, given the good approximation of the IC process by the Thomson formula at lower energies, can be estimated with the same Eq.~(\ref{eq:maximum_electron_injection_energy_thomson}) as for the maximum electron injection energy, just by substituting $E^{o}_{\gamma_{\mathrm{max}}}$ with $E^{o}_{\gamma_{\mathrm{peak}}}$. This gives us the initial guess $E^{\prime \mathrm{min}}_e \approx 10^8$~eV.

    Finally, for the last parameter --- electron injection luminosity --- we set $L^{\prime}_{e} = 2 \times 10^{43}$~erg/s as our initial guess, just comparing the model and observed SEDs by eye, trying several values of $L^{\prime}_{e}$.

    With the model parameters' values from our initial guess, we run the \texttt{AM$^{3}$} program with the total simulation time according to Eq. (\ref{eq:simulation_time}). The injected and cooled electron SEDs at the end of the simulation are shown in Fig. \ref{fig:electron_sed} along with the fit of the functional form of Eq. (\ref{eq:superexp}) to the high-energy part of the cooled electron SED, resulting in excellent agreement with the superexponential cutoff with $\gamma_b = 3.7$. The \textit{super}exponential cutoff of the \textit{cooled} electron SED in Fig. \ref{fig:electron_sed} results in the \textit{sub}exponential cutoff of the $\gamma$-ray SED with $\gamma_{b} = 0.49$ (see Fig.~\ref{fig:superexp_VS_quartic} and Table~\ref{table:game_of_slopes}), which is in agreement with the expected value from \citet{2012ApJ...753..176L} for a case in between the synchrotron ($\gamma_{b} = 3.7 / (3.7 + 4) = 0.48$) and Planckian ($\gamma_{b} = 3.7 / (3.7 + 2) = 0.65$) photons serving as the target photon fields \citep[][table 1]{2012ApJ...753..176L} for the IC in the Thomson regime.

    The results of the ``standing-feature'' leptonic model are presented in Sect.~\ref{sec:standing_blob_model_results}.

\subsection{Description of the ``moving-blob'' model}
\label{sec:moving_blob_model_description}

    We assume that the blob is launched at the distance $x_s$ from the SMBH with a Lorentz factor $\Gamma$ and an angle $\theta_{\mathrm{view}}$ between the blob velocity and the LoS. At the end of the simulation, the blob reaches distance $x_f$ from the SMBH. The total simulation time is $t^{\prime}_{\mathrm{sim}}$ in the blob rest frame, $t_{\mathrm{tot}}$ in the object's host galaxy rest frame, and $t^{o}_{\mathrm{tot}}$ in the observer's rest frame (equal the time between the receipt by the observer of the first photon emitted from the position $x_s$ and the last photon emitted from the position $x_f$). These are related as follows \citep[see, e.g.,][sect. 2.1]{2015PhR...561....1K}:
    \begin{equation}
        t_{\mathrm{tot}} = \frac{x_f - x_s}{\beta c},
    \label{eq:total_time_of_blob_motion}
    \end{equation}
    \begin{equation}
        t^{\prime}_{\mathrm{sim}} = \frac{t_{\mathrm{tot}}}{\Gamma},
    \label{eq:simulation_time_moving_blob}
    \end{equation}
    \begin{equation}
        t^{o}_{\mathrm{tot}} = (1+z) \frac{t^{\prime}_{\mathrm{sim}}}{\delta} = (1 + z) \frac{t_{\mathrm{tot}}}{\Gamma \delta} = (1+z) (1 - \beta \cos \theta_{\mathrm{view}}) t_{\mathrm{tot}}.
    \label{eq:relation_between_times}
    \end{equation}
    If observations start not exactly at the moment when the first photon emitted by the blob from $x_s$ has reached the Earth, but with some time shift $t^{o}_{\mathrm{shift}}$ between the moment of the beginning of the observations and the moment of the first photon reaching the Earth (albeit not registered due to the delayed observations), then
    \begin{equation}
        t^{o}_{\mathrm{tot}} = t^{o}_{\mathrm{shift}} + t^{o}_{\mathrm{obs}},
    \label{eq:time_shift}
    \end{equation}
    where $t^{o}_{\mathrm{obs}}$ is the total duration of the observations in the Earth frame. Fitting the model SED to the data, we keep $t^{o}_{\mathrm{shift}}$, $x_s$, $x_f$ as free parameters, whereas $\theta_{\mathrm{view}}$ is fixed, and $t^{o}_{\mathrm{obs}}$ is also fixed equal to the actual duration of observations over which the observed SED is averaged (7 d if not stated otherwise). Thus, $\beta$, hence $\Gamma$ and $\delta$ as well, become dependent variables following (see Eqs. \ref{eq:total_time_of_blob_motion}, \ref{eq:relation_between_times}, and \ref{eq:time_shift})
    \begin{equation}
        \beta = \left({\frac{c(t^{o}_{\mathrm{obs}} + t^{o}_{\mathrm{shift}})}{(1+z)(x_f - x_s)}  + \cos \theta_{\mathrm{view}}}\right)^{-1}.
    \end{equation}
    In the \texttt{AM$^{3}$} simulation, we set the number of time steps equal to $N_s = 100$. At each time step the current position of the blob is updated according to the equation (see Eqs. \ref{eq:total_time_of_blob_motion} and \ref{eq:simulation_time_moving_blob})
    \begin{equation}
        x(t^\prime) = x_s + \beta c \Gamma t^{\prime}.
    \label{eq:x_moving_blob}
    \end{equation}
    Eq. (\ref{eq:x_moving_blob}) is valid only in the case of the constant Lorentz factor $\Gamma$; we briefly discuss the case of a varying $\Gamma$ in Sect.~\ref{sec:moving_blob_model_results}.
    
    All the external parameters dependent on $x$ are updated at the beginning of each step. The photon fields are calculated in the same way as in Sect. \ref{sec:external_photons}. The escape timescales are calculated in the same way as in Sect. \ref{sec:escape}. The snapshot of the model SED$(E; t^{\prime}_{k})$ is stored at each $k$th time step. Then, for each $j$th day of the flare (from first to seventh), following Eqs. (\ref{eq:relation_between_times}) and (\ref{eq:time_shift}), we calculate the model SED averaged over all those times $t^{\prime}_k$ in the blob frame which satisfy the condition
    \begin{equation}
        j - 1 \leq \frac{t^{\prime}_k (1+z)/\delta - t^{o}_{\mathrm{shift}}}{1 \mathrm{\, d}} \leq j,
    \end{equation}
    where $k=1,2,...,N_s$. We assume the last observation period stops exactly at the moment when the photon emitted from the blob position at $x_f$ arrives at the Earth, i.e. there is no time shift between the end of the observations and the end of the simulation.

    We assume that electrons are injected into the blob at a rate
    \begin{equation}
    \left( \frac{d \dot{N}_e^\prime}{dE_e^\prime} \right)_{\mathrm{inj}}= 
    \begin{cases}
        \dot{N}_{e}^{\prime} \left( \frac{E^{\prime}_{e}}{E^{\prime}_{\mathrm{ref}}} \right)^{-s_e}, & E^{\prime{\mathrm{min}}}_{e} \leq  E^{\prime}_{e} \leq E^{\prime{\mathrm{max}}}_{e}, \\
        0, & \text{otherwise},
    \end{cases}
    \end{equation}
    where $\dot{N}_{e}^{\prime} \equiv dN_{e}^{\prime}/dt^{\prime}$, normalization factor $\dot{N}_{e}^{\prime}$ is adjusted corresponding to the value of the electron injection luminosity $L^{\prime}_{e}$ serving as a free parameter along with the minimum injection energy $E^{\prime{\mathrm{min}}}_{e}$, the spectral index $s_e$, and the maximum injection energy $E^{\prime{\mathrm{max}}}_{e}$, whereas the reference energy $E^{\prime}_{\mathrm{ref}}$ is fixed. We allow the injection electron spectrum to vary over time following
    \begin{equation}
        \dot{N}_{e}^{\prime}(t^\prime) =  \dot{N}_{e}^{\prime}(0)  \left( \frac{x(t^\prime)}{x_{e_{0}}} \right)^{\alpha_{e} \Theta(t^{\prime} - t^{\prime}_{e_0})},
    \end{equation}
    where $\Theta$ denotes the Heaviside step function, $t^{\prime}_{e_0}$ is the time when the injection starts changing, $\alpha_{e}$ is a free parameter, $x_{e_{0}}$ is the position of the blob when $t^{\prime} = t^{\prime}_{e_0}$. This parametrisation is chosen to allow the electron injection rate to adjust to changing external photon fields, with the increasing rate (in case $\alpha_{e} > 0$) to compensate for decreasing photon field density of the BLR photons in the blob frame, while the term $\Theta(t^{\prime} - t^{\prime}_{e_0})$ allows to adjust when (and if) the change of the electron injection rate happens.
    We also allow the blob radius to vary over time according to a similar equation
    \begin{equation}
        R^{\prime}_{b}(t^\prime) = R^{\prime}_{b}(0) \left( \frac{x(t^\prime)}{x_{b_{0}}} \right)^{\alpha_{b} \Theta(t^{\prime} - t^{\prime}_{b_0})},
    \end{equation}
    where $t^{\prime}_{b_0}$ denotes the time when the blob radius starts changing, $\alpha_{b}$ is a free parameter, $x_{b_{0}}$ is the position of the blob when $t^{\prime} = t^{\prime}_{b_0}$. This parametrisation is chosen to reflect the luminal blob expansion in case $\alpha_{b} = 1$ or subluminal expansion in case $\alpha_{b} < 1$ (cf., e.g, \citealp{2023A&A...669A.151Z}) while $\Theta(t^{\prime} - t^{\prime}_{b_0})$ allows to adjust when (and if) the expansion begins.
    
    In total, the ``moving-blob'' model has 13 free parameters: $L^{\prime}_{e}(0)$, $s_e$, $E^{\prime{\mathrm{min}}}_{e}$, $E^{\prime{\mathrm{max}}}_{e}$, $B^{\prime}$, $x_s$, $x_f$, $R^{\prime}_{b}(0)$, $t^{\prime}_{e_0}$, $t^{\prime}_{b_0}$, $t^{o}_{\mathrm{shift}}$, $\alpha_{e}$, and $\alpha_{b}$. The transformation of the SED from the blob frame into the observer frame for a single blob propagating along the jet is described by the same Eq. (\ref{eq:sed_transformation}) as for the ``standing-feature'' model satisfying Eq. (\ref{eq:number_of_blobs}) as shown by \citet[][appendix B]{1997ApJ...484..108S}.

    The results of the ``moving-blob'' leptonic model are presented in Sect.~\ref{sec:moving_blob_model_results}.

\section{Results of the leptonic models}\label{sec:results_leptonic}
    \subsection{Results of the ``standing-feature'' leptonic model}
    \label{sec:standing_blob_model_results}
    \begin{figure*}
    \centering
    \subfloat[MJD 55517--55518]{\includegraphics[width=0.47\textwidth]{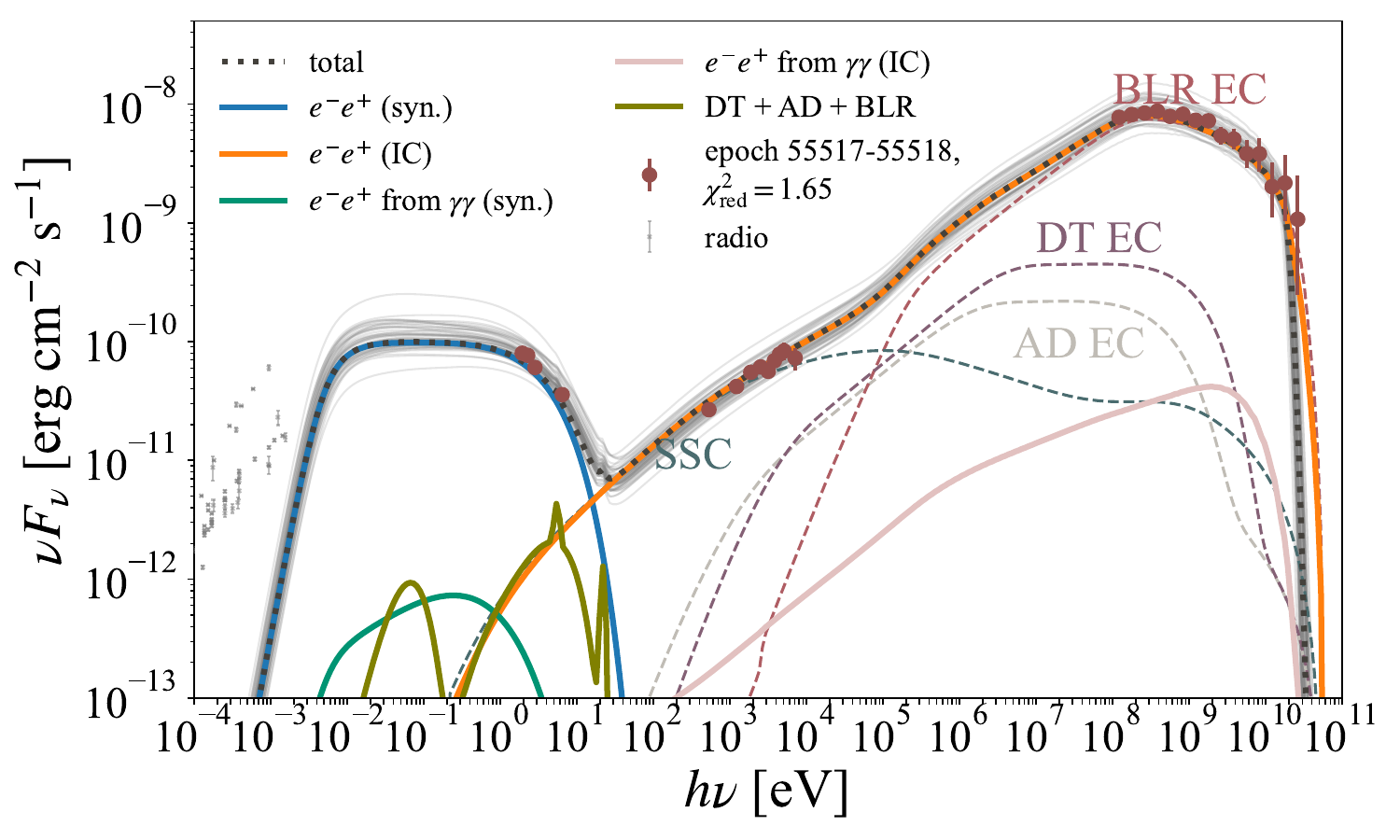}\label{fig:55517-55518}}
    \hfill
    \subfloat[MJD 55518--55519]{\includegraphics[width=0.47\textwidth]{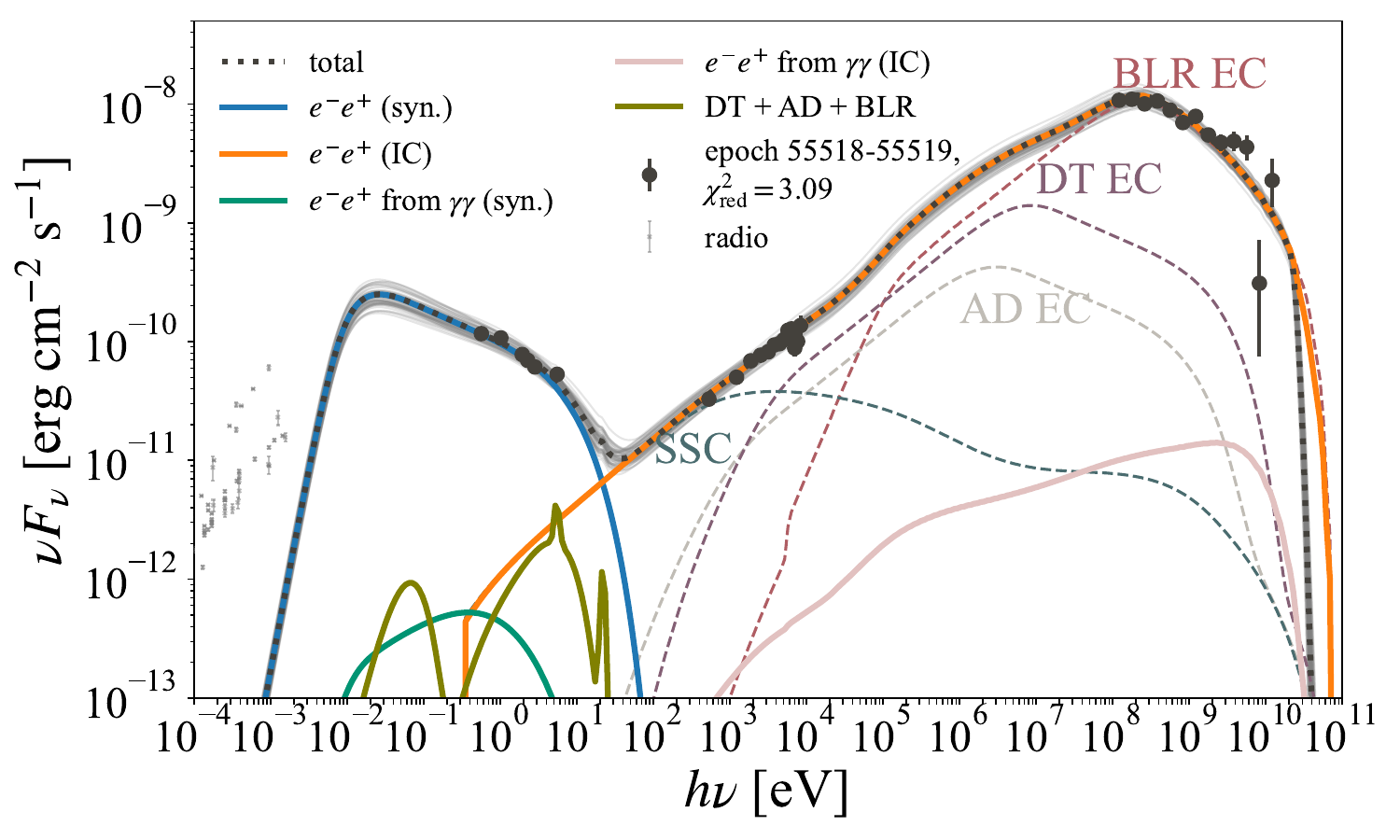}\label{fig:55518-55519}}
    
    \vspace{0.05cm} 
    
    \subfloat[MJD 55519--55520]{\includegraphics[width=0.47\textwidth]{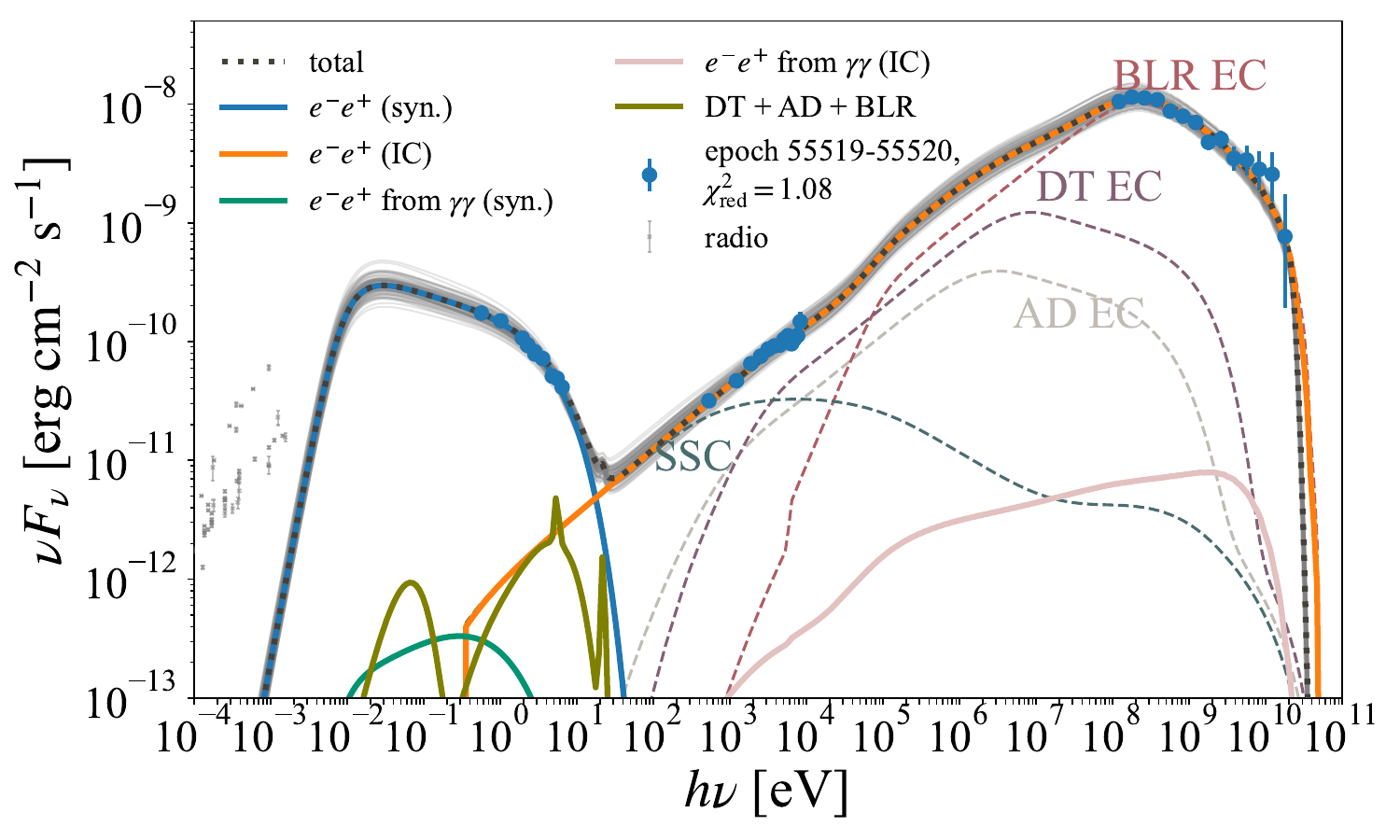}\label{fig:55519-55520}}
    \hfill
    \subfloat[MJD 55520--55521]{\includegraphics[width=0.47\textwidth]{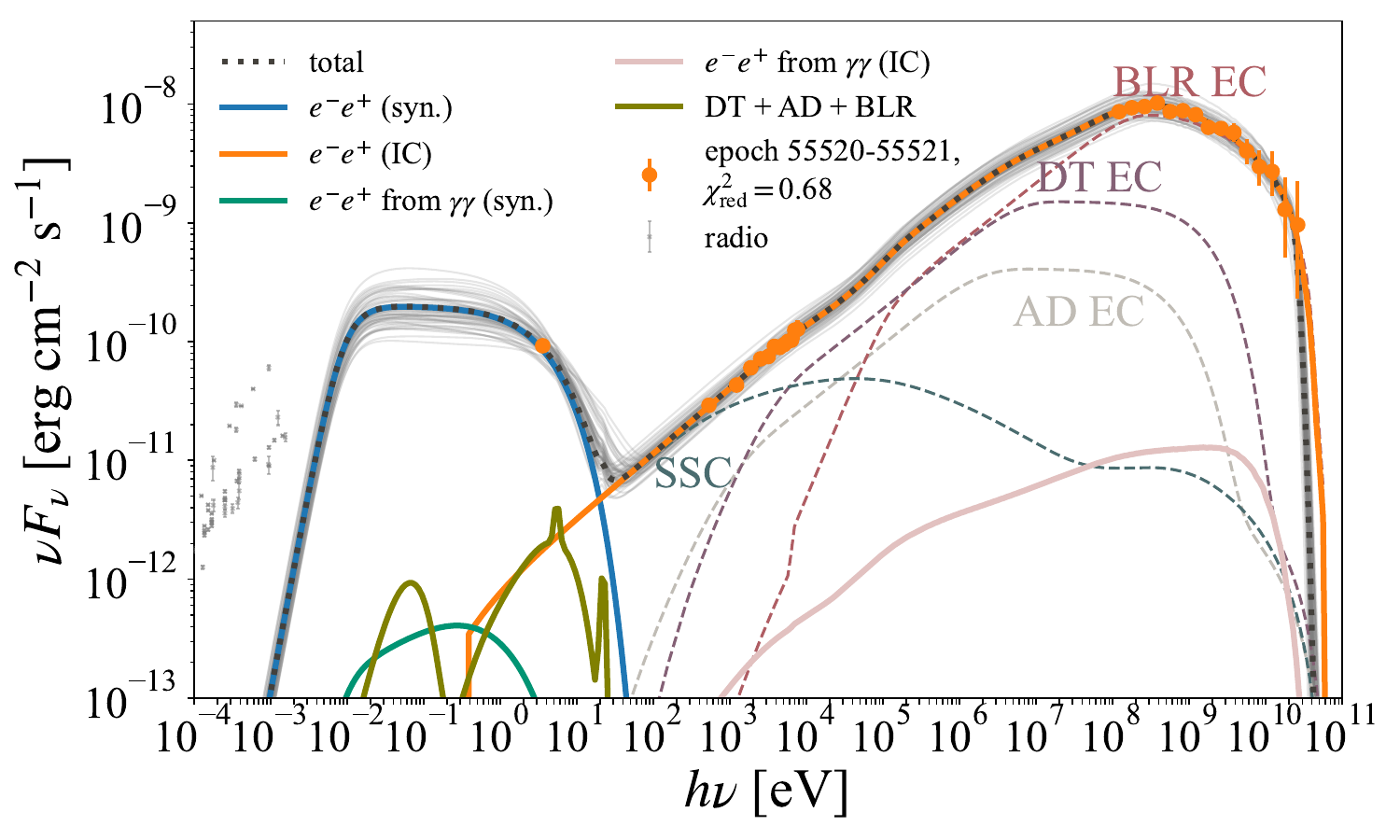}\label{fig:55520-55521}}
    
    \vspace{0.05cm} 
    
    \subfloat[MJD 55521--55522]{\includegraphics[width=0.47\textwidth]{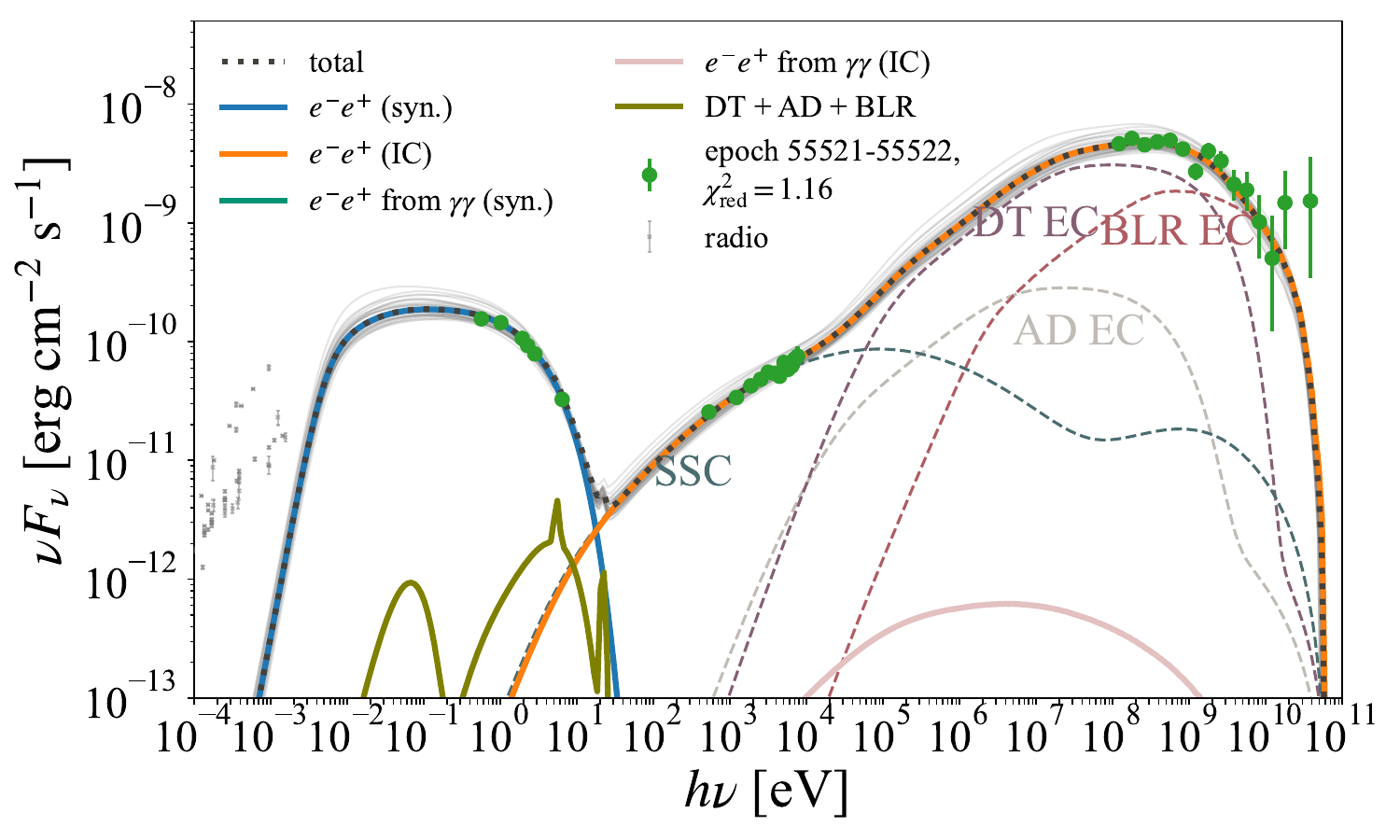}\label{fig:55521-55522}}
    \hfill
    \subfloat[MJD 55522--55523]{\includegraphics[width=0.47\textwidth]{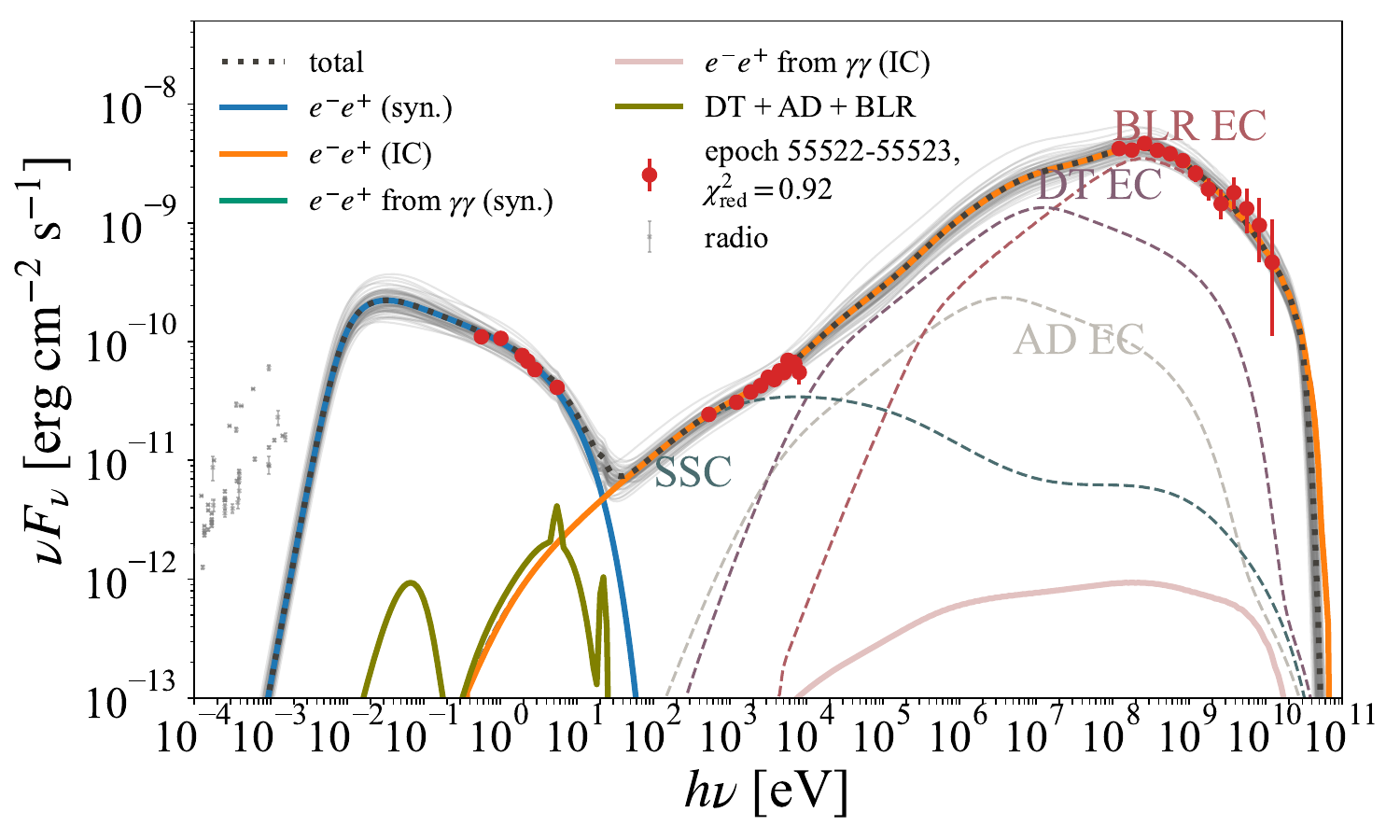}\label{fig:55522-55523}}
    
    \vspace{0.05cm} 
    
    \subfloat[MJD 55523--55524]{\includegraphics[width=0.47\textwidth]{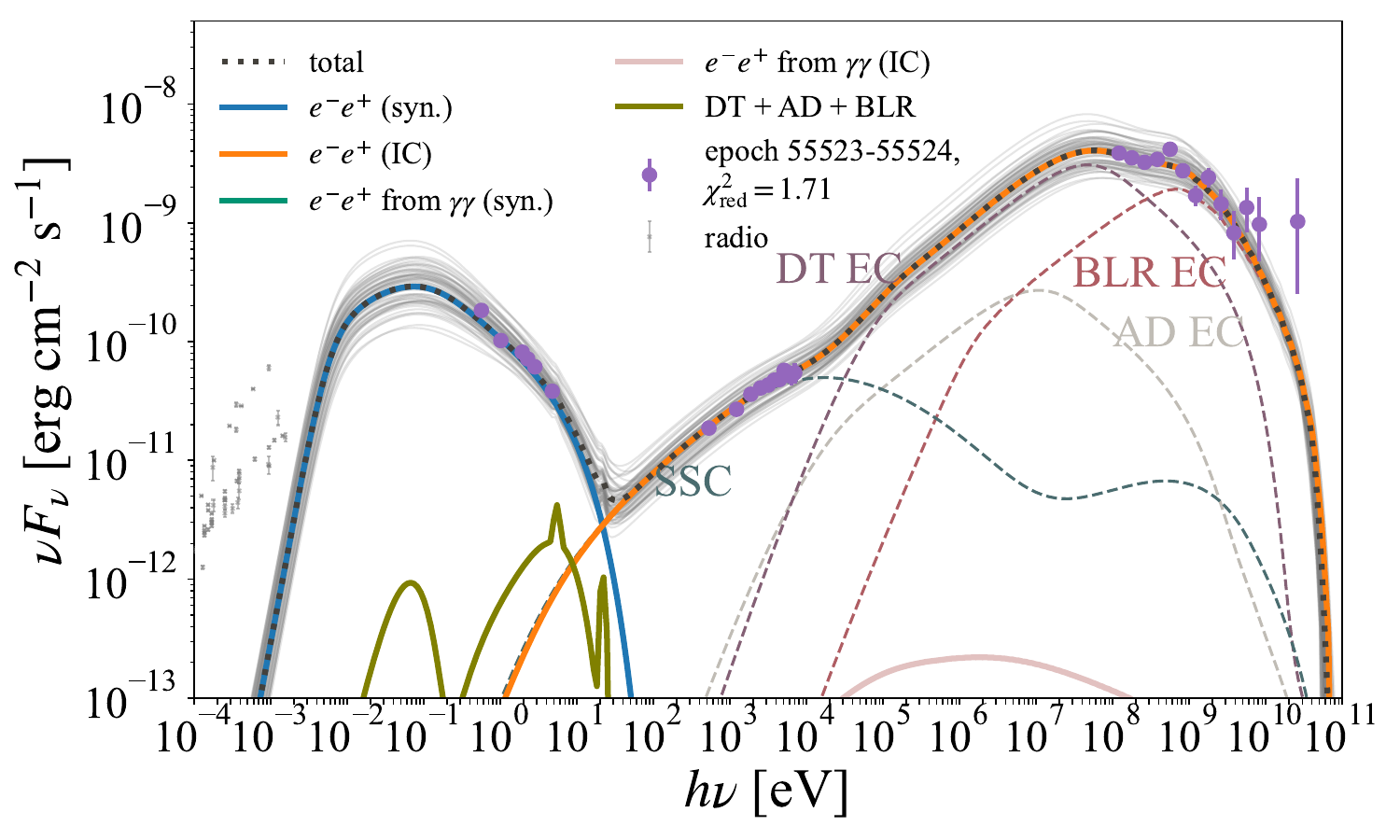}\label{fig:55523-55524}}
    \hfill
    \subfloat[MJD 55517--55524]{\includegraphics[width=0.47\textwidth]{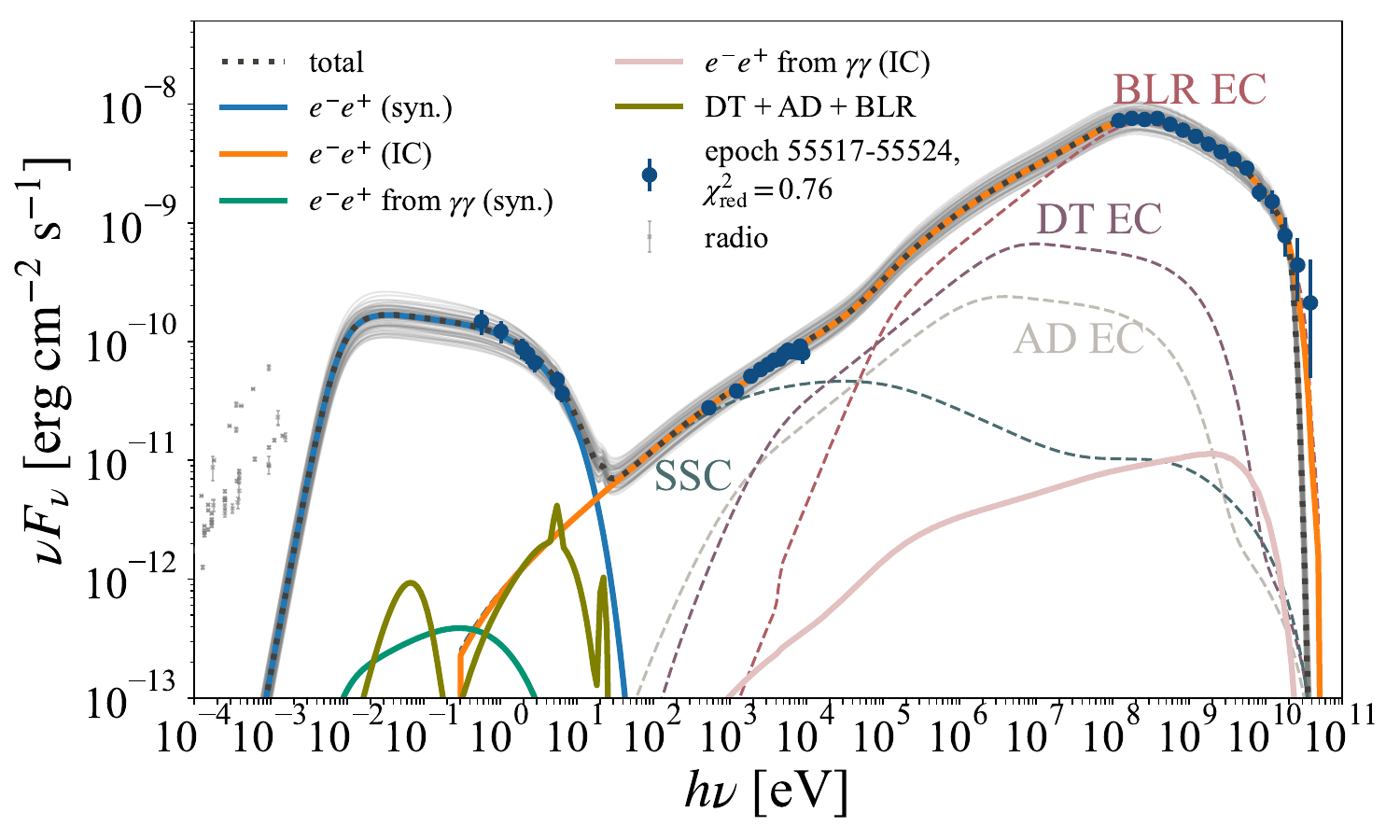}\label{fig:55517-55524}}
    
    \caption{SEDs averaged over 1-day periods (a to g) and SED averaged over the whole period of MJD 55517--55524 (h).}
    \label{fig:multi_panel}
    \end{figure*}
    \begin{figure}
         \includegraphics[width=0.975\columnwidth]{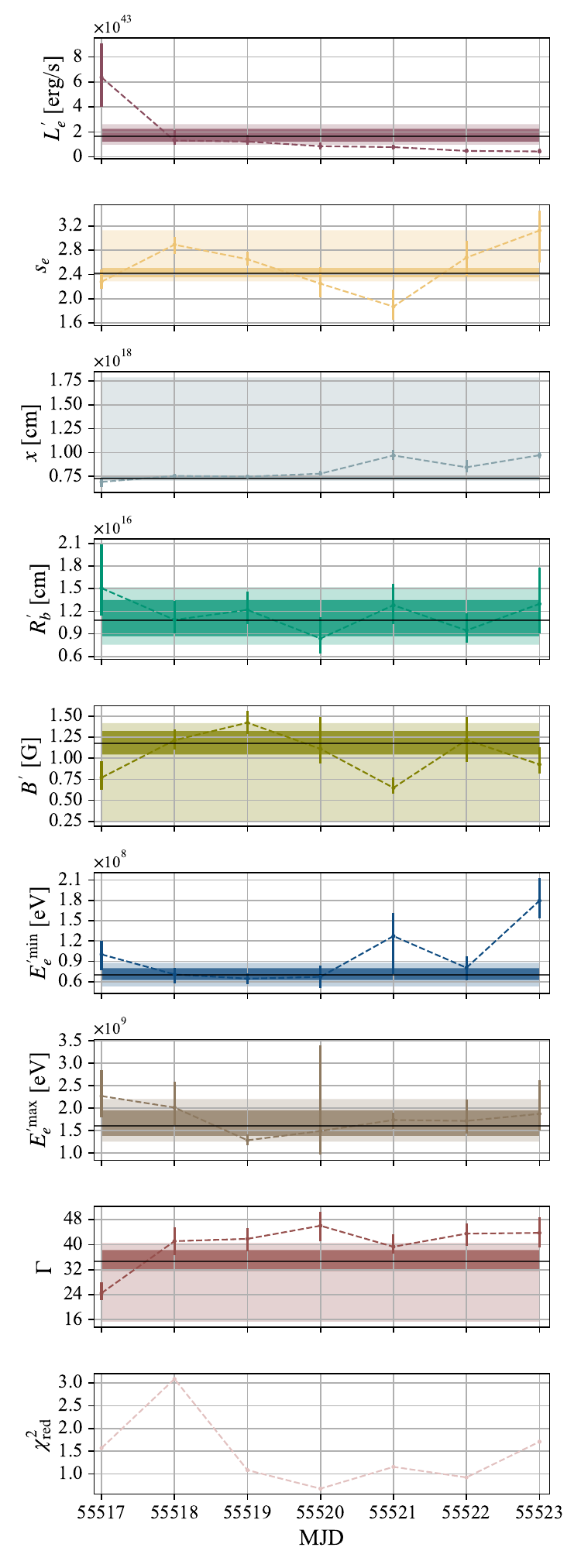}
         \caption{Best-fit values and their $1\sigma$ uncertainties of the free parameters obtained from the fit of the ``standing-feature'' model to one-day-long quasi-simultaneous SEDs observed during MJD 55517--55524, along with the reduced chi-squared values (dashed-line-connected dots with error bars). For the seven-day-averaged SED, the best-fit values are shown as dark horizontal lines with coloured shades denoting $1\sigma$ (darker) and $2\sigma$ (lighter) uncertainties.}
         \label{fig:sequence_of_standing_blobs}
    \end{figure}
    \begin{figure}
        \includegraphics[width=0.975\columnwidth]{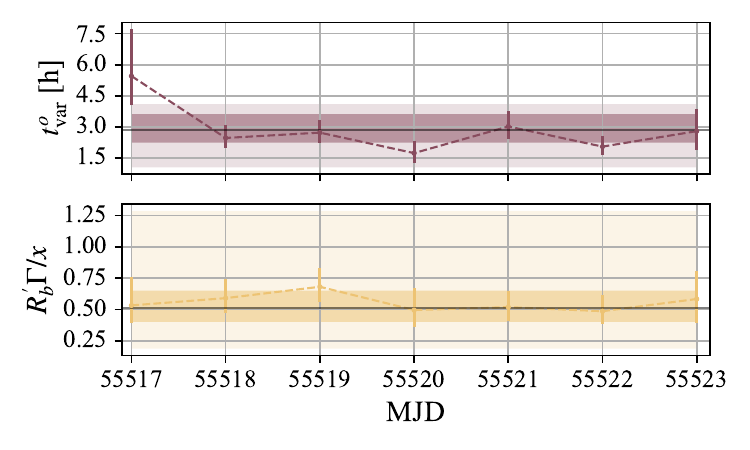}
        \caption{Upper panel: the variability timescale inferred from Eq.~(\ref{eq:decay_timescale}). Lower panel: the numerical values of the expression $R^{\prime}_{b} \Gamma / x$, which must be $\leq 1$ so that the blob cross-section does not exceed the cross-section of the jet in the conical approximation. The best-fit values and uncertainties (propagated from the uncertainties of the independent parameters shown in Fig.~\ref{fig:sequence_of_standing_blobs}) are depicted in the same manner as in Fig.~\ref{fig:sequence_of_standing_blobs}.}
        \label{fig:variability_timescale_and_check_conical_cross_section}
    \end{figure}
    
    The individual observed SEDs from each of the days of MJD 55517--55524 and the best-fit SED model curves are shown in Fig. \ref{fig:multi_panel} along with the fit to the whole-seven-day averaged dataset (Fig.~\ref{fig:55517-55524}) performed in the same manner as for the one-day-long SEDs with $t^{\prime}_{\mathrm{sim}}$ from Eq. (\ref{eq:simulation_time}) calculated for $t^{o}_{\mathrm{av}} = 1$\,d to keep this parameter fixed. Details of the fitting procedure using Monte-Carlo Markov Chain (MCMC) with \texttt{emcee} \citep{emcee} are presented in Appendix~\ref{appendix:corner_plot}. To demonstrate the uncertainty of the model SED, we draw 300 random MCMC states from the MCMC used for the parameter inference and select those MCMC samples which have their model parameter values lying between the 16th and 84th percentiles of each of their marginal posterior one-dimensional distributions. Then, for each of them, we calculate the model SED and superimpose it in Fig.~\ref{fig:multi_panel} as a thin solid grey curve. The contribution of each photon field to the observable emission due to the IC process is shown in Fig.~\ref{fig:multi_panel} with dashed lines with superimposed labels, where SSC stands for ``synchrotron self-Compton'', AD EC --- ``external Compton on the AD radiation'', DT EC --- ``external Compton on the DT radiation'', BLR EC --- ``external Compton on the BLR photon field''. The best-fit values with $1\sigma$ uncertainty of the parameters for all seven days are shown in Fig. \ref{fig:sequence_of_standing_blobs}. In coloured shades, we show $1\sigma$ (darker) and $2\sigma$ (lighter) uncertainties from the MCMC fitting for the seven-day-averaged period (see Appendix \ref{appendix:corner_plot}) while the MCMC best-fit values for the seven-day-averaged fit are shown as black solid horizontal lines.
    
    From Figs.~\ref{fig:multi_panel}-\ref{fig:sequence_of_standing_blobs} we conclude the following:
    \begin{enumerate}
        \item The ``standing-feature'' pure leptonic model provides a good fit to all one-day-averaged SEDs as well as the seven-day-averaged SED with the exception of MJD 55518--55519 (Fig.~\ref{fig:55518-55519}): due to the under-fluctuation in the second to last \textit{Fermi}-LAT bin, the quality of the fit worsens. If one were simply to remove that bin and recalculate the chi-squared (without re-optimisation), the reduced chi-squared value would become $\chi^{2}_{\mathrm{red}} = 2.33$ and a new optimisation could improve this value even further. Thus, we argue that for all seven days of the flare, the ``standing-feature'' pure leptonic model provides a good description of the observed SEDs.
        \item The IR, optical, and UV data are well described with the synchrotron emission from the relativistic electrons in the magnetic field with value not far from our initial guess (see Eq. \ref{eq:magnetic_field}) and fluctuations within $\approx 3\sigma$ from the median MCMC value for the seven-day-averaged SED.
        \item The radio data cannot be described with our model due to the compactness of the blob leading to synchrotron self-absorption between $10^{-3} - 10^{-2}$~eV ($\sim 10^{11}- 10^{12}$~Hz). This implies that an extended emission zone must exist to account for the enhanced radio emission during the flare \citep{2013ApJ...773..147J}.
        \item The contribution of the direct AD+BLR radiation to the optical and UV SED is subdominant.
        \item The X-ray data are well described as the sum of the electron SSC and EC emission on the DT and AD photon fields (similar to, e.g., the leptonic model of \citealp{Bottacini:2016xtg}).
        \item The $\gamma$-ray data are well described by IC emission of electrons with energies above the cooling break. The IC process uses the BLR photons as target photons and is in the Thomson regime (with the exception of the highest-energy bins, where some Klein-Nishina cross-section corrections may play a role). The \textit{Swift}-XRT and \textit{Fermi}-LAT data are connected with emission from the electron IC on the AD radiation coming from behind the blob in the host galaxy rest frame, and, with a greater contribution, the DT radiation, isotropic in the host galaxy rest frame.
        \item While the electron injection luminosity does not seem to change significantly within MJD 55517--55524, the Lorentz factor does change and shows an increasing trend throughout the flare. This can be explained as follows. During the last three days of the flare, the \textit{Fermi}-LAT observes higher values of the $\gamma$-ray SED in the two last energy bins, which implies that $x$ must increase in order to avoid $\gamma\gamma$ absorption. Increasing $x$ when it is greater than $R_{\mathrm{BLR}}$ leads to a significant drop in the density of the photon field. To compensate for this drop, the Lorentz factor $\Gamma$ should be increased accordingly.
        \item The asymmetry of the MCMC uncertainty of $x$ for MJD 55517---55524 is driven by the lower limit on $x$ due to $\gamma\gamma$ absorption (see Eq.~\ref{eq:maximum_observed_energy}).
        \item The maximum electron injection energy slightly varies but stays within the $2\sigma$ uncertainty of the MCMC value for MJD 55517--55524 and is close to the value estimated with Eq. (\ref{eq:maximum_electron_injection_energy}).
        \item The minimum electron injection energy $E^{\prime{\mathrm{min}}}_{e}$ appears to be well constrained, which is mainly driven by the fact that the peak of the $\gamma$-ray SED observed by \textit{Fermi}-LAT at a few hundred MeV is due to the break of the cooled electron spectrum happening at $E^{\prime{\mathrm{min}}}_{e}$ as we mentioned in Sect.~\ref{sec:injection}. Changing $E^{\prime{\mathrm{min}}}_{e}$ significantly would result in a poor description of the high-statistics lower-energy \textit{Fermi}-LAT $\gamma$-ray SED; besides, it would worsen the fit to the X-ray data as well due to the corresponding shift of the SED peak of the DT EC component.
        \item The obtained fit with a low value of the maximum electron injection energy $E^{\prime{\mathrm{max}}}_{e} \sim 10^9$~eV results in significant curvature of the observed $\gamma$-ray spectrum (see Sect. \ref{sec:injection}). In Fig. \ref{fig:timescales}, we can see that this maximum injection energy corresponds to the electron acceleration timescale $t^{\prime \mathrm{acc}}_{e} = \xi_{\mathrm{acc}} t^{\prime \mathrm{gyro}}_{e}$ with $ \xi_{\mathrm{acc}} \sim 10^{7}$, where the gyroperiod $t^{\prime \mathrm{gyro}}_{e}$ is defined in Eq. (\ref{eq:gyroperiod}). A model with a similar value of $\xi_{\mathrm{acc}}$ was suggested by \citet{1996ApJ...463..555I} to explain early observations of another FSRQ: 3C 279. They interpreted $\xi_{\mathrm{acc}} \sim 10^{7}$ as a sign that the ratio of the energy of the background magnetic field to the energy of the magnetic turbulence is as large.
        \item For each day, the best-fit value of the blob size stays within the $2\sigma$ MCMC uncertainty of the value obtained for the seven-day-averaged dataset (lighter shade in Fig.~\ref{fig:sequence_of_standing_blobs}). The inferred variability timescale (see Eq. \ref{eq:decay_timescale} and the upper panel of Fig.~\ref{fig:variability_timescale_and_check_conical_cross_section}) is approximately from $2$~h to $5$~h. These values are close to the values of $2.6 \pm 1.0$~h obtained by \citet{2019ApJ...877...39M}.
        \item For all of the days of the flare and the seven-day-averaged fit, the condition $R^{\prime}_{b} \Gamma / x \leq 1$ that the blob cross-section does not exceed the cross-section of the jet in the conical approximation\footnote{We note that during the fit such a constraint was not imposed.} is satisfied for values within $\pm 1\sigma$ around the best-fit values (see the lower panel of Fig.~\ref{fig:variability_timescale_and_check_conical_cross_section}).
    \end{enumerate}

    \subsection{Results of the ``moving-blob'' leptonic model}
    \label{sec:moving_blob_model_results}
    \begin{figure*}
        \centering
         \includegraphics[width=1.45\columnwidth]{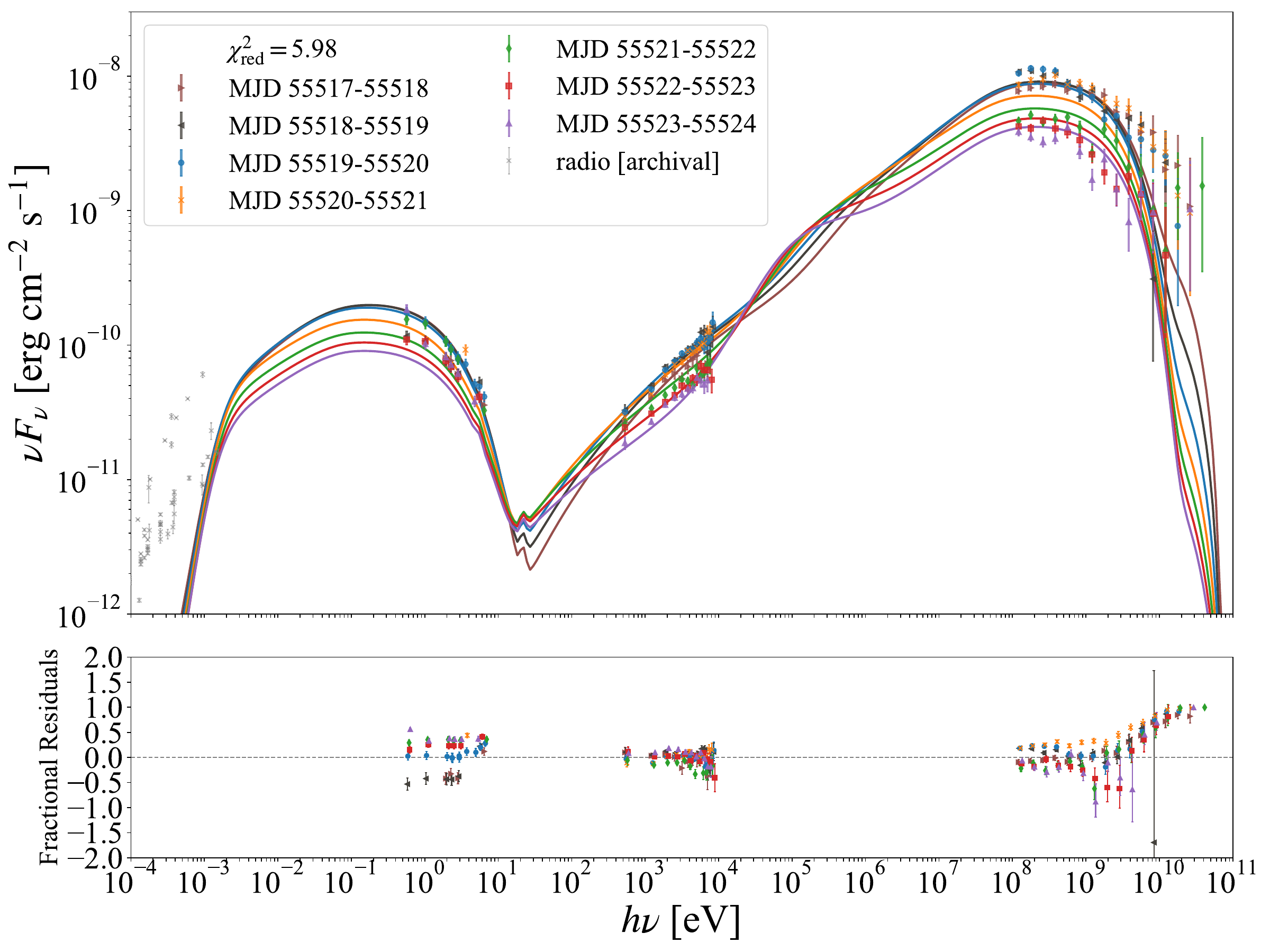}
         \caption{Upper panel: ``moving-blob'' leptonic model SEDs averaged over one day each from the fit to the seven observed one-day-averaged SEDs for MJD 55517--55524. Lower panel: fractional residuals, defined as the differences between the observed and model SEDs divided by the observed SEDs (data points for different dates are slightly shifted in the horizontal direction for better visibility).}
    \label{fig:moving_blob_sed}
    \end{figure*}
    \begin{figure}
         \centering
         \includegraphics[width=0.94\columnwidth]{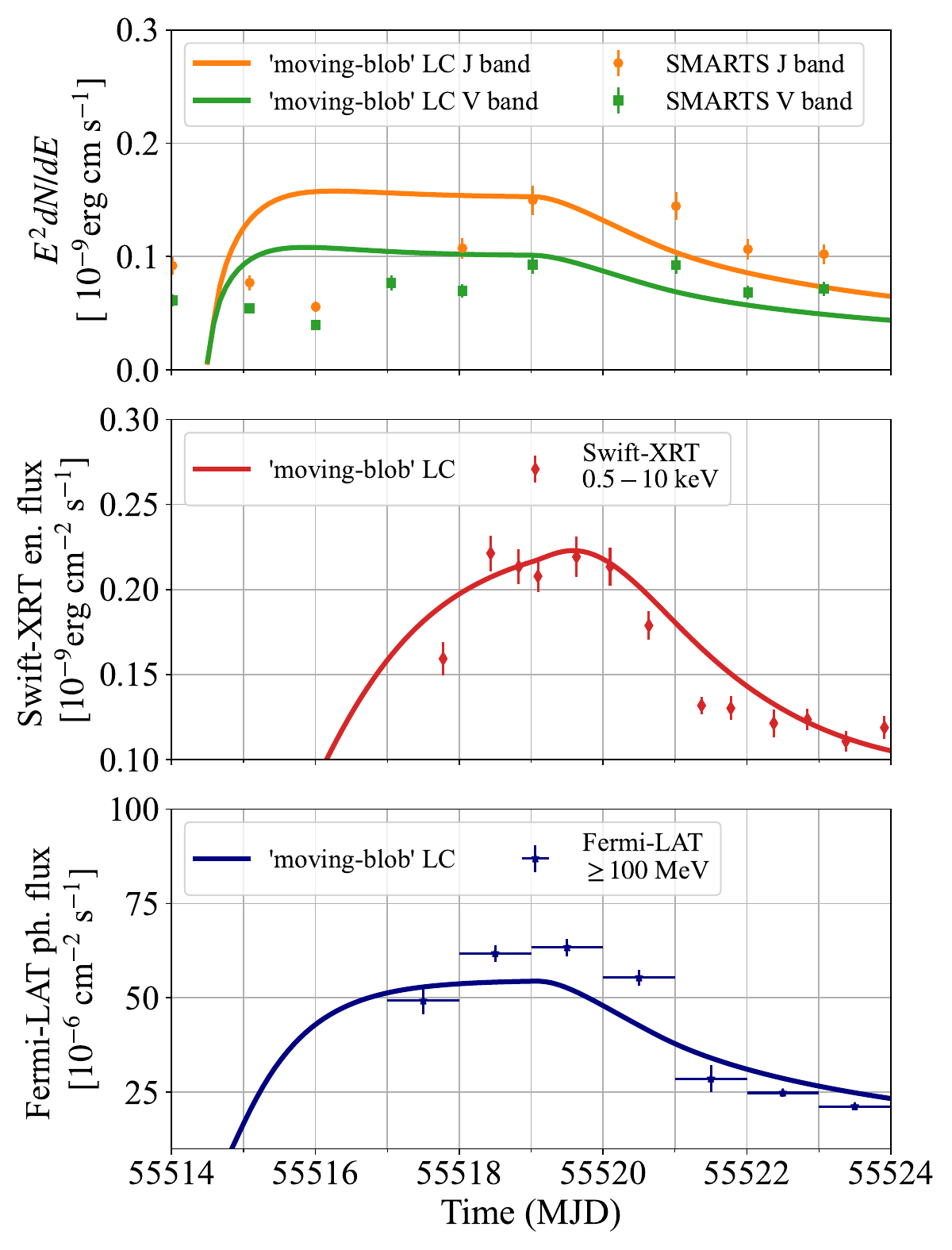}
         \caption{The observed (data points with error bars) and ``moving-blob'' model (solid curves) light curves in the optical (upper panel), X-ray (middle panel), and $\gamma$-ray (lower panel) energy bands. See the legend and the text in Sect.~\ref{sec:moving_blob_model_results} for details.}
         \label{fig:multiwavelength_lightcurves}
    \end{figure}
    
    Using \texttt{iMINUIT}, we fit the SEDs of the ``moving-blob'' model to the data from seven one-day-averaged observed SEDs using first the \texttt{SIMPLEX} algorithm and \texttt{MIGRAD}. The SEDs resulting from the fit are shown in Fig.~\ref{fig:moving_blob_sed}. The total quality of the obtained fit for the union of the seven SEDs is rather poor: $\chi^2_{\mathrm{red}} \approx 6$. Due to the significantly larger number of free parameters in the ``moving-blob'' model than in the ``standing-feature'' model (13 VS 8) and seven times longer simulation timescale, it is computationally expensive to run an MCMC parameter inference for the former, so, given the poor quality of the fit as well, we limit ourselves only to the \texttt{iMINUIT} minimisation.
    
    Given that the ``moving-blob'' model is time-dependent, this allows us to plot measured light curves of 3C 454.3 during the flare and compare them to the model ones in Fig.~\ref{fig:multiwavelength_lightcurves}. In the upper panel of Fig.~\ref{fig:multiwavelength_lightcurves}, the differential time-dependent SED values measured in two optical bands by \textit{SMARTS} (see Sect.~\ref{sec:optical}) are shown as data points with error bars. ``Moving-blob'' model light curves (solid orange (J band) and green (V band) curves) are in poor agreement with the optical data. In the middle panel of Fig.~\ref{fig:multiwavelength_lightcurves}, the integral energy flux measured by \textit{Swift}-XRT in the energy range $0.5-10.0$~keV is shown as red circles with error bars. The X-ray light curve is taken from the ``Open Universe'' database \citep{Giommi:2019jqg, 2021MNRAS.507.5690G} and corrected for the X-ray absorption with the seven-day-averaged value of the hydrogen column density $\kappa$ from Table~\ref{table:swift_xrt_data_analysis}. In this range of the electromagnetic spectrum, the ``moving-blob'' model (see the solid red curve) clearly shows the best agreement with the data comparing to the other bands, albeit it still fails to describe the sharp drop in the X-ray energy flux after MJD 55521, which might be related to the large size of the blob radius in the model (see below). In the lower panel of Fig.~\ref{fig:multiwavelength_lightcurves}, the \textit{Fermi}-LAT light curve from our analysis (see Sect.~\ref{sec:fermi} and Table~\ref{table:fermi_lat_data_analysis}) is shown as one-day-averaged photon flux ($E^{o}_{\gamma} \geq 100$~MeV) data points with error bars. The ``moving-blob'' model (blue solid curve) reproduces the trend of the observed $\gamma$-ray light curve, although it does not manage to describe the amplitude of the peak of the flare as well as it does for the X-ray data.
    
    While the ``moving-blob'' model manages to describe the changes in the $X$-ray energy range, it fails to do so in the IR-optical-UV range (significantly over-predicting the optical flux at the earlier times and under-predicting at later times) and in the higher-energy part of the \textit{Fermi}-LAT spectrum. The latter is mainly due to the movement of the blob far away from the BLR photon field resulting in the decrease of the target photon field energy (see Eq. \ref{eq:line_transformation}) and density which leads to the under-prediction of the number of the $\gamma$-ray photons at higher energies at later days of the flare (cf. Fig.~\ref{fig:multi_panel}, where, due to nearly constant $x$, the higher-energy part of the \textit{Fermi}-LAT SED is described by the BLR EC). Moreover, to compensate for the decrease in the external photon density, $L^{\prime}_{e}$ is increased accordingly. This would overshoot the X-ray data which are explained by SCC with the flux proportional to $L^{\prime 2}_{e}$. To mitigate overshooting of the X-ray data, the fitter increases the initial blob radius $R^{\prime}_{b}(0)$ up to $6 \times 10^{16}$ cm to decrease the synchrotron photon density in the blob frame (decreasing as $R^{\prime -3}_{b}$) which provides a good description of the X-ray data but is in tension with the values expected from the typical flare decay timescale of a few hours (see Sect. \ref{sec:blob_radius}). Moreover, the large size of the blob requires a significant amount of time for electrons to produce synchrotron photons, serving as the target for the SSC process. This results in blob being launched around $t^{o}_{\mathrm{shift}} \approx 2.5$~d earlier than MJD 55517 --- the beginning of the period of interest; this causes the over-prediction of the optical flux visible in the upper panel of Fig.~\ref{fig:multiwavelength_lightcurves}.
    
    Consideration of a time-dependent (accelerating) blob Lorentz factor $\Gamma$, the hint of which could be seen in Fig.~\ref{fig:sequence_of_standing_blobs} in the ``standing-feature'' model, does not seem to be able to mitigate the tension between the observed SEDs and the ``moving-blob'' model SEDs: while in the case of the ``standing-feature'' the main photon field on which the IC scattering occurs is the deboosted BLR ($x > R_{\mathrm{BLR}}$), the density of which drops drastically with a slight increase of $x$ in later days of the flare, and $\Gamma$ is increased to compensate for this drop, as discussed in Sect.~\ref{sec:standing_blob_model_results} (see Fig.~\ref{fig:sequence_of_standing_blobs}), the main target for the IC scattering in case of the ``moving-blob'' model is the DT radiation (the blob propagates from $x_{s} \approx 1.3 R_{\mathrm{BLR}}$ to $x_{f} \approx 25 R_{\mathrm{BLR}}$), and the blob does not leave the outer radius $R_{\mathrm{DT}} = 25 R_{\mathrm{BLR}}$ within the timescale of the flare, i.e. the DT photon density remains unchanged if $\Gamma$ is constant. Thus, the accelerating $\Gamma$ would cause shifts of the SEDs to higher energies and an increase of their amplitudes over time, which is not observed. Therefore, we limited ourselves to the case of the constant $\Gamma$ in the ``moving-blob'' model.
    
    The failure to describe the flare in the framework of the ``moving-blob'' model implies that the observed emission during the flare originates in a quasi-stationary ``standing feature'' in the jet, which does not move along with the jet plasma bulk motion and into which multiple blobs are injected as was modelled in Sect. \ref{sec:standing_blob_model_description}. We discuss possible interpretations of the prevalence of the ``standing feature'' in Sect. \ref{sec:standing_feature_discussion}.

\section{Description of the leptohadronic model}
\label{sec:leptohadronic}
    \begin{figure}
         \includegraphics[width=1.0\columnwidth]{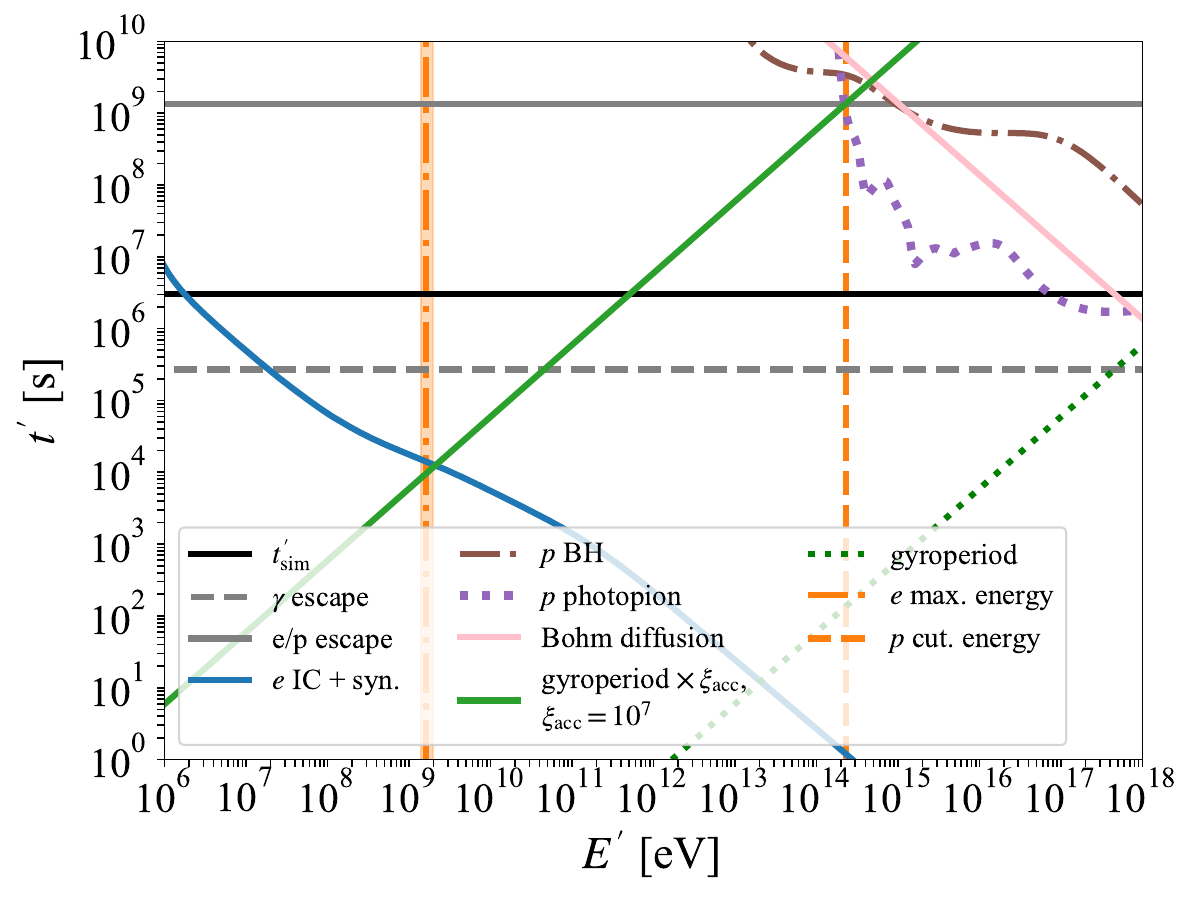}
         \caption{The characteristic energy-loss timescales for the ``standing-feature'' leptohadronic model for the averaged dataset MJD 55517--55524 with $t^{\prime}_{\mathrm{sim}}$ corresponding to one day in the Earth rest frame.}
         \label{fig:timescales}
    \end{figure}
    
    In this subsection, we obtain constraints on the maximum contribution of protons to the observed SED allowed by the data assuming the hadronic contribution in the model serves as a subdominant component of the model SED, in addition to the already obtained leptonic fit, similar to previous studies by, e.g., \citet{2018ApJ...864...84K,2019MNRAS.483L..12C,2019NatAs...3...88G}.

    For the best fit of the ``standing-feature'' leptonic model averaged over MJD 55517--55524, we show the characteristic timescales in the blob frame in Fig. \ref{fig:timescales}. To minimise the number of free parameters, we assume that the protons are co-accelerated with electrons in a process with an acceleration timescale
    \begin{equation}
        t^{\prime \mathrm{acc}}_{p} (E^{\prime}_{p}) = \xi_{\mathrm{acc}} t^{\prime \mathrm{gyro}}_{p},
    \label{eq:acceleration_timescale}
    \end{equation}
    where $t^{\prime \mathrm{gyro}}_{p} = t^{\prime \mathrm{gyro}}_{e}$ is defined in Eq. (\ref{eq:gyroperiod}). We assume for simplicity that $\xi_{\mathrm{acc}}$ is the same for both electrons and protons (see Appendix \ref{appendix:xi} for an investigation of other values of $\xi_{\mathrm{acc}}$ for protons). To estimate the value of $\xi_{\mathrm{acc}}$, we use the best-fit value of the maximum electron injection energy $E^{\prime \mathrm{max}}_e$ obtained from fitting the ``standing-feature'' leptonic model SED to the observed seven-day-averaged data (see Sect.~\ref{sec:standing_blob_model_results}) and find in Fig. \ref{fig:timescales} the interception of the vertical orange dashed-dotted line corresponding to $E^{\prime \mathrm{max}}_e$ (with the pale orange band representing its $1\sigma$ uncertainty) and the blue solid curve of the total electron energy-loss timescale, and superimpose the dependence of the gyroperiod on the energy (green dashed line) multiplied by $\xi_{\mathrm{acc}} = 10^{7}$ (green solid line), which results in the self-consistent estimate of $E^{\prime \mathrm{max}}$ as the energy when the electron acceleration and energy-loss timescales become equal. This heuristically obtained approximate value of $\xi_{\mathrm{acc}} = 10^{7}$ serves as the benchmark value for the leptohadronic model.
    
    To determine the maximum proton energy, we assume that the acceleration started long before the flare\footnote{Otherwise, the maximum proton energy would be determined by the interception of $t^{\prime \mathrm{acc}}_{p}$ with the simulation timescale $t^{\prime}_{\mathrm{sim}}$ and would be a few hundred GeV.} and look at the interception of the green solid line of $t^{\prime \mathrm{acc}}_{p} (E^{\prime}_{p})$ with the proton escape timescale (grey solid line), as well as the proton energy-loss timescale $t^{\prime \mathrm{p\gamma}}$ due to photopion production (purple dotted line) having almost the same value at this energy (see Fig. \ref{fig:timescales}):
    \begin{equation}
        t^{\prime \mathrm{acc}}_{p} ( E^{\prime{\mathrm{max}}}_{p}) = \min \{ t^{\prime p\gamma}_{p}(E^{\prime{\mathrm{max}}}_{p}); t^{\prime \mathrm{esc}}_{p} \},
    \label{eq:proton_max_energy}
    \end{equation}
    where $t^{\prime \mathrm{esc}}_{p} = t^{\prime \mathrm{esc}}_{e}$ is defined in Eq. (\ref{eq:electron_escape}). This results in $E^{\prime{\mathrm{max}}}_{p} \approx 2 \times 10^{14} \mathrm{\, eV}$ for MJD 55517--55524, where $E^{\prime{\mathrm{max}}}_{p}$ is taken as the energy of the exponential cutoff of the proton spectrum (shown as the orange dashed line in Fig.~\ref{fig:timescales}). We also check how this value varies from one day to another throughout the seven days: for a fixed value of $\xi_{\mathrm{acc}} = 10^{7}$, the variation of $E^{\prime{\mathrm{max}}}_{p}$ is less than $30$\%. The optimistic assumption that the proton acceleration started long before the flare is used, since we are focused on estimating the \textit{upper} limit of the neutrino yield. The assumption implies that the observable neutrino energy should gradually rise in time from the beginning of the acceleration. However, due to the total expected number of observed neutrinos from one source $\ll 1$, this prediction is hardly testable even with the next generation neutrino telescopes. It is not common to expect that protons are accelerated long before electrons (because the acceleration timescale is proportional either to the particle energy or mass, \citealp[e.g.,][]{Weidinger:2014yya}), although it is not excluded since the protons, due to their significantly less efficient energy losses, are effectively decoupled from fast-cooling electrons and are not required to be continuously injected into the system as electrons. This assumption is not critical in our study, since it is used only in the heuristic way to find $E^{\prime{\mathrm{max}}}_{p}$, and a wide range of other possible values of the proton cutoff energy is considered in Appendix~\ref{appendix:xi}. In an alternative assumption, when protons begin their acceleration at the beginning of the flare, $E^{\prime{\mathrm{max}}}_{p}$ could be defined as the energy at which the acceleration and flare (simulation) timescales become equal, in which case a different range of $\xi_{\mathrm{acc}}$ can be considered to cover the same range of $E^{\prime{\mathrm{max}}}_{p}$.

    The minimal proton energy is fixed at $E^{\prime \mathrm{min}}_{p} = 2 \times 10^9$~eV, and their power-law spectral index fixed at $s_{p} = 2.0$. In principle, the power-law spectral index $s_{p}$ of the proton distribution should follow the electron spectral index $s_{e}$, but, as can be seen in Fig.~\ref{fig:sequence_of_standing_blobs}, $s_{e}$ varies noticeably from day to day, so we choose the standard value of $s_{p} = 2$ for simplicity. Besides, in some acceleration scenarios the difference of $0.1-0.4$ between $s_{e}$ and $s_{p}$ can be expected \citep[e.g.,][]{Diesing:2019lwm}. The uncertainty of the value of $\xi_{\mathrm{acc}}$ has a more significant impact on the neutrino SED than the uncertainty of $s_{p}$\footnote{E.g., \citet{2018ApJ...864...84K} show how different values of $s_p$ affect the neutrino SED.}, because $\xi_{\mathrm{acc}}$ defines the peak energy of the neutrino SED as shown in Appendix~\ref{appendix:xi}.

    We choose the constant escape timescale with the value of $\eta_{p} =\eta_{e} \sim 5 \times 10^{3}$ to avoid the dependence of the derived proton cutoff energy on a particular mechanism of the proton acceleration: in a case of a more efficient proton acceleration with $\xi_{\mathrm{acc}} < 10^{7}$ the cutoff proton energy will be defined by the proton energy-loss timescale. We also verify that the most optimistic Bohm-limit diffusion escape timescale described by Eq.~(\ref{eq:bohm}) and shown as the solid pink line in Fig.~\ref{fig:timescales} lies somewhat higher than the proton energy-loss timescale, i.e. our approach of the heuristic determination of $E^{\prime{\mathrm{max}}}_{p}$ is not affected by a particular mechanism of the proton acceleration if $\xi_{\mathrm{acc}} \leq 10^{7}$. We explore how the observed electromagnetic and neutrino SEDs change if the proton acceleration is more efficient with $\xi_{\mathrm{acc}} < 10^{7}$ in Appendix~\ref{appendix:xi}.
    
    From Fig.~\ref{fig:timescales}, one can see that within the simulation timescale corresponding to one day in the observer's frame, most of the electrons effectively cool and a continuous injection of freshly accelerated electrons is required to maintain the blob radiation. On the other hand, protons with energies up to $10^{16}$~eV have an energy-loss timescale much greater than the simulation timescale. Thus, given the maximum energy of protons is $E^{\prime{\mathrm{max}}}_{p} \approx 2 \times 10^{14} \mathrm{\, eV}$, no continuous proton injection is required, and the initially predetermined steady-state proton SED remains almost unchanged throughout the simulation.
    
    We found that for the whole period of MJD 55517--55524, if we set the steady-state proton SED with energy density in the blob frame equal to the energy density of the injected electrons $u_p^{\prime} = u_e^{\prime}$, this results in approximately the same value of the $\chi^{2}$ as for the purely leptonic model. We use $u_p^{\prime} = u_e^{\prime}$ as a starting point. Further in a loop, we increase the normalization of the proton SED by $5\%$ each step until the Bayes factor \citep{BayesFactors}
    \begin{equation}
        BF = \exp {\left(- \frac{\chi^{2}(u^{\prime}_{p}) - \chi^{2}_{\mathrm{leptonic}}}{2} \right)}
    \label{eq:bayes_factor}
    \end{equation}
    drops to $BF = 10^{-2}$ ($\approx 3 \sigma$ Gaussian equivalent). We denote the corresponding value of the proton energy density $u^{\prime\mathrm{UL}}_{p}$ as the upper limit (UL). The obtained ratio between the UL proton energy density and injected electron energy density is (see Sect.~\ref{sec:results_leptohadronic})
    \begin{equation}
        \rho_{p/e} \equiv \frac{u^{\prime\mathrm{UL}}_{p}}{u^{\prime}_{e}} \approx 1.3 \times 10^{2}.
    \label{eq:rho}
    \end{equation}

    The results obtained with the ``standing-feature'' leptohadronic model are presented in Sect.~\ref{sec:results_leptohadronic}. Since the leptonic model within the ``moving-blob'' approach resulted in a poor description of the data (see Sect.~\ref{sec:moving_blob_model_results}), we do not develop its leptohadronic counterpart.

    \subsection{Method of estimating the expected muon neutrino yield from 3C 454.3 in \textit{IceCube}}\label{sec:neutrino_yield_estimates}

    In this section, we describe our procedure for calculating the \textit{IceCube} muon neutrino yield from 3C 454.3 for any given moment of time assuming a leptohadronic model with the constant factor $\rho_{p/e}$ as the one obtained in Eq. (\ref{eq:rho}) for MJD 55517--55524.

    \begin{enumerate}
        \item We select four other additional time periods of the 3C~454.3 observations when quasi-simultaneous data from X-ray and IR/optical/UV telescopes (among \textit{Swift}-UVOT, \textit{SMARTS} and \textit{Steward} Observatory) are available. The selected periods are MJD 54700--54701; 54827--54833; 55180--55181; 55471--55472 and cover different values of observed $\gamma$-ray fluxes from 3C~454.3.
        \item For each of the additionally selected periods, we repeat the data analysis as described in Sect. \ref{sec:data} and fitting within the ''standing-feature'' leptonic model as described in Sect. \ref{sec:standing_blob_model_description}.
        \item For each of the additionally selected periods, having obtained the best-fit values of the ''standing-feature'' leptonic model's parameters, we turn on hadronic processes injecting protons with the energy density $\rho_{p/e} = 1.3 \times 10^{2}$ times greater than the injected electron energy density for the period under consideration. The energy of the exponential cutoff of the proton injection spectrum is defined by the interception of the acceleration timescale with $\xi_{\mathrm{acc}} = 10^{7}$ and escape timescale (see Eq. \ref{eq:proton_max_energy}).
        \item The obtained with \texttt{AM$^{3}$} observable neutrino $\mathrm{SED}_{\nu}(E^{o}_{\nu}) = E^{o \, 2}_{\nu} dN^{o}_{\nu}/dE^{o}_{\nu}$ for each of the five considered periods is used to calculate the integral neutrino particle flux $\Phi^{o}_{\nu}$ for observable neutrino energies $E^{o}_{\nu} \geq 100$~GeV.
        \item We obtained (see Appendix~\ref{appendix:linearity}) a good linear fit in the form of $\Phi^{o}_{\nu} = a \Phi^{o}_{\gamma}$, where $a$ is a constant, and $\Phi^{o}_{\gamma}$ is the integral model photon flux for observable $\gamma$-ray energies $E^{o}_{\gamma} \geq 100$~MeV.
        \item Assuming that the 3C 454.3 neutrino SED obtained for MJD 55517--55524 can be used (after scaling to the corresponding value of $\Phi^{o}_{\gamma}$) for each time bin $t_{i}$ of a light curve from the \textit{Fermi}-LAT light curve repository (LCR) curated by \citet{pyLCR}, we calculate the neutrino SED for a given time bin $t_{i}$ as follows:
        \begin{equation}
            \mathrm{SED}_{\nu}(E^{o}_{\nu}; t_{i}) = \frac{\Phi^{\mathrm{o}}_{\gamma}(t_{i}) \mathrm{SED}_{\nu}(E^{o}_{\nu};t_{0})}{\Phi^{\mathrm{o}}_{\gamma}(t_{0})},
        \label{eq:neutrino_gamma_sed_scaling}
        \end{equation}
        where $t_{0}$ represents the interval of MJD 55517--55524. If at time $t_{i}$ in the LCR there is only an UL on the flux value (the source is not detected with the test statistic $TS > 4$), we assume $\Phi^{\mathrm{o}}_{\gamma}(t_{i}) = 0$ for that time bin.
        \item For each time bin $t_{i}$, muon (+ anti-muon) \textit{IceCube} neutrino expected daily rate $\dot{N}_{\nu}^{\mathrm{UL}}$ for observable neutrinos with energies $E^{o}_{\nu} \geq 100$~TeV is calculated using $\mathrm{SED}_{\nu}(E^{o}_{\nu}, t_{i})$ and the \textit{IceCube} muon neutrino effective area $A_{\mathrm{eff}}(E^{o}_{\nu})$ at the declination of 3C~454.3 $16.15^{\circ}$ \citep{IceCubeDataRelease, 2021arXiv210109836I, 2020PhRvL.124e1103A}. We use the threshold of $E^{o}_{\nu} \geq 100$~TeV since approximately at these energies the IceCube collaboration issues alerts \citep{2023ApJS..269...25A} reporting neutrino-like events with the probability of the signal being of astrophysical origin $\geq 0.3$ (\texttt{BRONZE} events) or $\geq 0.5$ (\texttt{GOLD} events) and the expected background counts become low \citep[][fig. 18]{IceCube:2016umi}. When converting all-flavour neutrino (and anti-neutrino) intensity into the muon one we use the factor of $1/3$ due to neutrino oscillations in vacuum.
        \item Having obtained the light curve of the \textit{IceCube} muon neutrino yield $\dot{N}_{\nu}^{\mathrm{UL}}(t_{i})$, we average it over the whole range of the \textit{Fermi}-LAT observation time to obtain the mean yearly expected yield.
    \end{enumerate}

    The obtained \textit{IceCube} neutrino light curve and expected mean neutrino yield are presented in Sect.~\ref{sec:neutrino_light_curve}.

    \subsection{Method of estimating the contribution of \textit{Fermi}-LAT FSRQs to the \textit{IceCube} neutrino flux}\label{sec:fsrqs_contribution_estimates}

    This section describes our method to roughly estimate the expected \textit{IceCube} muon neutrino yield at energies $\geq 100$~TeV from all \textit{Fermi}-LAT FSRQs within the framework of the single-zone model used to describe the brightest blazar flare. We use the fourth \textit{Fermi}-LAT source catalogue 4FGL, version \texttt{gll\_psc\_v35.fit} \citep{2020ApJS..247...33A} and select all objects with flags \texttt{fsrq} or \texttt{FSRQ}. For each of the FSRQs, we extract its light curve $\Phi^{\mathrm{o}}_{\gamma}(t_{i})$ from the LCR or use the yearly-binned light curve from the catalogue if the object is not present in the LCR. Then we repeat steps (vi), (vii), and (viii) described in the previous section using light curves $\Phi^{\mathrm{o}}_{\gamma}(t_{i})$ of the objects of interest instead of the light curve of 3C 454.3 and the \textit{IceCube} effective area at the declination corresponding to the object of interest but keeping the shape of the neutrino SED and $\nu-\gamma$ integral flux proportionality constant $a$ from that of 3C 454.3. This method can give only a rough estimate of the expected neutrino yields from FSRQs assuming their neutrino SED shape follows the one obtained for 3C 454.3 for MJD 55517--55524 and it does not take into account that for different objects the dissipation radius $x$, Doppler factor $\delta$, proton-to-electron energy density ratio $\rho_{p/e}$, and acceleration parameter $\xi_{\mathrm{acc}}$ can vary as well as the eight parameters of the leptonic model and parameters defining the photon fields.
    
    To estimate the neutrino flux observed by \textit{IceCube} at energies $E^{o}_{\nu} \geq 100$~TeV, we sum up all signalnesses of all the \textit{IceCube} alert tracks\footnote{excluding those giving a signal in the \textit{IceTop} detector} from the public \textit{IceCat-1} catalogue \citep{2023ApJS..269...25A}. This results in $\approx 12$ ``pure'' astrophysical muon neutrinos observed by IceCube at energies $E^{o}_{\nu} \geq 100$~TeV.

    The estimated contribution of the \textit{Fermi}-LAT FSRQs to the IceCube neutrino flux is presented in Sect.~\ref{sec:IceCat1}.

\section{Results of the leptohadronic model}\label{sec:results_leptohadronic}

    \begin{figure*}
        \centering
         \includegraphics[width=1.5\columnwidth]{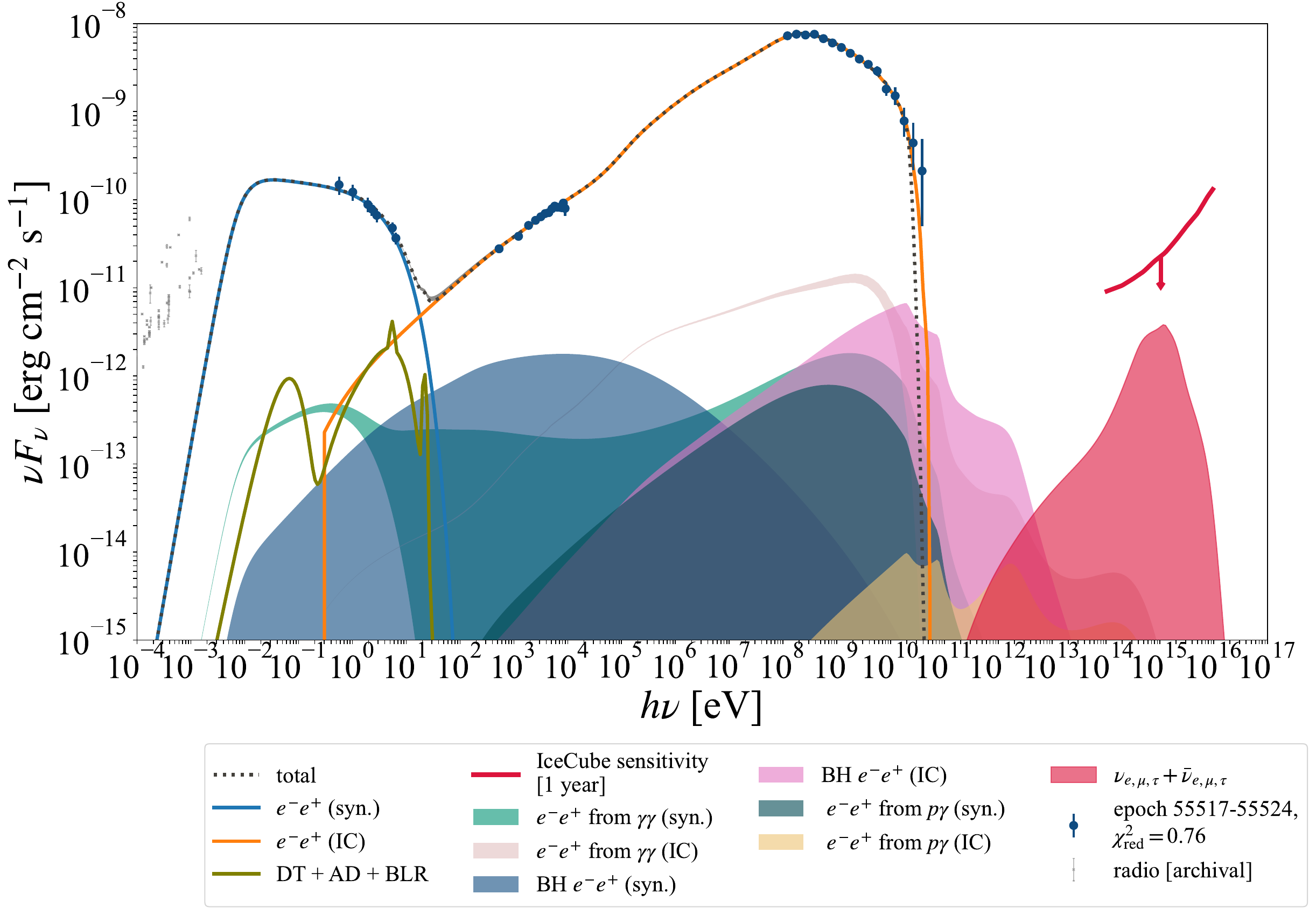}
         \caption{The photon and neutrino SED with model curves for the leptohadronic ``standing-feature'' model with a fit to the observed SED averaged over MJD 55517--5524 with proton acceleration parameter $\xi_{\mathrm{acc}} = 10^{7}$. Shaded regions correspond to the spread of proton luminosities $0 \leq u^{\prime}_{p} \leq u^{\prime\mathrm{UL}}_{p}$, where $u^{\prime\mathrm{UL}}_{p}$ is the UL corresponding to the Bayes factor with respect to the best-fit leptonic model $BF = 10^{-2}$ ($\approx 3 \sigma$ Gaussian equivalent). The crimson solid curve with a downward arrow shows the \textit{IceCube} one-year sensitivity (see Eq.~\ref{eq:icecube_sensitivity}) for the UL of 2.44 muon neutrinos between $E^{o \, \mathrm{min}}_{\nu} = 10^{14}$~eV and $E^{o \, \mathrm{max}}_{\nu} = 10^{16}$~eV.}
         \label{fig:leptohadronic}
    \end{figure*}
    
    \begin{figure}
         \includegraphics[width=0.90\columnwidth]{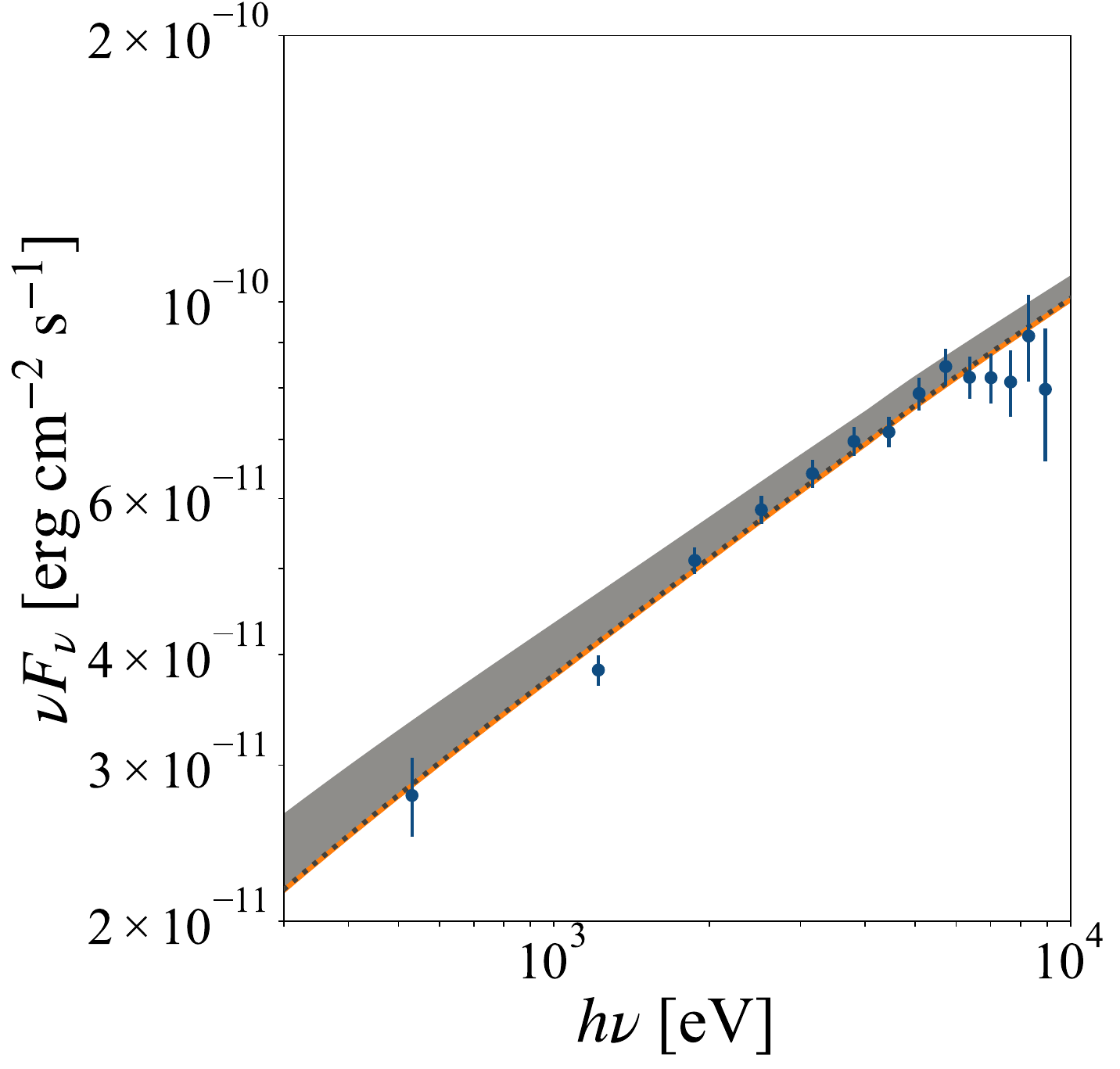}
         \caption{The inset of Fig.~\ref{fig:leptohadronic} focusing on the X-ray energy range. The solid orange curve with black dots shows the leptonic fit to the data, while the grey band shows the additional hadronic contribution due to the Bethe-Heitler synchrotron photons and the synchrotron and IC radiation by $e^{+}e^{-}$ pairs produced in $\gamma\gamma$-initiated cascades. The upper boundary of the grey band corresponds to the case of proton energy density tripled w.r.t. the one we considered as the upper limit: $u^{\prime}_{p} = 3 \times u^{\prime \mathrm{UL}}_{p}$.}
         \label{fig:X_ray_inset}
    \end{figure}
    
    \begin{figure*}
         \includegraphics[width=2\columnwidth]{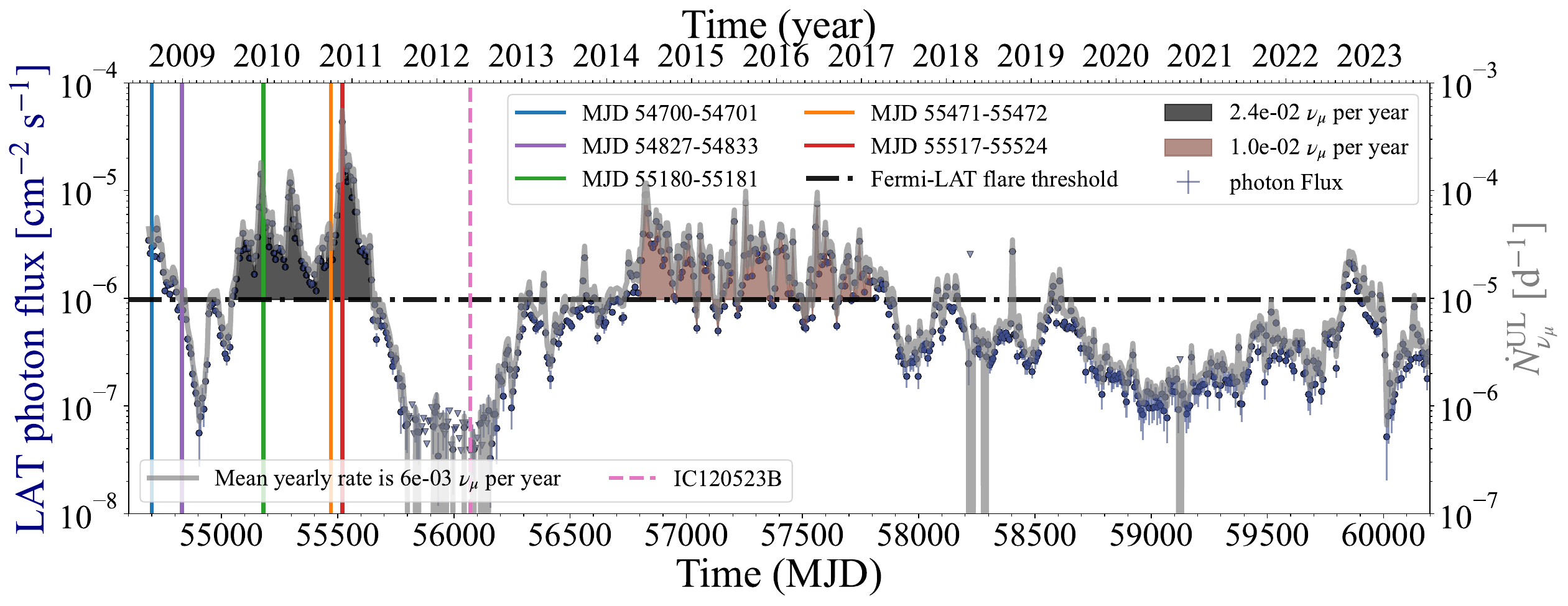}
         \caption{The light curve of 3C 454.3 based on the \textit{Fermi}-LAT $\geq 100$~MeV photon flux (blue data points with error bars for detections if the source-detection test statistic $TS \geq 4$ and triangles for ULs otherwise, left ordinate axis) from the \textit{Fermi}-LAT LCR \citep{pyLCR}, and the corresponding \textit{IceCube} daily muon neutrino yield for energies $E^{o}_{\nu} \geq 100$~TeV (grey curve, right ordinate axis). The \textit{Fermi}-LAT flare threshold is defined as the 68\% percentile of the distribution of the \textit{Fermi}-LAT weekly-binned photon fluxes. The mean yearly \textit{IceCube} muon neutrino rate for $E^{o}_{\nu} \geq 100$~TeV is shown in the legend along with the rates averaged over the periods of outbursts highlighted with grey and brown shaded regions. The \textit{IceCube} muon neutrino alert IC120523B is shown as the dashed vertical curve.}
         \label{fig:light_curve}
    \end{figure*}
    
    The plot with the leptohadronic ``standing-feature'' model fitting the observed SED for MJD 55517--55524 is shown in Fig. \ref{fig:leptohadronic}. In the leptohadronic model, all relevant processes responsible for the production of secondaries are considered: BH pair production, decay of particles produced in $p\gamma$ collisions, $\gamma\gamma$ absorption with the subsequent synchrotron and IC emission of the $e^+e^-$ pairs produced in these processes. Photon and all-flavour neutrino SEDs produced as a result of these processes are shown in Fig.~\ref{fig:leptohadronic} as colourful shades (see the legend) with their upper boundary corresponding to the UL proton energy density $u^{\prime \mathrm{UL}}_{p} = \rho_{p/e} u^{\prime}_{e}$ with $\rho_{p/e} = 1.3 \times 10^{2}$ and the lower boundary corresponding to the purely leptonic model with $u^{\prime}_{p} = 0$. Proton and muon synchrotron and IC emissions are completely negligible. As in the case of TXS~0506+056 \citep[e.g.,][]{2018ApJ...864...84K}, it is X-ray data which drive the constraints on $\rho_{p/e}$ where at higher values of $\rho_{p/e}$ the synchrotron emission from BH electrons as well as the synchrotron and IC radiation by $e^{+}e^{-}$ pairs produced in $\gamma\gamma$-initiated cascades start overshooting the observed X-ray SED. In Fig.~\ref{fig:leptohadronic}, however, this cascade contribution to the X-ray flux seems to be subdominant, and it appears to the naked eye that an even higher $u^{\prime \mathrm{UL}}_{p}$ would still be allowed by the data. Similar to how it has been done by \citet{2018ApJ...864...84K}, we demonstrate in the zoomed-in X-ray area of the SED plot, how the model SED looks like in case $u^{\prime}_{p} = 3 \times u^{\prime \mathrm{UL}}_{p}$ as shown in Fig.~\ref{fig:X_ray_inset}. As seen in the X-ray inset, in the case when the proton energy density is three times greater than the one we consider as the upper limit $u^{\prime \mathrm{UL}}_{p}$, the model SED exceeds $1\sigma$ error bars in $12$ out of $14$ data points, clearly over-predicting the X-ray flux.

    The obtained proton energy density $u^{\prime \mathrm{UL}}_{p} = \rho_{p/e} u^{\prime}_{e} = 1.1 \times 10^{2}$~erg/cm$^3$ corresponds to apparent jet power of $P^{\mathrm{UL}}_p = \pi c \beta \Gamma^2 R^{\prime 2}_b u^{\prime \mathrm{UL}}_{p} = 1.4 \times 10^{48}$~erg/s while the Eddington luminosity is $L_{\mathrm{Edd}} = 2.1 \times 10^{47}$~erg/s. The fact that the apparent proton jet power exceeds the Eddington luminosity by an order of magnitude implies that these parameters cannot be at this level for a long time if we do not assume the possibility of super-Eddington accretion. Such high power is mainly attributed to a high value of the Lorentz factor $\Gamma = 35$ during the flare and is lower during quieter periods of activity (see Appendix~\ref{appendix:4SEDs}). Alternatively, the value of the proton-to-electron-energy density can be lower than the UL $\rho_{p/e} = 1.3 \times 10^{2}$ or the SMBH mass is higher than we assumed. The derived value of $\rho_{p/e}$ is, however, two orders of magnitude lower than those obtained in some models for TXS~0506+056 \citep[e.g.,][]{2019MNRAS.483L..12C} and of the same order of magnitude as those obtained by \citet[][model LMBB2b]{2018ApJ...864...84K}.

    We superimpose the \textit{IceCube} sensitivity in Fig.~\ref{fig:leptohadronic} obtained as follows. For the neutrino energy range from $E^{o \, \mathrm{min}}_{\nu} = 10^{14}$~eV to $E^{o \, \mathrm{max}}_{\nu} = 10^{16}$~eV, $i = 0,1,...,N_{\mathrm{bins}}$, we calculate the factor
    \begin{equation}
        m_{i} = \frac{N_{\mathrm{max}}^{90\%\mathrm{\,CL}}}{N_{\mathrm{bins}}} \left[ \int_{E^{o}_{i-1}}^{E^{o}_{i}} \frac{1}{3} \frac{dN_{\nu}^{o}}{dE^{o}_{\nu}} A_{\mathrm{eff}}(E^{o}_{\nu};\delta_{\mathrm{obj}}) dE^{o}_{\nu} \times {\Delta T} \right]^{-1}
    \label{eq:icecube_sensitivity}
    \end{equation}
    by which we multiply the upper boundary of the neutrino SED presented in Fig.~\ref{fig:leptohadronic} assessed at the intermediate energy bins $E^{o}_{i}$ between $E^{o \, \mathrm{min}}_{\nu}$ and $E^{o \, \mathrm{max}}_{\nu}$ with the number of bins $N_{\mathrm{bins}} = 30 \times \lg(E^{o \, \mathrm{max}}_{\nu} / E^{o \, \mathrm{min}}_{\nu})$. $A_{\mathrm{eff}}(E^{o}_{\nu};\delta_{\mathrm{obj}})$ is the \textit{IceCube} muon neutrino effective area at the declination of 3C~454.3 $\delta_{\mathrm{obj}} = 16.15^{\circ}$ \citep{IceCubeDataRelease, 2021arXiv210109836I, 2020PhRvL.124e1103A}. The number $N_{\mathrm{max}}^{90\%\mathrm{\,CL}} = 2.44$ (muon neutrinos per whole energy range from $E^{o \, \mathrm{min}}_{\nu}$ to $E^{o \, \mathrm{max}}_{\nu}$) is the UL at 90\% confidence level from \citet[][table XII]{1998PhRvD..57.3873F} in case of zero background and zero observed events. The time period considered is $\Delta T = \mathrm{1 \, yr}$. The factor of $1/3$ accounts for neutrino oscillations in vacuum. Thus, the crimson solid curve with a downward arrow in Fig.~\ref{fig:leptohadronic} shows a one-year \textit{IceCube} sensitivity [90\% confidence UL] for the 3C 454.3 neutrino spectrum presented as the crimson shaded region in Fig.~\ref{fig:leptohadronic}.
    
\subsection{\textit{IceCube} expected neutrino yield and light curve of 3C 454.3}
\label{sec:neutrino_light_curve}

    The obtained \textit{IceCube} neutrino light curve superimposed on the \textit{Fermi}-LAT one for 3C 454.3 is presented in Fig. \ref{fig:light_curve}. The four other periods used to obtain the $\nu-\gamma$ relation between integral fluxes are indicated with vertical lines in Fig. \ref{fig:light_curve}, their SEDs and best-fit parameters are presented in Appendix~\ref{appendix:4SEDs}. The average \textit{IceCube} neutrino yield for 3C 454.3 for $E^{o}_{\nu} \geq 100$~TeV is $\sim 6 \times 10^{-3}$ muon neutrinos per year.

    \citet{Capel:2022cnm} found that, to be consistent with a single coincident observation of a neutrino alert and flaring $\gamma$-ray blazar \citep[i.e. the association of the flare in TXS 0506+056 with the neutrino alert IC170922A,][]{2018Sci...361.1378I}, for FSRQs, the ratio $Y_{\nu \gamma}$ (see their eq. 18 and fig. 8) of the neutrino energy flux in the range $10$~TeV--$100$~PeV over the $\gamma$-ray energy flux in the range $1$--$100$~GeV must be $10^{-5} \lesssim Y_{\nu \gamma} \lesssim 10^{-2}$ if neutrinos are produced in both quite and flaring states of FSRQs, or $10^{-4} \lesssim Y_{\nu \gamma} \lesssim 10^{0}$ if neutrinos are produced only during flares. In our fiducial leptohadronic model for MJD 55517--55524, we find $Y_{\nu \gamma} \approx 8 \times 10^{-4}$, which is in good agreement with both constraints by \citet{Capel:2022cnm}.

\subsubsection{On the \textit{IceCube} neutrino alert IC120523B associated with 3C~454.3}\label{sec:IC120523B}

    In \textit{IceCat-1} \citep{2023ApJS..269...25A}, there is one muon neutrino event with estimated energy $E^{o}_{\nu} = 168$~TeV and signalness (probability of the neutrino being astrophysical) $0.49$ positionally associated with 3C~454.3 albeit not contemporaneous with the studied flare\footnote{In the original version of the catalogue presented in the article by \citet{2023ApJS..269...25A} this event due to a typo was called IC120523A, having the same name as another alert. However, in the updated version of \textit{IceCat-1} \citep{IceCat-1-dataset}, the alert is called IC120523B.}. Taking the mean rate of $6 \times 10^{-3}$ muon neutrinos per year from our estimations for 3C~454.3 and the duration of \textit{IceCube} observations of $\approx 10$ years, we obtain the Poisson probability of detecting one or more muon neutrinos with energies $\geq 100$~TeV from 3C~454.3 to be $\left(1 - \exp[-6\times 10^{-3} \times 10] \right) \approx 0.06$. The obtained value indicates that on a long-period average, our model agrees with the single positional association of a muon neutrino with 3C 454.3, especially in light of the possible Eddington bias for cosmic neutrino sources \citep{2019A&A...622L...9S}. The fact that the neutrino event was on MJD 56070.6 during the global minimum of the $\gamma$-ray activity (see Fig.~\ref{fig:light_curve}) may indicate (assuming the neutrino is not atmospheric) that the neutrino emission in blazars is lagging behind the leptonic electromagnetic emission (e.g., due to slowness of photopion processes, see Fig.~\ref{fig:timescales}). If we assume the protons continue to lose their energy under the same conditions as during MJD 55517--55524, then 99\% of the neutrino fluence would be emitted in less than 450 days after the peak of the flare, while the alert came 551 days after the flare, which makes the association of this neutrino with the flare improbable. However, if the Lorentz factor of the emitting region drops to $\Gamma \approx 10$ (typical for more quiescent states, see Appendix~\ref{appendix:4SEDs}), 99\% of the neutrino fluence would be emitted in $\sim 1300$ days, which covers well the neutrino-flare lag of 551 days. On the other hand, the expected neutrino yield from a source with a lower Lorentz factor is lower. If there is a time lag between neutrino and $\gamma$-ray emission, it may explain why 3C~454.3 was among the three brightest neutrino candidate sources in the search performed by \citet{2023ApJ...954...75A} with $\gamma$-ray average flux weights (in agreement with our long-term average estimate) but was not among the best ten candidates in their search with monthly flux weights since they did not take into account a possible $\nu-\gamma$ time lag. The detailed investigation of the properties of delayed neutrino emission is beyond the scope of the paper.

\subsection{Contribution of the 4FGL FSRQs to the \textit{IceCube} neutrino flux at energies $E^{o}_{\nu} \geq 100$~TeV}\label{sec:IceCat1}
    The estimated average neutrino yield in \textit{IceCube} from all 4FGL FSRQs is $\sim 6 \times 10^{-2}$ muon neutrinos per year given the linear relation between the $\nu$ and $\gamma$-ray integral fluxes and Eq.~(\ref{eq:neutrino_gamma_sed_scaling}) for linking neutrino SEDs and \textit{Fermi}-LAT light curves from the LCR. Given $\approx 12$ ``pure'' muon neutrinos per year observed by \textit{IceCube}, it implies the 4FGL FSRQs contribute at the level up to $\sim 0.5$ \% of the \textit{IceCat-1} yearly muon neutrino rate at energies $E^{o}_{\nu} \geq 100$~TeV. Our model-dependent upper limit is in agreement with (i) the work by the \textit{IceCube} Collaboration \citep{IceCube:2016qvd}, where it was shown that $<27\%$ ($<50\%$) of the diffuse neutrino flux in the energy range between $\sim 10$~TeV and $\sim 2$~PeV assuming its power-law spectral index of $-2.5$ ($-2.2$) can be produced by the \textit{Fermi}-LAT blazars from the second AGN catalogue; (ii) the work by \citet{2018ApJ...865..124M}, which predicts that the blazar flares contribute up to $1\%-10\%$ of the sub-PeV neutrino diffuse flux; (iii) the paper by \citet{Hooper:2018wyk}, which constrained the \textit{Fermi}-LAT FSRQs' contribution to the neutrino flux to be at the level of $\lesssim 10\%$; (iv) the study by \citet[][see their fig. 12]{2023ApJ...954..194Y}, where, for the linear relation between the $\gamma$-ray and $\nu$ fluxes, the \textit{Fermi}-LAT 4LAC FSRQs were found to contribute up to $\lesssim 100\%$ of the \textit{IceCube} diffuse high-energy neutrino flux. Thus, our rough model-driven estimate of the maximum contribution of \textit{Fermi}-LAT FSRQs to the \textit{IceCube} neutrino flux is lower by one to two orders of magnitude than the upper limits obtained in the references above and agrees with the conclusion by \citet{2023ApJ...954...75A} that a small fraction, $<1\%$, of bright AGNs emits neutrinos that pass the \textit{IceCube} alert criteria.

    \subsubsection{The case of TXS 0506+056}\label{sec:TXS0506+056}
        Although TXS 0506+056 is not classified as an FSRQ in the 4FGL catalogue \citep{2020ApJS..247...33A}, it still may belong to this class as pointed out by \citet{2019MNRAS.484L.104P} showing that this source may host BLR and DT external photon fields and a model similar to the one used for 3C 454.3 may describe the geometry of this source. We apply the procedure described in Sects. \ref{sec:neutrino_yield_estimates} and \ref{sec:fsrqs_contribution_estimates} to TXS 0506+056. This results in $\sim 4 \times 10^{-4}$ \textit{IceCube} muon neutrinos per year on long-term average and in $\sim 10^{-3}$ muon neutrinos per year within $\pm 180$ days around the IceCube-170922A arrival time (at energies $E^{o}_{\nu} \geq 100$~TeV) which lies within the range of different model predictions of \citet[][table 9]{2018ApJ...864...84K} who performed dedicated modelling for this source specifically with various parametrization of the external photon field.

    \subsubsection{Expected neutrino yields in other neutrino telescopes}\label{sec:future_neutrino_telescopes}

        Other neutrino telescopes under construction, such as the Pacific Ocean Neutrino Experiment \citep[\textit{P-ONE},][]{P-ONE:2020ljt}, the Baikal Gigaton Volume Detector \citep[\textit{Baikal-GVD},][]{Avrorin:2011zzc}, and the Cubic Kilometre Neutrino Telescope \citep[\textit{KM3NeT},][]{KM3Net:2016zxf} will have the expected yearly muon neutrino rate from \textit{Fermi}-LAT FSRQs comparable to the one in \textit{IceCube}. Their combined statistics with \textit{IceCube} will result in the detection of a muon neutrino with energy $\geq 100$~TeV from \textit{Fermi}-LAT FSRQs every 4-5 years. The projected Tropical Deep-sea Neutrino Telescope \citep[\textit{TRIDENT},][]{TRIDENT:2022hql} with its effective area a factor of $\sim 10$ larger than that of \textit{IceCube} will detect on average $\sim 1$ muon neutrino with energy $\geq 100$~TeV from flares in \textit{Fermi}-LAT FSRQs every other year. A similar detection rate can be expected from the proposed \textit{IceCube-Gen2} \citep{IceCube-Gen2:2023vtj}. Therefore, in our model, next-generation neutrino telescopes are expected to detect approximately one multimessenger ($\gamma + \nu_{\mu}$) flare per year from \textit{Fermi}-LAT FSRQs.

\section{Discussion}
\label{sec:discussion}

    \subsection{Location of the $\gamma$-ray emitting region beyond $R_{\mathrm{BLR}}$}
    In contrast to the results of \citet{2011MNRAS.410..368B, 2015RAA....15.1455H,2021MNRAS.504.5074S} who also modelled the SED of 3C 454.3 in its flaring states, we could not obtain a fit with the dissipation radius $x < R_{\mathrm{BLR}}$ lying within the BLR due to the $\gamma\gamma$ absorption as explained in Sect. \ref{sec:dissipation_radius}. On the other hand, our conclusion $x > R_{\mathrm{BLR}}$ that the dissipation radius lies beyond the outer radius of the BLR but not very far from it is in agreement with recent studies by \citet{2017ApJS..228....1Z,2018MNRAS.477.4749C,2019ApJ...877...39M,Kundu:2025nbd}. \citet{2022PhRvD.105b3005W} suggested that multi-wavelength flares (contemporaneous optical, $X-$, and $\gamma$-ray) should occur at intermediate distances from the SMBH as in the case for the flare of 3C~454.3 we considered, however this may be not the case for all blazars: e.g., FSRQ PKS 1424-418 shows evidence of $\gamma\gamma$ absorption inside the BLR during flares \citep{Agarwal:2024vld}. Different conclusions about the location $x$ of the $\gamma$-ray emitting region may be explained by different assumptions about the AD luminosity (see Table \ref{table:disc}) and the BLR size and/or geometry \citep{2017MNRAS.464..152A,2021ApJ...920...30N,2024Univ...10...29N,2024arXiv240811292Z,Hopkins24a}. A systematic study of the influence of the uncertainties of the BLR and AD parameters on the estimation of $x$ is beyond the scope of the current work.

    \subsection{``Standing feature'' vs. ``moving blob''}
    \label{sec:standing_feature_discussion}
    
    The fact that the dissipation radius $x$ does not change much during the modelling and the model with a ``moving blob'' fails to describe the seven consecutive SEDs can be interpreted in favour of a model like the turbulent, extreme multi-zone (TEMZ) model by \citet{2014ApJ...780...87M} or other models invoking standing or slowly moving bright features in blazar jets like standing shocks \citep{1985ApJ...298..114M,1988ApJ...334..539D} or reconnection layers \citep{2016MNRAS.462...48S,2016MNRAS.462.3325P}. The latter model can also describe the day-to-day change in the parameters of the leptonic model stemming from an injection of a new plasmoid with parameters different from each previous one since the generation of new plasmoids is expected to be stochastic. Thus, works focusing rather on the statistical properties of blazar variability as a stochastic process might be applicable \citep{2024PhRvD.110d3027D,2024ApJ...965..104F,2024ApJ...974....1T}.

    The stochastic nature of blazar variability may hinder attempts to model the variability with regular changes of one-zone model parameters and/or invoking of a ``moving blob'' as we tried to do in Sect.~\ref{sec:moving_blob_model_description}. Attempts of \citet{2016ApJ...826...54D,2016PhDT........66D} in changing some of the parameters with its profile obeying a functional form of a rapid increase with an exponential decay did not result in a good description of the data. \citet{2016ApJ...830...94F} concluded that he cannot describe the 3C 454.3 variability with a blob moving from $x < R_{\mathrm{BLR}}$ to $x > R_{\mathrm{BLR}}$ as well as with a model of a blob changing its radius \citep{2024FrASS..1184234F}. On the other hand, models invoking a single moving and expanding blob may successfully explain the time lag between radio and $\gamma$-ray light curves \citep{2022A&A...657A..20B,2022A&A...658A.173T,2023A&A...669A.151Z} observed in many blazars \citep[e.g.,][]{2022MNRAS.510..469K}.

    \subsection{Contribution of \textit{Fermi}-LAT FSRQs to the neutrino flux}
    \label{sec:poor_neutrino}
    
    Neutrinos with observable energies $E^{o}_{\nu}$ generally come from protons with observable energies $E^{o}_{p} \sim 20 E^{o}_{\nu}$ \citep[for detailed studies of the $E^{o}_{p}$-$E^{o}_{\nu}$ relation see][]{2021JCAP...03..050R}. Given $E^{o}_{p} = \delta E^{\prime}_{p} / (1 + z)$, Lorentz factor $\Gamma \approx 25$, and the redshift $z = 0.859$ for 3C 454.3, we can get that the proton energy in the blob frame roughly equals the observed neutrino energy $E^{\prime}_{p} \sim E^{o}_{\nu}$. Thus, if the neutrino SED peaks at energies between $10^{14}-10^{16}$ eV (as in Fig. \ref{fig:leptohadronic}), it is produced by the protons with energies $E^{\prime}_{p} \lesssim 10^{16}$ eV in the blob frame. In our leptohadronic model, for protons at these energies, as can be seen in Fig.~\ref{fig:timescales}, the timescale of photopion losses is much greater than the timescale of the flare implying that the $p\gamma$ process in the blob is optically thin $\tau_{p\gamma} \ll 1$.

    If one wants to increase $\tau_{p\gamma}$, the most efficient way to do so is to move the production zone inside the BLR since beyond $R_{\mathrm{BLR}}$ the BLR photon field density falls vary rapidly with increasing $x$ according to \citet{2009MNRAS.397..985G} and the apparent energies of photons decrease according to Eq. (\ref{eq:line_transformation}). By moving the dissipation radius $x$ into the BLR one can increase the expected neutrino flux by a factor of $\sim 7 \times 10^1$ (on average) but this makes it impossible to describe the observed $\gamma$-rays due to severe $\gamma\gamma$ absorption as elaborated in Sect.~\ref{sec:dissipation_radius} (see also \citealp{2016PhRvL.116g1101M,2020PhRvL.125a1101M}).

    Protons can have an acceleration timescale different from the one of electrons. By means of a simple scan (see Appendix~\ref{appendix:xi}), we obtained that for $10^{4} \lesssim \xi_{\mathrm{acc}} \lesssim 10^{5}$ neutrino production reaches maximum and is $\sim 3$ times more efficient than for the case of $\xi_{\mathrm{acc}} \sim 10^{7}$. Thus, for a model with a separate neutrino production zone inside the BLR and optimal acceleration timescale, one can increase the neutrino yield by a factor of $\sim 2 \times 10^2$, raising the potential contribution of 4FGL FSRQs to the \textit{IceCat-1} neutrinos up to $100$\%. This is just a rough estimate and does not take into account possible changes in other model parameters. The consideration of a model with several emission zones is beyond the scope of the current paper.

    \subsection{A possible delay between neutrino emission and electromagnetic flares in blazars}\label{sec:neutrino_lag}

    For neutrino SEDs peaking at a few hundred TeV~--~a few PeV, the energy-loss time scale of proton photopion production exceeds the timescale of a typical multi-wavelength flare by almost three orders of magnitude. Depending on the Doppler factor, dissipation radius and parameters of the external photon field, if the conditions for $p\gamma$ interactions hold for sufficiently long, the time window within which a significant neutrino fluence is expected can be as wide as $\sim 10^{3}$~d implying the possibility of a substantial months-to-years lag between a potential neutrino detection and the corresponding multi-wavelength flare. This circumstance can be accounted for in future searches for correlations between high-energy neutrinos and their electromagnetic counterparts.

\section{Conclusions}
\label{sec:conclusions}

We studied in detail the brightest \textit{Fermi}-LAT blazar flare in FSRQ 3C 454.3 that occurred in November 2010 having obtained seven consecutive multi-wavelength SEDs of each day of the flare around its peak (MJD 55517--55524). By modelling the flare SEDs with \texttt{AM$^{3}$}, we showed that the seven consecutive SEDs cannot be described with the synchrotron and IC emission from a single blob moving down the jet whereas the ``standing-feature'' model in which multiple blobs are passing through a quasi-stationary feature located beyond the outer radius of the BLR describes the data well. A pure leptonic model is, in principle, sufficient to describe the seven observed consecutive SEDs.

If protons are co-accelerated with electrons, the X-ray data constrain the proton energy density to be at maximum $\rho_{p/e} \approx 1.3 \times 10^2$ times greater than the electron energy density. By applying the same approach to four other periods of activity of 3C~454.3, we determined that the integral neutrino and $\gamma$-ray fluxes are approximately linearly proportional. Applying the obtained $\nu-\gamma$ scaling to the whole period of \textit{Fermi}-LAT observations using the light curve from the \textit{Fermi} LCR we constructed a neutrino light curve of 3C~454.3. We estimate that on average 3C~454.3 produces $6 \times 10^{-3}$ \textit{IceCube} muon neutrinos per year at energies $E^{o}_{\nu} \geq 100$~TeV in the framework of the single-zone ``standing-feature'' leptohadronic model. The proton maximum energy is not directly constrained by the electromagnetic flaring SED of 3C~454.3. Under more optimistic assumptions about the maximum proton energy (i.e.\ larger values $E^{\prime \mathrm{max}}_{p}$, lower values $\xi_{\mathrm{acc}}$), the maximum neutrino yield is similar or a factor of a few different from that in our benchmark model but the peak neutrino energy can reach $\sim 100$~PeV, consistent with the energy of the recently discovered $>$100 PeV neutrino KM3-230213A~\citep{KM3NeT2025}.

Extrapolation of the 3C 454.3 model to the whole list of \textit{Fermi}-LAT 4FGL FSRQs shows that they can account for up to $\sim[0.5 (\rho_{p/e}/130)]$\% of the \textit{IceCat-1} neutrino alerts at energies $E^{o}_{\nu} \geq 100$~TeV implying that FSRQs during multi-wavelength flares are not the main source of $\geq 100$~TeV neutrinos observed by \textit{IceCube}. The total expected yield from \textit{Fermi}-LAT FSRQs in the planned \textit{TRIDENT} and \textit{IceCube-Gen2} neutrino telescopes will be about $\dot{N}_{\mathrm{yr}}(E^{o}_{\nu} \geq 100 \mathrm{\, TeV}) \sim 1 \times (\rho_{p/e}/10^2)$ muon neutrinos per year. Next-generation neutrino telescopes are thus expected to detect approximately one multimessenger ($\gamma$ + $\nu_{\mu}$) flare per year from FSRQs. We have shown that the electromagnetic flare peak can precede the neutrino arrival by months to years and that the Doppler factor of the jet plays a crucial role in determining this time delay.

\section*{Acknowledgements}

We thank Karri Koljonen for help with the \texttt{ISIS} software, D. Ehlert for sharing his functions for calculating photon fields from the AD, and M. Petropoulou, X. Rodrigues, M. Kachelrieß, D. Khangulyan, T. Dzhatdoev, M. Turchetta, J. Simpson, and B. Sen for helpful discussions. We thank the anonymous Referee for their comprehensive review, which allowed us to improve the quality of the manuscript.

This work made use of data supplied by the UK Swift Science Data Centre at the University of Leicester. Part of this work is based on archival data, software or online services provided by the Space Science Data Center -- ASI. This research made use of the \textit{Fermi}-LAT data provided by NASA Goddard Space Flight Center. This paper has made use of up-to-date SMARTS optical/near-infrared light curves that are available at \url{www.astro.yale.edu/smarts/glast/home.php}. Data from the \textit{Steward} Observatory spectropolarimetric monitoring project were used. This research has made use of the NASA/IPAC Infrared Science Archive, which is funded by the National Aeronautics and Space Administration and operated by the California Institute of Technology. This work made use of the following software packages: \texttt{python} \citep{python}, \texttt{astropy} \citep{astropy:2013, astropy:2018, astropy:2022}, \texttt{Jupyter} \citep{2007CSE.....9c..21P, kluyver2016jupyter}, \texttt{matplotlib} \citep{Hunter:2007}, \texttt{numpy} \citep{numpy}, \texttt{scipy} \citep{2020SciPy-NMeth, scipy_4718897}, \texttt{astroquery} \citep{2019AJ....157...98G, astroquery_10799414}, \texttt{pandas} \citep{pandasA,pandasB}, \texttt{\href{https://github.com/fermi-lat/Fermitools-conda/wiki}{fermitools 2.2.0}}, \texttt{fermipy} (v. 1.2.2) \citep{Wood2017}, \texttt{Xspec} \citep{1996ASPC..101...17A}, \texttt{iMINUIT} \citep{iminuit, James:1975dr}, \texttt{snakemake} \citep{Snakemake}, \texttt{pyLCR} \citep{pyLCR}\footnote{\url{https://github.com/dankocevski/pyLCR}}, \texttt{numba} \citep{numba:2015, Numba_7198926}, and \texttt{tqdm} \citep{tqdm_8233425}. This research has made use of NASA's Astrophysics Data System and Pathfinder \citep{pathfinder}. Software citation information aggregated using \texttt{\href{https://www.tomwagg.com/software-citation-station/}{The Software Citation Station}} \citep{software-citation-station-zenodo,software-citation-station-paper}.

\section*{Data Availability}


All data used for obtaining the observed SEDs are publicly available at the corresponding web resources of the instruments cited in Sect.~\ref{sec:data}. The \texttt{AM$^{3}$} program for modelling the observable SEDs is publicly-available \citep{2023ascl.soft12031K}\footnote{\url{https://gitlab.desy.de/am3/am3}}. The derived observational SED data points and our modelling codes are available at the \textit{Zenodo} repository \citep{PodlesnyiOikonomou2025aZenodo}.



\bibliographystyle{mnras}
\bibliography{bib}



\appendix

\section{On the systematic uncertainty of the UV, optical, and IR spectral-energy distributions}
\label{appendix:SMARTS_VS_Steward}
In this appendix, we compare \textit{Steward} and \textit{SMARTS} apparent magnitudes of 3C 454.3 measured within the same nights over the whole history of their joint observations in the R and V bands. The corresponding scatter plots are shown in Figs. \ref{fig:R_band_uncertainty} and \ref{fig:V_band_uncertainty} respectively. The root-mean-square difference (i.e. the square root of the mean squared difference of the magnitude) and the mean difference between \textit{Steward} and \textit{SMARTS} magnitudes are displayed in the titles of Figs. \ref{fig:R_band_uncertainty} and \ref{fig:V_band_uncertainty}. There is a systematic shift in the values of magnitudes between Steward and SMARTS ($0.02-0.04$), which is subdominant to the root-mean-square scatter ($0.08-0.09$). An investigation of the reasons behind these discrepancies is beyond the scope of the paper. We take into account the observed discrepancy as an additional $8\%$ relative systematic uncertainty added quadratically to all IR, UV, and optical SEDs as stated in Sect.~\ref{sec:systematics}.

\begin{figure}
    \includegraphics[width=1\columnwidth]{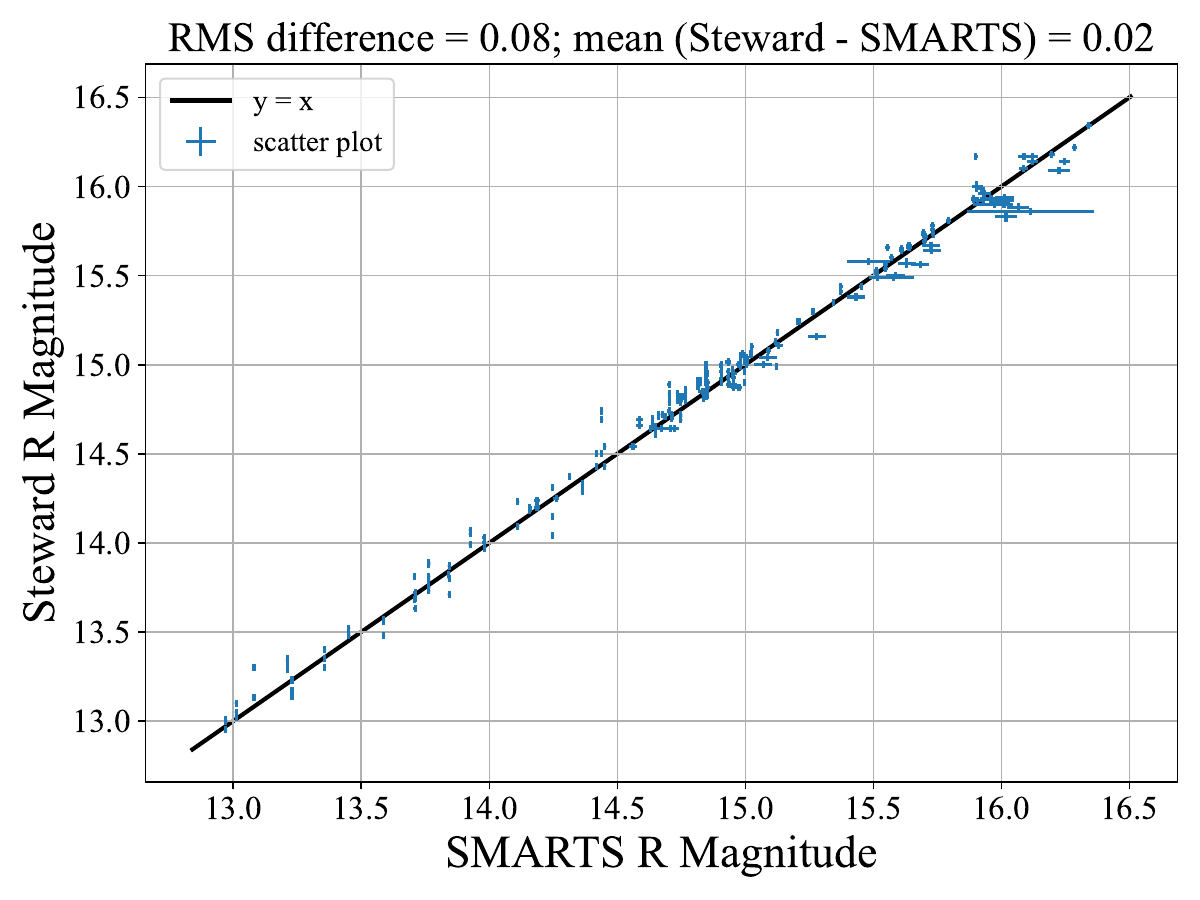}
    \caption{Scatter plot of 3C 454.3 Vega apparent magnitudes measured by \textit{Steward} Observatory and \textit{SMARTS} within same nights in the R band.}
    \label{fig:R_band_uncertainty}
\end{figure}
\begin{figure}
    \includegraphics[width=1\columnwidth]{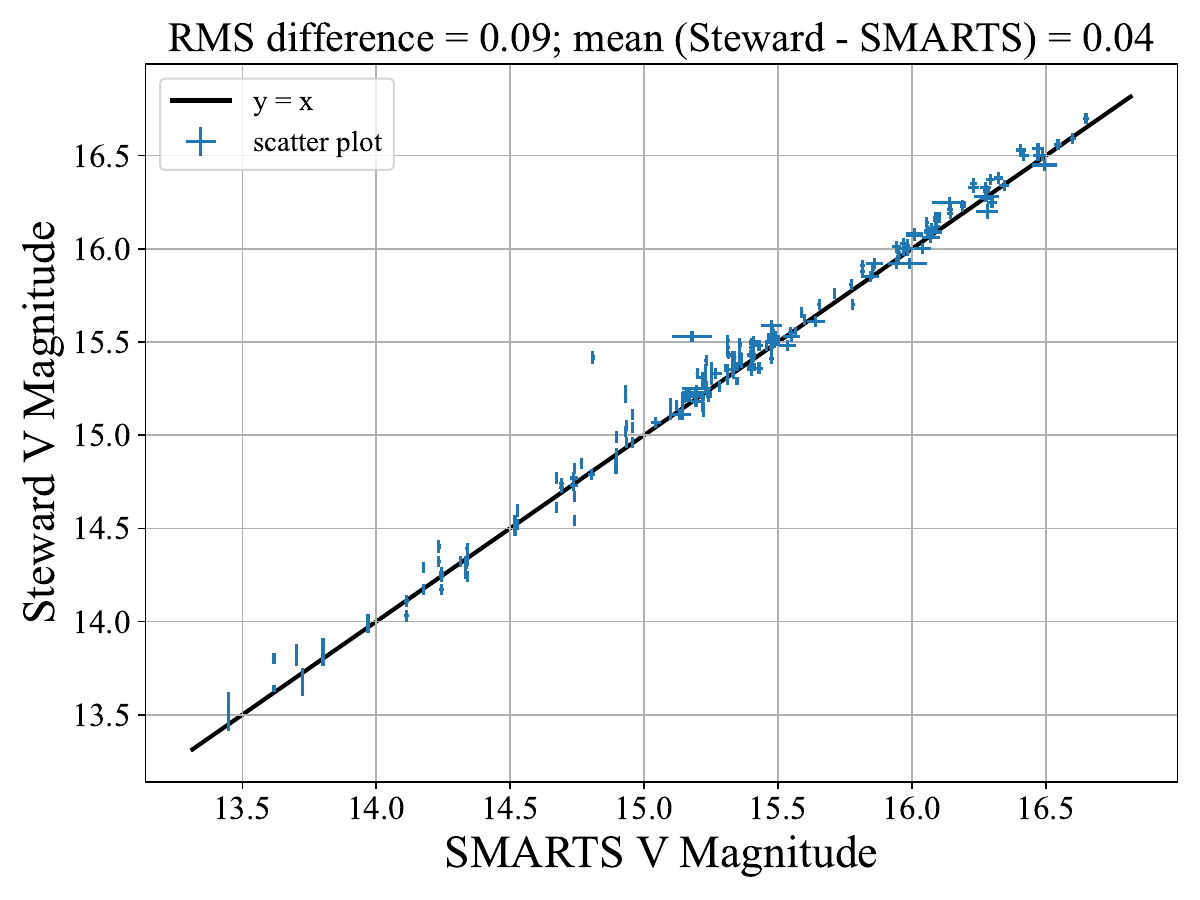}
    \caption{Scatter plot of 3C 454.3 Vega apparent magnitudes measured by \textit{Steward} Observatory and \textit{SMARTS} within same nights in the V band.}
    \label{fig:V_band_uncertainty}
\end{figure}

\section{On the deabsorbed X-ray SED and its uncertainty}\label{appendix:X_ray_deabsorption}
To account for the possible photoelectric absorption of X rays along the LoS, the solar abundance is adopted from \citet{2000ApJ...542..914W} and the photoionization cross-section table $\sigma_{\mathrm{photoel}}(E_i)$ is from \citet{1996ApJ...465..487V}. The initial guess for the hydrogen column density $\kappa$ is set to $\kappa = 6.81 \times 10^{20}$ cm$^{-2}$ according to the average value calculated with the use of the web tool\footnote{\url{https://heasarc.gsfc.nasa.gov/cgi-bin/Tools/w3nh/w3nh.pl?}} within the $0.1^\circ$-cone with the centre at the position of 3C 454.3 on \textit{H} I maps obtained by \citet{2016A&A...594A.116H}, although the column density is left free along with the normalization $N_{X}$, break energy $E_{\mathrm{br}}$, and the two spectral indices ($\Gamma_1$ and $\Gamma_2$) of the broken-power-law X-ray spectrum. Using \texttt{fit}, \texttt{plot\_unfold}, and \texttt{write\_plot} \texttt{ISIS} functions, we extract both the observed $\mathrm{SED}_o(E_i)$ and the deabsorbed $\mathrm{SED}_d(E_i)$ from the best-fit model and take into account the uncertainty related to the deabsorption correction in the following manner:
\begin{equation}
    k_a(E_i) = \frac{\mathrm{SED}_o(E_i)}{\mathrm{SED}_d(E_i)} = \exp{\left(-\kappa \sigma_{\mathrm{photoel}}(E_i)\right)},
\end{equation}
\begin{equation}
    \varsigma_{d}(E_i) = \sqrt{ \left( \frac{\varsigma_{o}(E_i)}{k_a(E_i) }\right)^{2} +  \left( \sigma_{\mathrm{photoel}}(E_i) \mathrm{SED}_d(E_i)  \varsigma_{\kappa} \right)^2},
\end{equation}
\begin{equation}
    \sigma_{\mathrm{photoel}}(E_i) = - \frac{\ln k_a(E_i)}{\kappa},
\end{equation}
where $\varsigma_{d}(E_i)$ is the final uncertainty of the deabsorbed SED after taking into consideration the uncertainty of the hydrogen column density $\varsigma_{\kappa}$, and $\varsigma_{o}(E_i)$ is the statistical uncertainty of the observed $\mathrm{SED}_o(E_i)$ in the $i$th energy bin $E_i$. It is the deabsorbed $\mathrm{SED}_d(E_i)$ and its total uncertainty $\varsigma_{d}(E_i)$ which is used further in modelling the multi-wavelength electromagnetic emission of 3C 454.3 $\varsigma_{\kappa}$ is estimated using \texttt{conf\_loop} \texttt{ISIS} function calculating 68\% confidence interval for each of the free parameters after having obtained the best-fit with the absorbed broken-power-law model. After obtaining $\kappa = 7.7^{\, +1.8}_{-2.4} \times 10^{20}$~cm$^{-2}$ for MJD 55517--55524, we perform the same fit as described in this paragraph for each of the seven one-day-averaged consecutive SEDs allowing $\kappa$ to vary only within the 68\% confidence interval obtained above. This constraint is required to ensure the convergence of the fits for each of the individual days, since otherwise some fits fail to converge. The obtained best-fit values for each of the individual one-day-averaged SEDs as well as for the seven-day-averaged SED are shown in Table~\ref{table:swift_xrt_data_analysis}. The inferred values of $\kappa$ agree with the ones obtained by \citet{2016A&A...594A.116H}.

\section{Spectral-energy distributions for four other periods}
\label{appendix:4SEDs}

Spectral-energy distributions for four other periods indicated with colourful vertical lines in Fig.~\ref{fig:light_curve} are shown in Fig.~\ref{fig:4SEDs}. Table~\ref{table:leptonic_parameters_table} presents the best-fit values of the parameters of the leptonic model for these four periods and MJD 55517--55524 as well.

\begin{table*}
    \centering
    \begin{tabular}{|l|c|c|c|c|c|}
    \toprule
    Parameter & MJD 54700--54701 & MJD 54827--54833 & MJD 55180-55181 & MJD 55471--55472 & MJD 55517--55524 \\
    \midrule
    $L^{\prime}_{e}$ [erg/s] & $6.5 \times 10^{43}$ & $2.3 \times 10^{43}$ & $1.5 \times 10^{43}$ & $8.2 \times 10^{42}$ & $1.7 \times 10^{43}$ \\
    $s_{e}$ & $2.7$ & $2.4$ & $3.3$ & $2.6$ & $2.4$ \\
    $x$ [cm] & $7.3 \times 10^{17}$ & $7.3 \times 10^{17}$ & $7.3 \times 10^{17}$ & $6.6 \times 10^{17}$ & $7.3 \times 10^{17}$ \\
    $R^{\prime}_{b}$ [cm] & $2.9 \times 10^{16}$ & $1.8 \times 10^{16}$ & $1.5 \times 10^{16}$ & $5.8 \times 10^{15}$ & $1.1 \times 10^{16}$ \\
    $B^{\prime}$ [G] & $1.0$ & $9.6 \times10^{-1}$ & $1.7$ & $2.2$ & $1.2$ \\
    $E^{\prime \mathrm{min}}_{e}$ [eV] & $2.2 \times 10^{8}$ & $1.7 \times 10^{8}$ & $2.1 \times 10^{8}$ & $1.1 \times 10^{8}$ & $7.0\times 10^{7}$ \\
    $E^{\prime \mathrm{max}}_{e}$ [eV] & $3.6 \times 10^{9}$ & $6.9 \times 10^{9}$ & $2.8 \times 10^{9}$ & $4.2 \times 10^{9}$ & $1.6 \times 10^{9}$ \\
    $\Gamma$ & $13$ & $11$ & $22$ & $19$ & $35$ \\
    $\chi^{2}_{\mathrm{red}}$ & $1.73$ & $1.95$ & $1.38$ & $0.70$ & $0.76$ \\
    \bottomrule
    \end{tabular}
    \caption{Values of the ``standing-feature'' model parameters providing the best fit to the observed SEDs in the corresponding periods. In all cases $t^{\prime}_{\mathrm{sim}}$ is derived from Eq. (\ref{eq:simulation_time}) for $t^{o}_{\mathrm{av}} = 1$ d. \label{table:leptonic_parameters_table}}
\end{table*}

\begin{figure*}
    \centering
    \subfloat[MJD 54700--54701]{\includegraphics[width=0.50\textwidth]{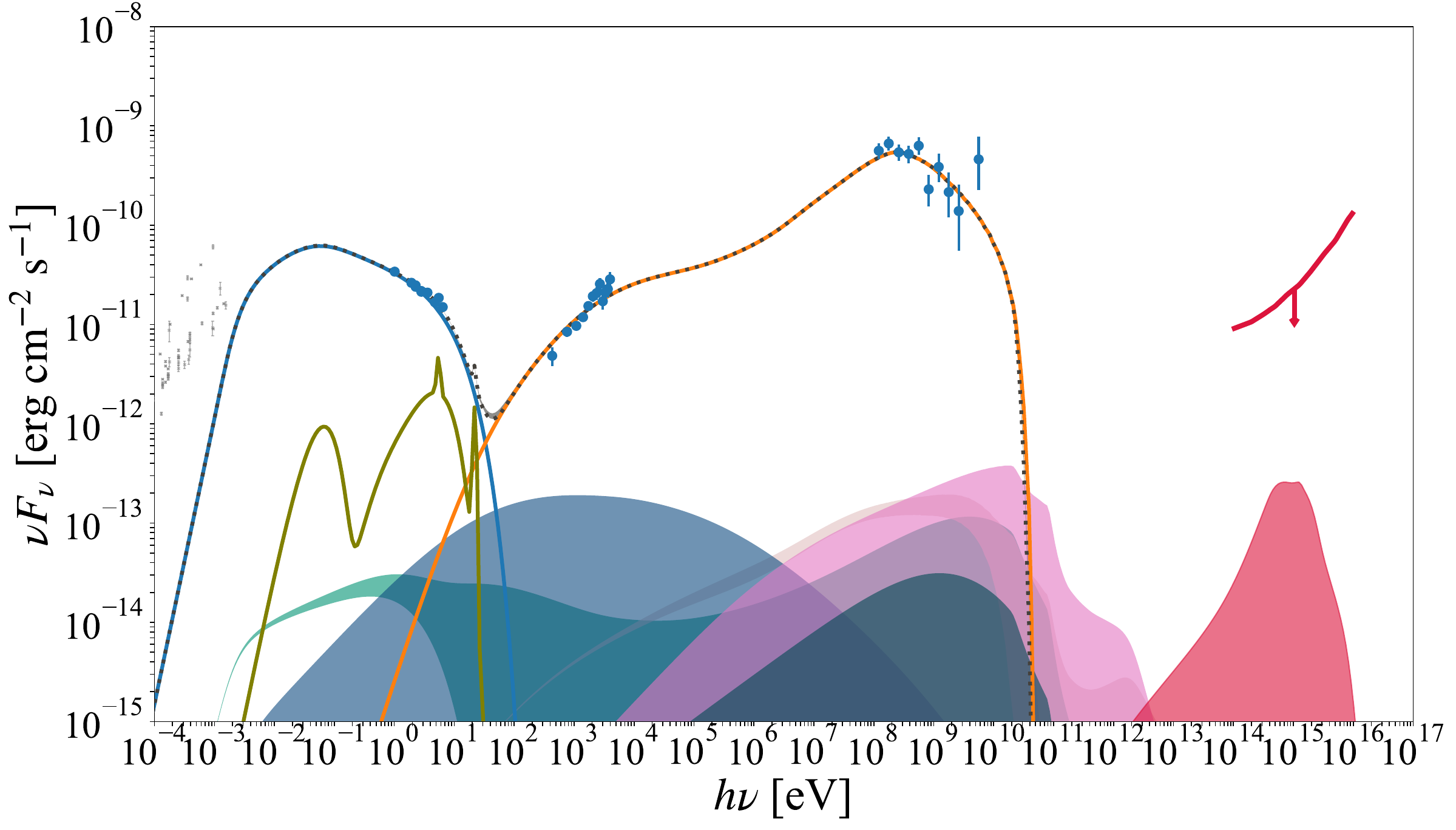}\label{fig:54700-54701}}
    \hfill
    \subfloat[MJD 54827--54833]{\includegraphics[width=0.50\textwidth]{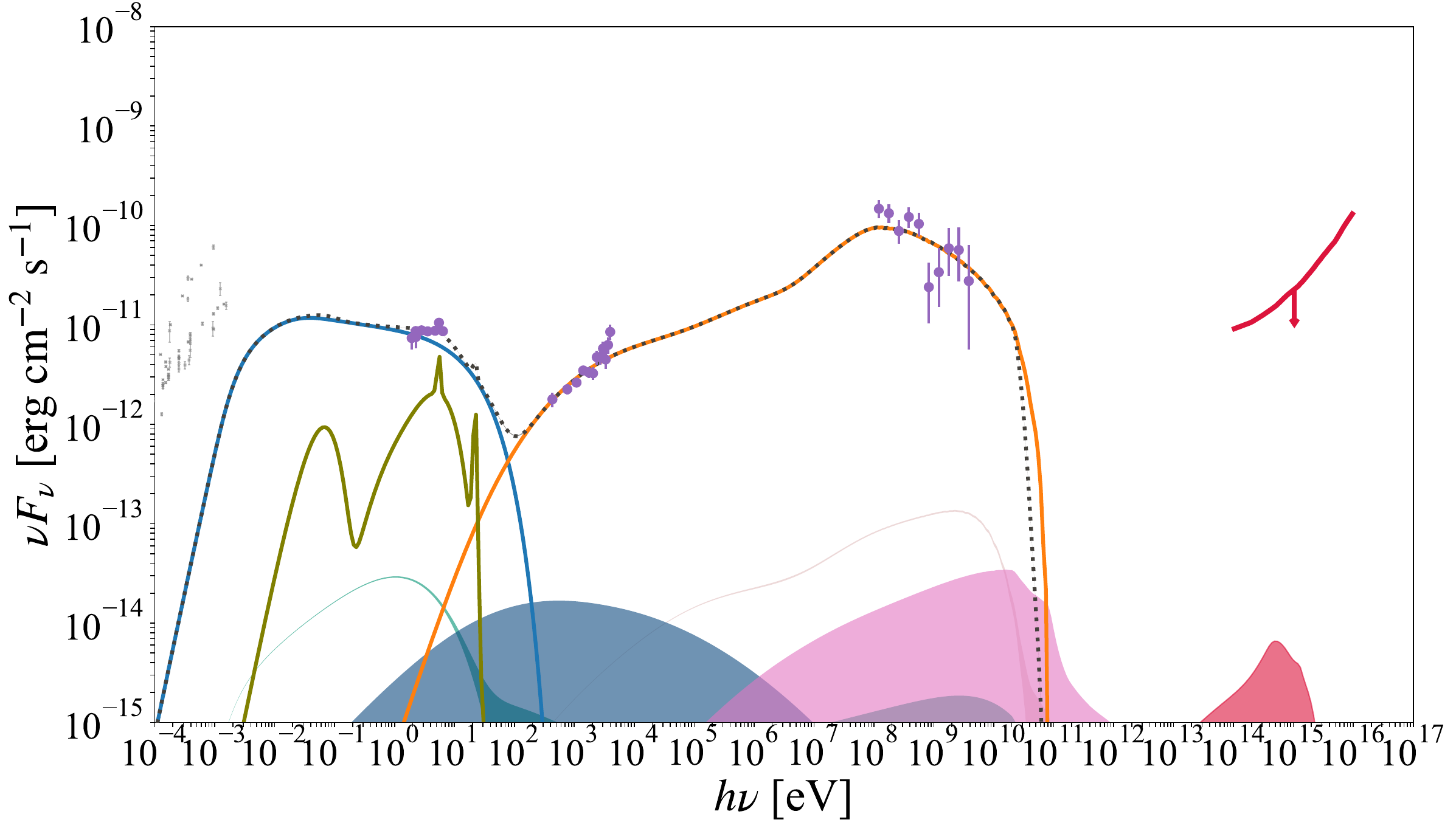}\label{fig:54827-54833}}
    
    \vspace{0.05cm} 
    
    \subfloat[MJD 55180--55181]{\includegraphics[width=0.50\textwidth]{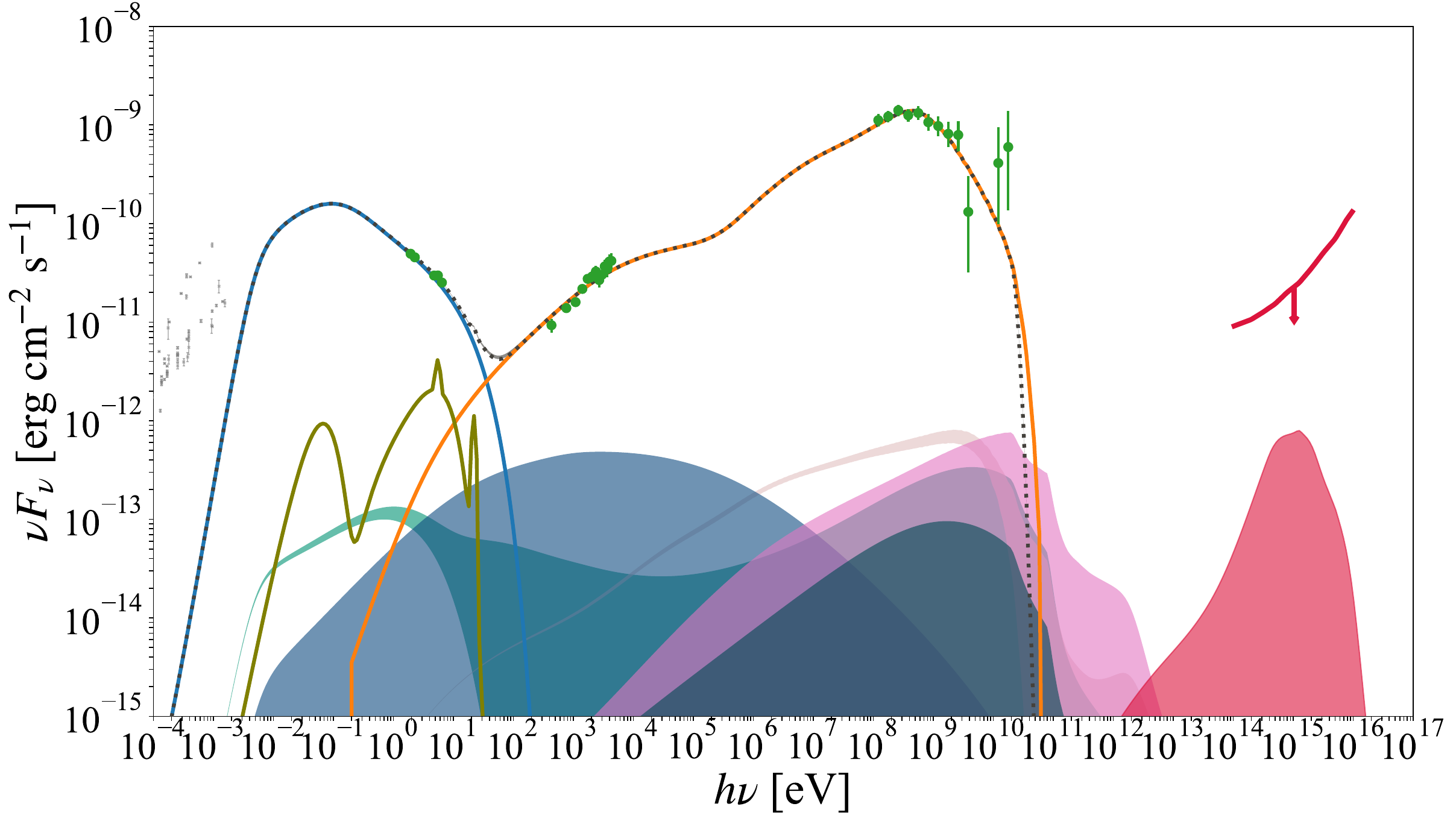}\label{fig:55180-55181}}
    \hfill
    \subfloat[MJD 55471--55472]{\includegraphics[width=0.50\textwidth]{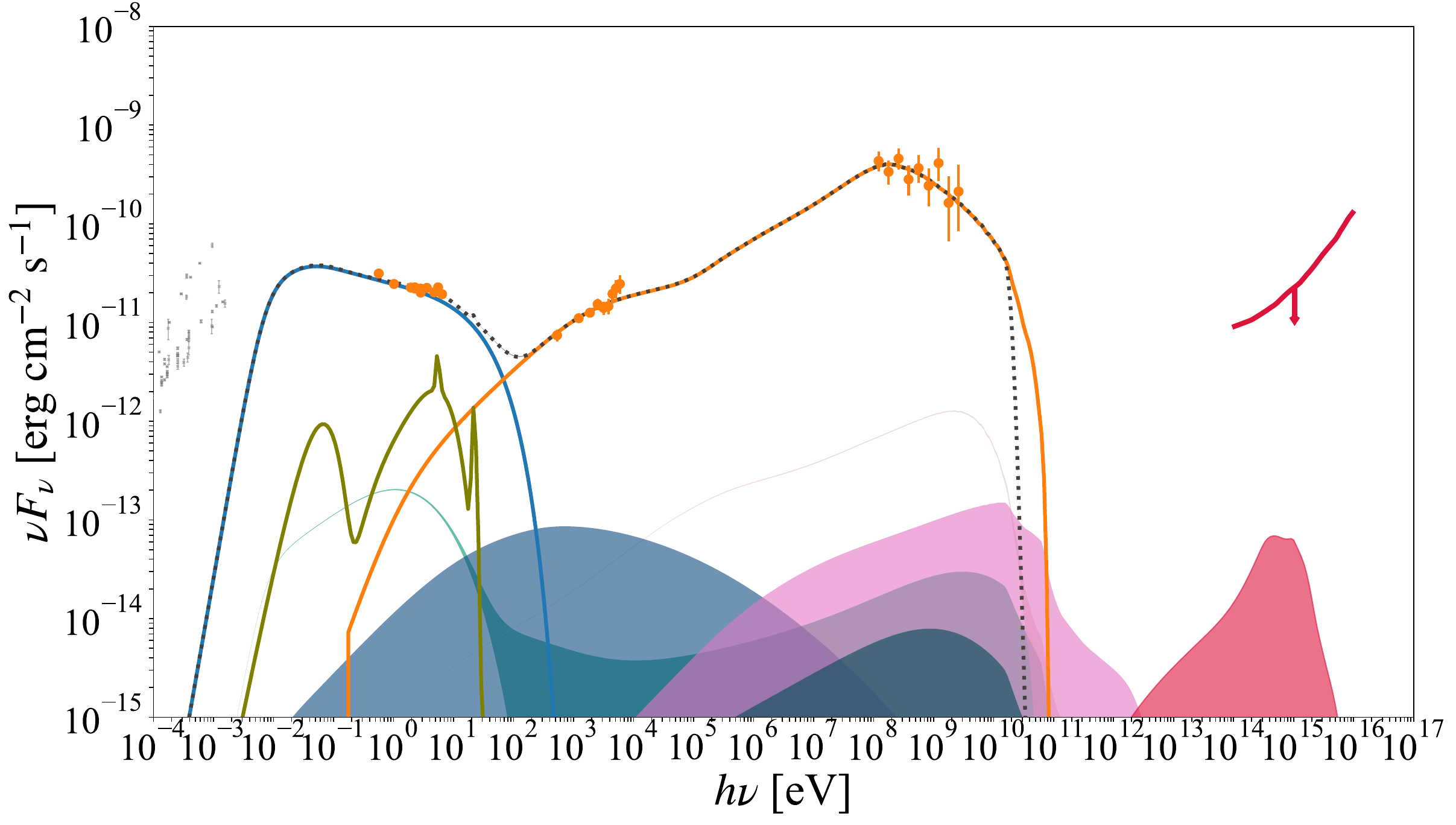}\label{fig:55471-55472}}

    \subfloat[legend]{\includegraphics[width=0.80\textwidth]{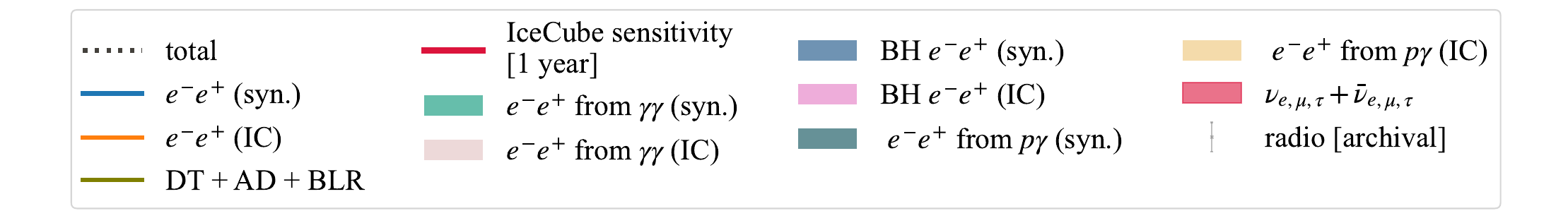}\label{fig:leptohadronic_legend_only}}
    
    \caption{SEDs averaged over MJD 54700--54701 (a), MJD 54827--54833 (b), MJD 55180--55181 (c), MJD 55471--55472 (d), fitted with the leptohadronic model (cf. Fig \ref{fig:leptohadronic}). The values of the model parameters are shown in Table \ref{table:leptonic_parameters_table}. The ratio of proton and electron energy densities is fixed to $\rho_{p/e} = 1.3 \times 10^{2}$.}
    \label{fig:4SEDs}
\end{figure*}

\section{Linear relation between integral $\nu$ and $\gamma$ fluxes}\label{appendix:linearity}
\begin{figure}
     \includegraphics[width=1\columnwidth]{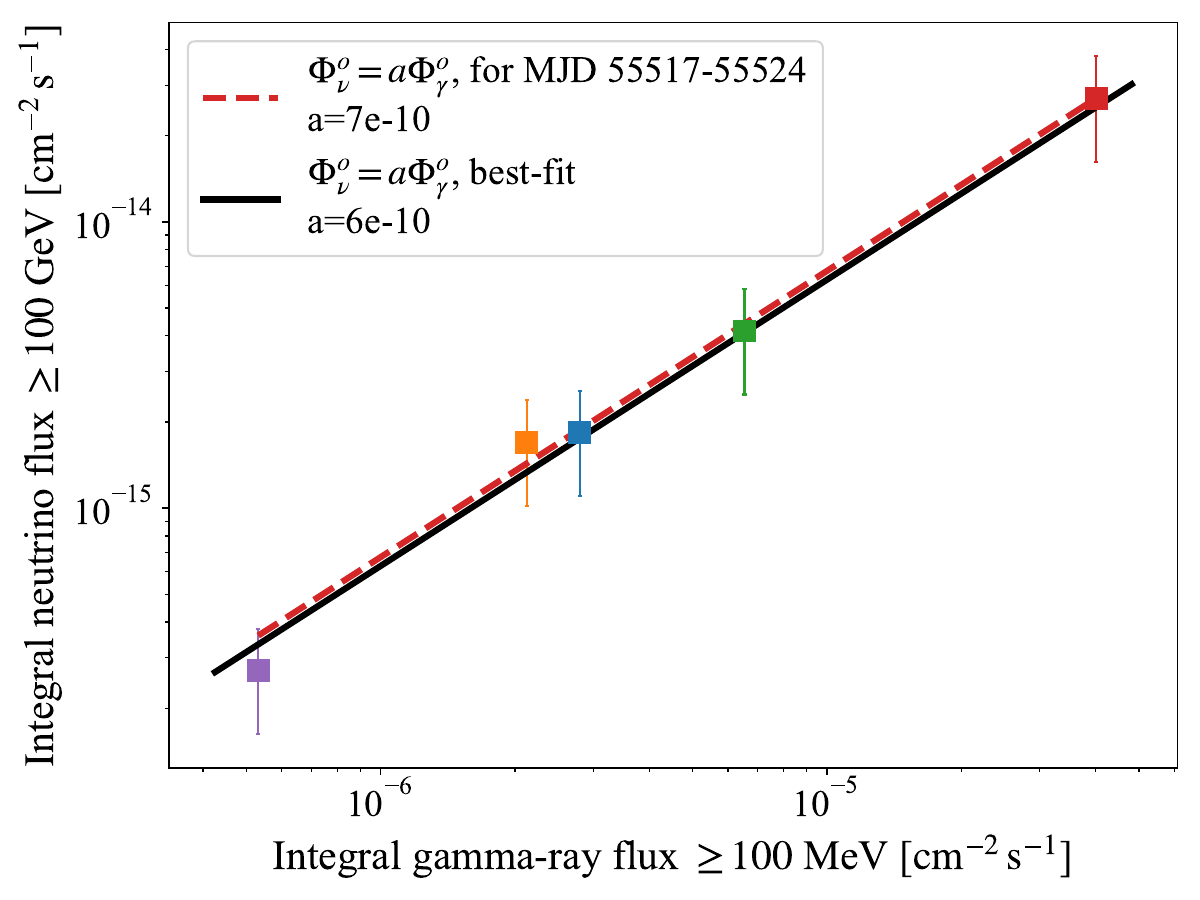}
     \caption{Illustration of the linear relation between the integral all-flavour $\nu$ flux of particles with energies $E^{o}_{\nu} \geq 100$~GeV and $\gamma$-ray flux of photons with energies $E^{o}_{\gamma} \geq 100$~MeV. The colours of data points correspond to the colours of the periods highlighted accordingly in Fig.~\ref{fig:light_curve}. The error bars for neutrino fluxes represent the numerical uncertainty $\sim 40$\% of normalisation of spectra of secondaries in hadronic processes \citep{cerruti2024comprehensivehadroniccodecomparison}}
     \label{fig:linearity}
\end{figure}

As mentioned in Sect.~\ref{sec:neutrino_yield_estimates}, we obtained electromagnetic and neutrino SEDs of 3C~454.3 covering four other periods of the source's activity. These periods are highlighted as vertical lines in Fig.~\ref{fig:light_curve}, and their SEDs are shown in Fig.~\ref{fig:4SEDs}. In Fig.~\ref{fig:linearity}, we plot the integral particle fluxes of neutrinos and $\gamma$-rays for the five considered periods. The fluxes turn out to be linearly related. We explain this linearity as follows. As can be seen from Table~\ref{table:leptonic_parameters_table}, despite the change of the integral $\gamma$-ray photon flux of 3C~454.3 by two orders of magnitude, the electron injection luminosity changes by less than one order of magnitude, and most of the $\gamma$-ray flux change is driven by the change of the Lorentz factor $\Gamma$ due to the very high Compton dominance of 3C~454.3. We find that both $\gamma$ and $\nu$ fluxes are approximately proportional to $\delta^{5}$. At energies $E^{o}_{\nu} \gtrsim 100$~TeV, neutrinos begin to be directly produced on the BLR photon field, and \textit{Fermi}-LAT $\gamma$ rays are produced via the IC scattering on the same photon field and the DT radiation boosted in the blob rest frame. At energies $E^{o}_{\nu} < 100$~TeV, we find that the main target for the neutrino production is the $\sim$~MeV $\gamma$-rays produced via external Compton scattering on the DT photon field (with some contribution of the direct radiation from the AD, see Fig.~\ref{fig:multi_panel}). We confirm this by turning off the IC process in the simulation and observing no neutrinos with energies below $100$~TeV, while switching off the synchrotron radiation almost does not affect the neutrino production (and neutrinos with energies $E^{o}_{\nu} \gtrsim 100$~TeV, as expected, are not affected by these switches because they are produced on the external BLR photon field). The dependence of $\nu$ flux on the fifth power of the Doppler factor (having fixed other parameters) is expected \citep{Dermer:2012rg} and can potentially explain the recently observed evidence that radio-bright AGNs, spatially associated with high-energy neutrinos, have their estimated Doppler factors systematically larger than all radio-bright AGNs \citep{Plavin:2025pjt}.

\section{How do various values of proton acceleration efficiency affect the expected neutrino yield?}\label{appendix:xi}

To investigate the impact of our choice of the acceleration parameter $\xi_{\mathrm{acc}} = 10^{7}$ for protons, the same as for electrons (obtained from fitting the electromagnetic SED), we show how the \textit{IceCube} muon neutrino daily yield from 3C 454.3 changes if we assume a different value of $\xi_{\mathrm{acc}}$ for protons in Fig.~\ref{fig:xi} (left axis, black circles). For each value of $\xi_{\mathrm{acc}}$, we determine the maximum proton energy by finding the intersection of the proton acceleration timescale $ t^{\prime \mathrm{acc}}_{p}$ (Eq.~\ref{eq:acceleration_timescale}) with the proton energy-loss timescale (dominated by the photopion losses, see the purple dotted curve in Fig.~\ref{fig:timescales}). Then we repeat the procedure for finding $\rho_{p/e}$ (Eq. \ref{eq:rho}) as was described in Sect.~\ref{sec:leptohadronic} until the Bayes factor (Eq. \ref{eq:bayes_factor}) drops to $10^{-2}$. The corresponding values of $\rho_{p/e}$ are shown in the same plot in Fig.~\ref{fig:xi} (right axis, blue squares). Since at a fixed proton energy density, SEDs of secondary photons increase with increasing proton maximum energy (i.e. decreasing $\xi_{\mathrm{acc}}$), the overshooting of the X-ray data happens at a lower normalization of the proton SED. Hence, $\rho_{p/e}$ decreases when $\xi_{\mathrm{acc}}$ decreases. In Figs.~\ref{fig:leptohadronic_sed_xi=5e+04} and \ref{fig:leptohadronic_sed_xi=10}, we show photon and neutrino SEDs for $\xi_{\mathrm{acc}} = 5 \times 10^4$ and $\xi_{\mathrm{acc}} = 10$ respectively. One can see that the peak energy of the neutrino SED shifts to higher values with decreasing $\xi_{\mathrm{acc}}$ (compare also to Fig.~\ref{fig:leptohadronic}). The values of $10^{4} \lesssim \xi_{\mathrm{acc}} \lesssim 10^{5}$ are optimal in terms of maximising the neutrino yield. For $\xi_{\mathrm{acc}} \leq 10^3$, $\gamma\gamma$-initiated cascades start to dominate over BH-initiated cascades, and $\rho_{p/e}$ must be significantly decreased not to overshoot the X-ray data while the peak energy of the neutrino SED continues shifting to higher values resulting in the lower expected \textit{IceCube} neutrino yield. Values $\xi_{\mathrm{acc}} \leq 10^3$ are required to produce neutrinos with energies similar to the energy of the recently discovered $\gtrsim$100 PeV neutrino KM3-230213A~\citep{KM3NeT2025}.

\begin{figure}
    \includegraphics[width=1\columnwidth]{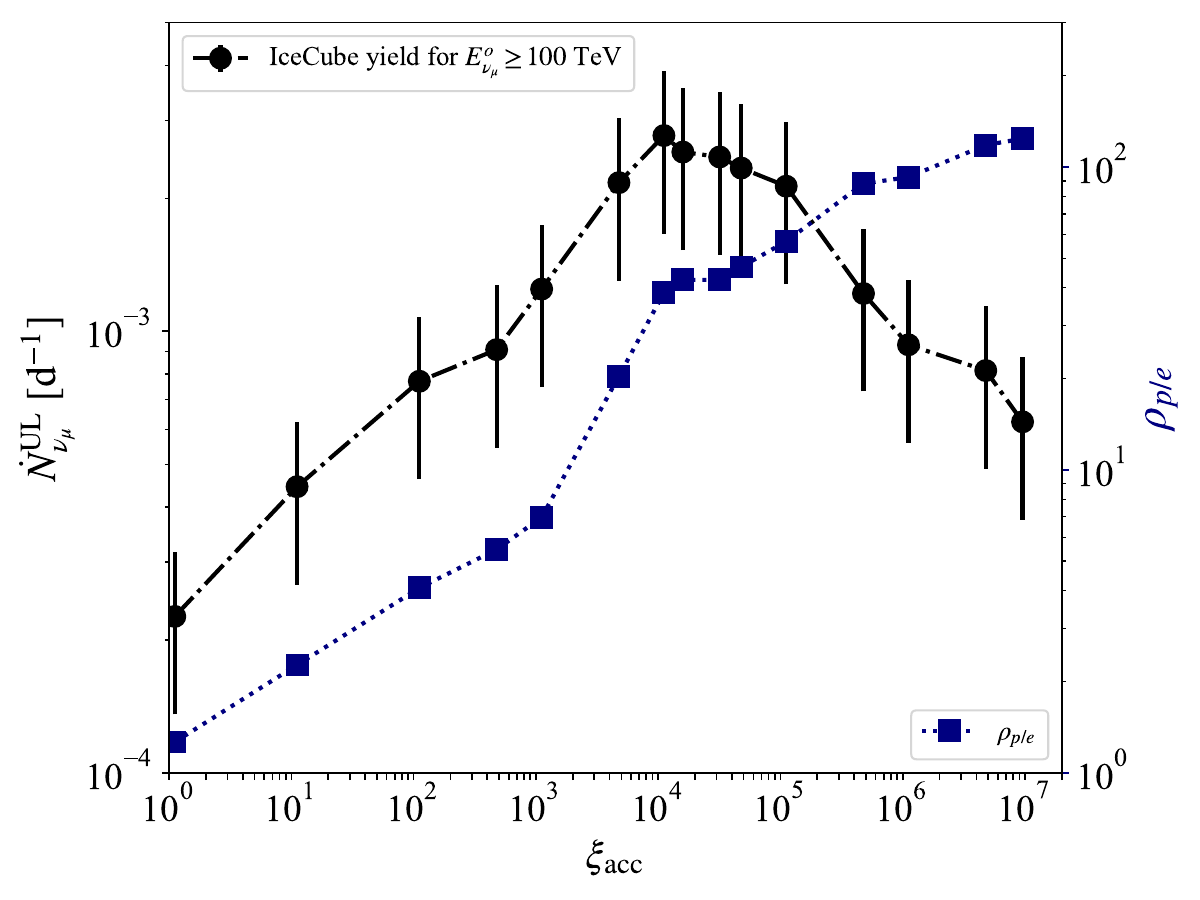}
    \caption{\textit{IceCube} daily muon neutrino yield from 3C 454.3 averaged over MJD 55517--55524 for different values of the proton acceleration parameter $\xi_{\mathrm{acc}}$ (black circles, left ordinate axis) along with the UL on the proton-to-electron energy density ratio $\rho_{p/e}$ as a function of $\xi_{\mathrm{acc}}$ (blue squares, right ordinate axis). The error bars for neutrino yields represent the numerical uncertainty $\sim 40$\% of normalisation of spectra of secondaries in hadronic processes \citep{cerruti2024comprehensivehadroniccodecomparison}.}
    \label{fig:xi}
\end{figure}
\begin{figure*}
         \includegraphics[width=1.667\columnwidth]{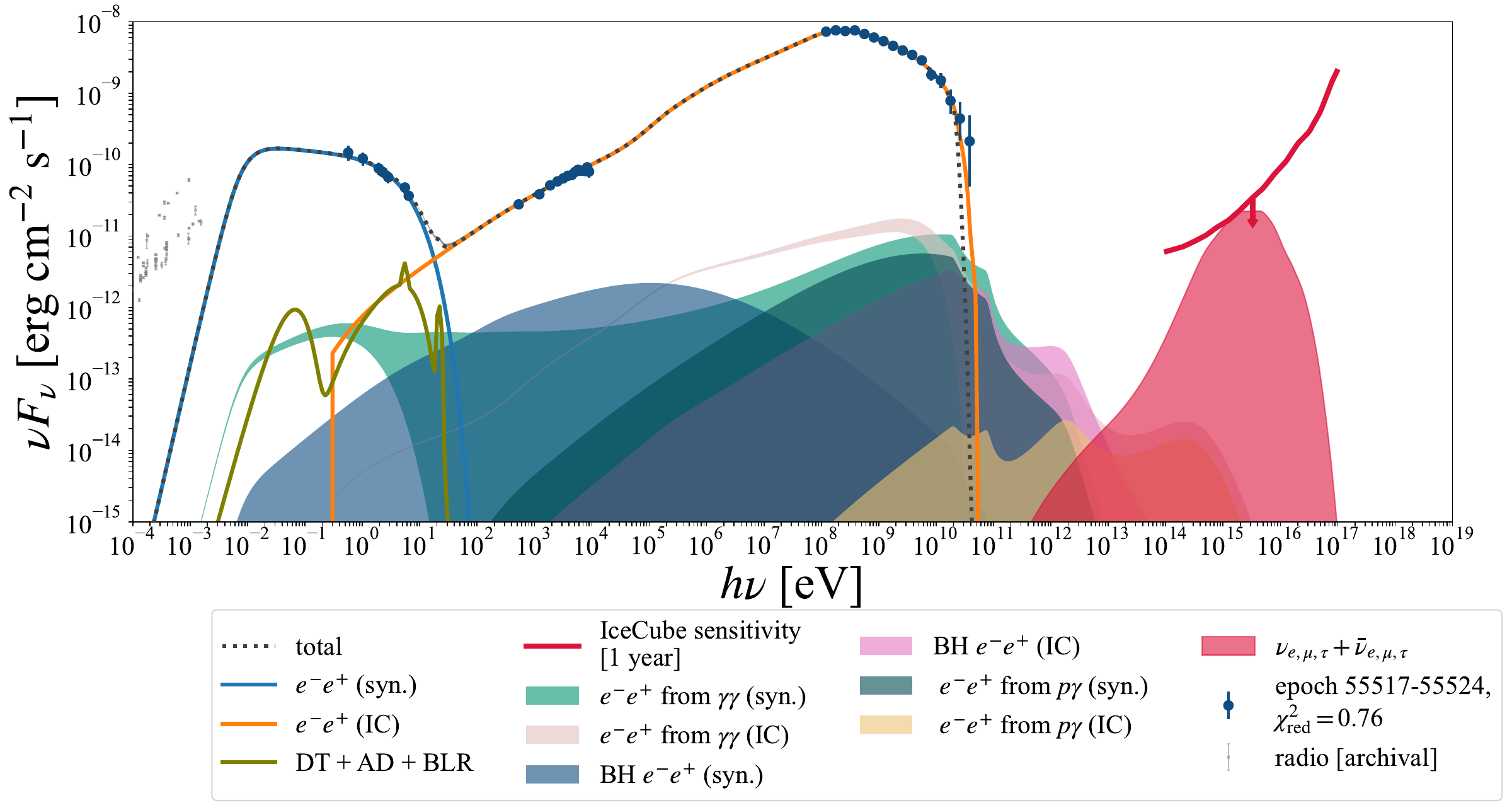}
         \caption{The photon and neutrino SED with model curves for the ``standing-feature'' leptohadronic model with a fit to the observed SED averaged over MJD 55517--5524 for proton acceleration parameter $\xi_{\mathrm{acc}} = 5 \times 10^{4}$. Shaded regions correspond to the spread of proton luminosities $0 \leq u^{\prime}_{p} \leq u^{\prime\mathrm{UL}}_{p}$, where $u^{\prime\mathrm{UL}}_{p}$ is the UL corresponding to the Bayes factor with respect to the best-fit leptonic model $BF = 10^{-2}$ ($\approx 3 \sigma$ Gaussian equivalent). The crimson solid curve with a downward arrow shows the \textit{IceCube} one-year sensitivity (see Eq.~\ref{eq:icecube_sensitivity}) for the UL of 2.44 muon neutrinos between $E^{o \, \mathrm{min}}_{\nu} = 10^{14}$~eV and $E^{o \, \mathrm{max}}_{\nu} = 10^{17}$~eV.}
         \label{fig:leptohadronic_sed_xi=5e+04}
\end{figure*}
\begin{figure*}
         \includegraphics[width=1.667\columnwidth]{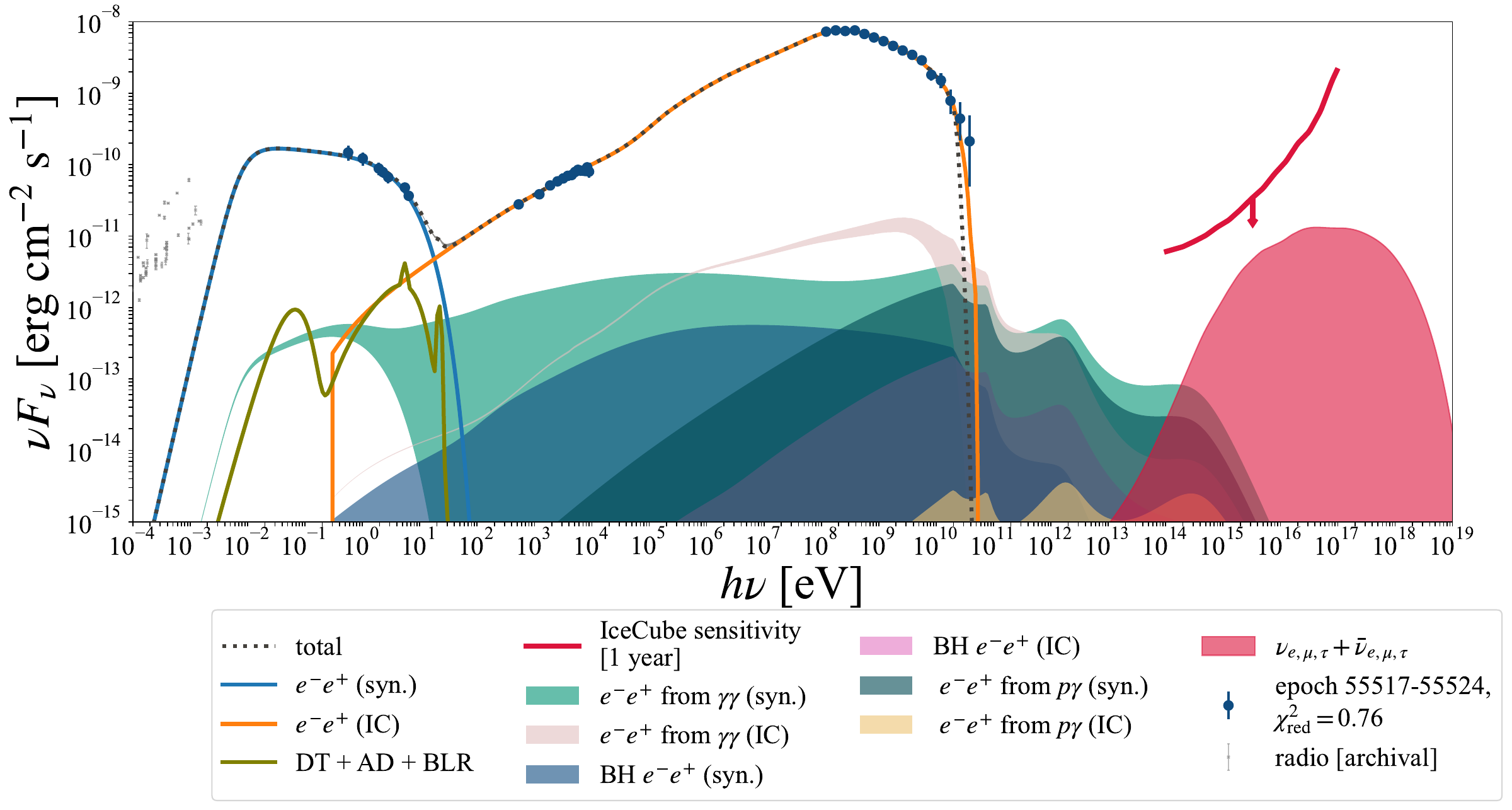}
         \caption{Same as Fig.~\ref{fig:leptohadronic_sed_xi=5e+04} but for $\xi_{\mathrm{acc}} = 10$.}
         \label{fig:leptohadronic_sed_xi=10}
\end{figure*}

\section{MCMC fits to the spectral-energy distributions during the flare}
\label{appendix:corner_plot}

To estimate the uncertainty of the parameter values describing the seven-day averaged SED (MJD 55517--55524) around the peak of the flare as well as the SEDs of each individual day within the period, we first run the \texttt{MIGRAD} routine using \texttt{iMINUIT} to obtain the values of parameters delivering the minimum of $\chi^{2}$. Then, to explore the parameter space around the obtained minimum, for each of the seven days and the seven-day-averaged period, we initialise a Monte-Carlo Markov Chain (MCMC) \citep{emcee} with $
24$ walkers. The initial guess for each walker for each parameter $w_{i}$ is chosen as $\left[ \hat{w}_{i} \times (1 \pm r N\{0, 1\}) \right]$, where $\hat{w}_{i}$ is the best-fit value of the $i$th parameter from the \texttt{MIGRAD} optimisation, $r = 0.25$ (0.1 for MJD 55523--55524 due to a failed first attempt of the fit), and $N\{0, 1\}$ stands for the standard normal distribution. The priors for the parameters are chosen to be flat within the following segments (see the description of the model parameters in Sect.~\ref{sec:standing_blob_model_description}):
\begin{enumerate}
        \item $[0.1\hat{L}_{e}^{\prime}; 10\hat{L}_{e}^{\prime}]$;
        \item $[0.5\hat{s}_{e}; 2\hat{s}_{e}]$;
        \item $[0.02\hat{E}_{e}^{\prime{\mathrm{min}}}; 3\hat{E}_{e}^{\prime{\mathrm{min}}}]$;
        \item $[0.333\hat{E}_{e}^{\prime{\mathrm{max}}}; 30\hat{E}_{e}^{\prime{\mathrm{max}}}]$;
        \item $[0.333\hat{R}_{b}^{\prime}; 3\hat{R}_{b}^{\prime}]$;
        \item $[0.2\hat{B}^{\prime}; 5\hat{B}^{\prime}]$;
        \item $[0.1\hat{\Gamma}; 2\hat{\Gamma}]$;
        \item $[0.5\hat{x}; 3\hat{x}]$.
    \end{enumerate}

The flat priors might bias sampling towards higher parameter values for those parameters that span more than one order of magnitude. To investigate the effect of the priors, for the seven-day-averaged dataset, we run a new MCMC fit changing the priors for parameters spanning more than one order of magnitude from uniform to log-uniform. We obtain results very close to the ones with the uniform priors, with all posterior parameter credible intervals remaining close to those with uniform priors, within $\pm 10$\% variation. Thus, we conclude that the posterior distributions around the minimum obtained with \texttt{iMINUIT} are driven by the likelihood function and do not distinguish between uniform and log-uniform priors for the chosen number of MCMC steps. The results below are reported for the flat priors.

The best-fit values of the model parameters are defined as the medians of the marginal posterior one-dimensional parameter distributions obtained after running the MCMCs, while their uncertainties are defined as the differences between the median and the $16$th and $84$th percentiles\footnote{following the tutorial at \url{https://emcee.readthedocs.io/en/stable/tutorials/line}}. The best-fit parameter values with their uncertainties are presented in Fig.~\ref{fig:sequence_of_standing_blobs}, while corresponding corner plots are shown in Figs.~\ref{fig:corner_plot}, \ref{fig:corner_plot_55517_55518}, \ref{fig:corner_plot_55518_55519}, \ref{fig:corner_plot_55519_55520}, \ref{fig:corner_plot_55520_55521}, \ref{fig:corner_plot_55521_55522}, \ref{fig:corner_plot_55522_55523}, \ref{fig:corner_plot_55523_55524}. A qualitative description of the results is presented in Sect.~\ref{sec:standing_blob_model_results}.

\begin{figure*}
    \includegraphics[width=2\columnwidth]{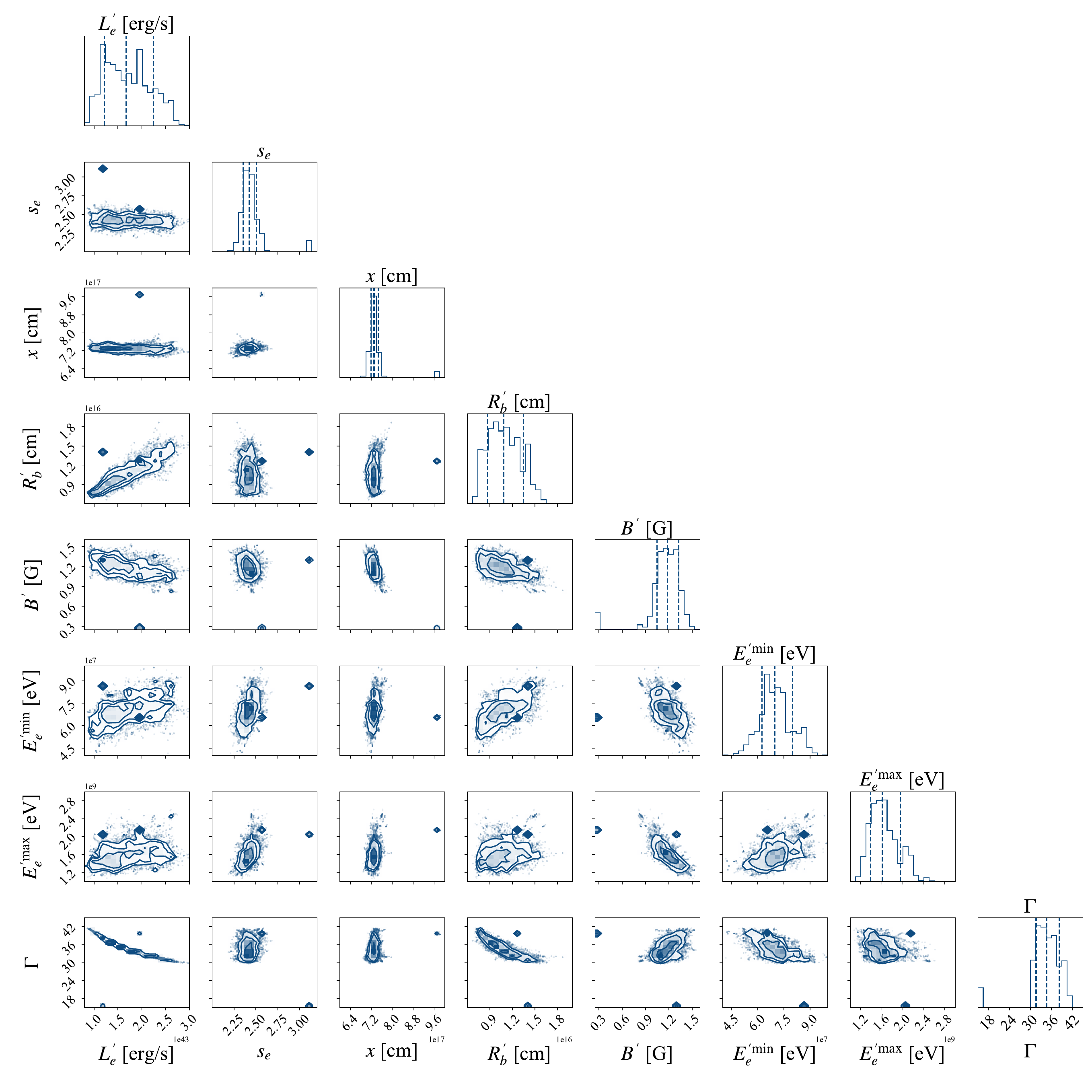}
    \caption{The corner plot \citep{corner} obtained from the MCMC fitting \citep{emcee} of the seven-day-averaged SED (MJD 55517--55524) with a ``standing-feature'' leptonic model. $t^{\prime}_{\mathrm{sim}}$ here is the same as for the one-day-averaged SEDs with $t^{o}_{\mathrm{av}} = 1$ d (see Eq. \ref{eq:simulation_time}). The corner plot is obtained from an MCMC containing $3,000$ steps per walker with $3\times8=24$ walkers after combining chains from all the walkers (discarding $1,000$ initial steps of each walker as a burn-in) and applying thinning of $4$. Dashed vertical lines in the one-dimensional marginal histograms represent $16\%$, $50\%$, and $84\%$ percentiles of the posterior marginal distributions. Contour lines in the 2D marginal histograms represent $11.8\%$, $39.3\%$, $67.5\%$, and $86.4\%$ 2D containment regions.}
    \label{fig:corner_plot}
\end{figure*}

\begin{figure*}
    \includegraphics[width=2\columnwidth]{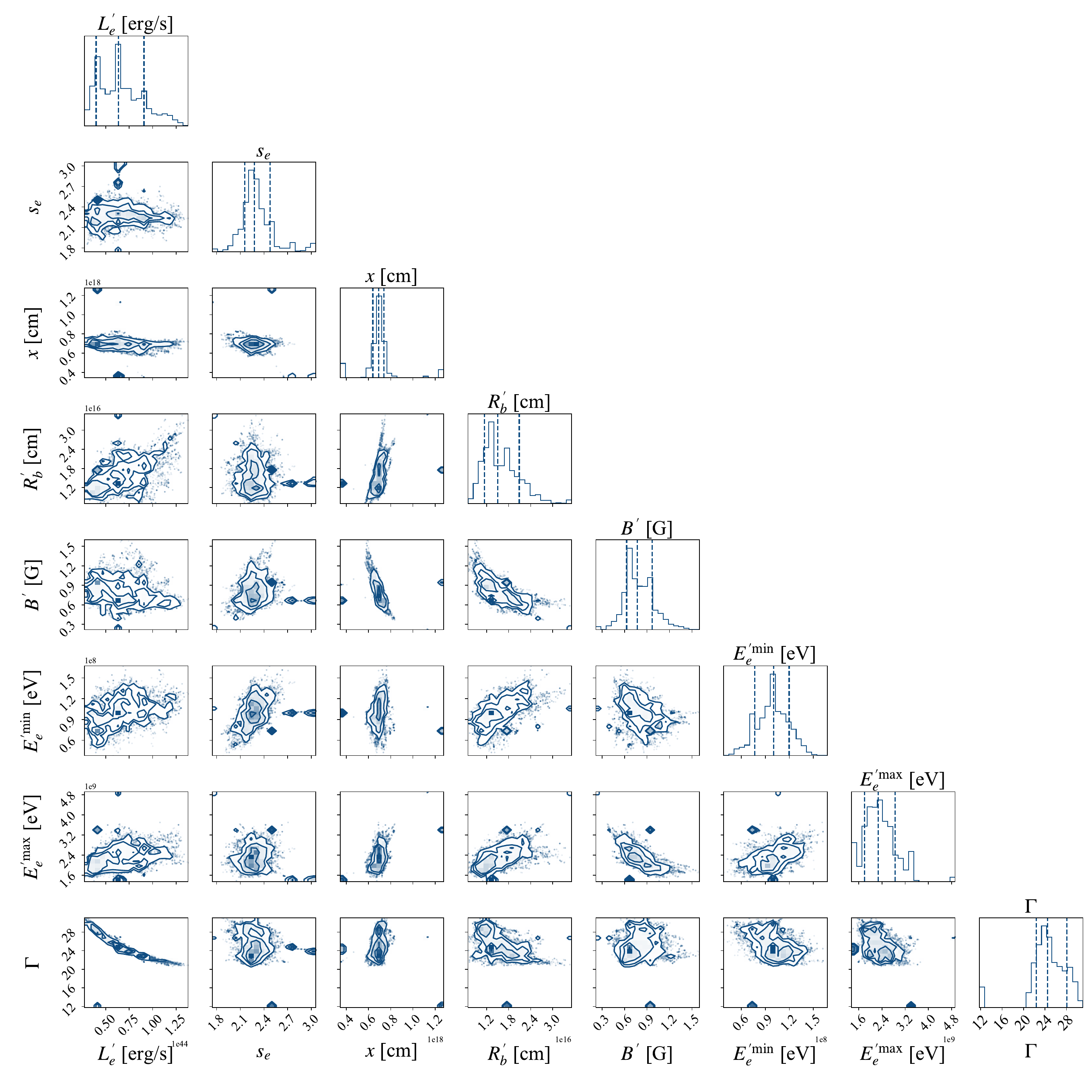}
    \caption{The same as Fig. \ref{fig:corner_plot} but for MJD 55517--55518.}
    \label{fig:corner_plot_55517_55518}
\end{figure*}

\begin{figure*}
    \includegraphics[width=2\columnwidth]{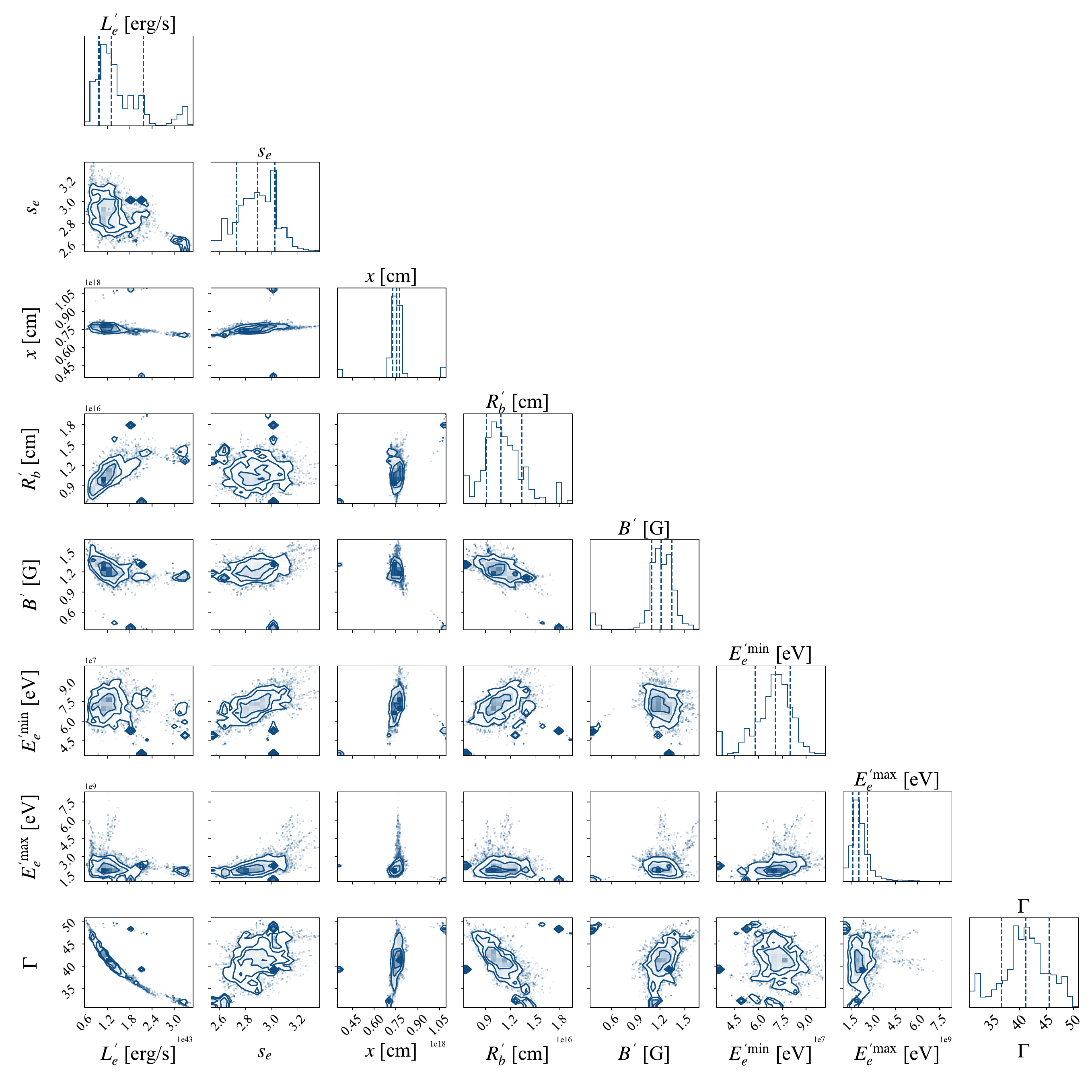}
    \caption{The same as Fig. \ref{fig:corner_plot} but for MJD 55518--55519.}
    \label{fig:corner_plot_55518_55519}
\end{figure*}

\begin{figure*}
    \includegraphics[width=2\columnwidth]{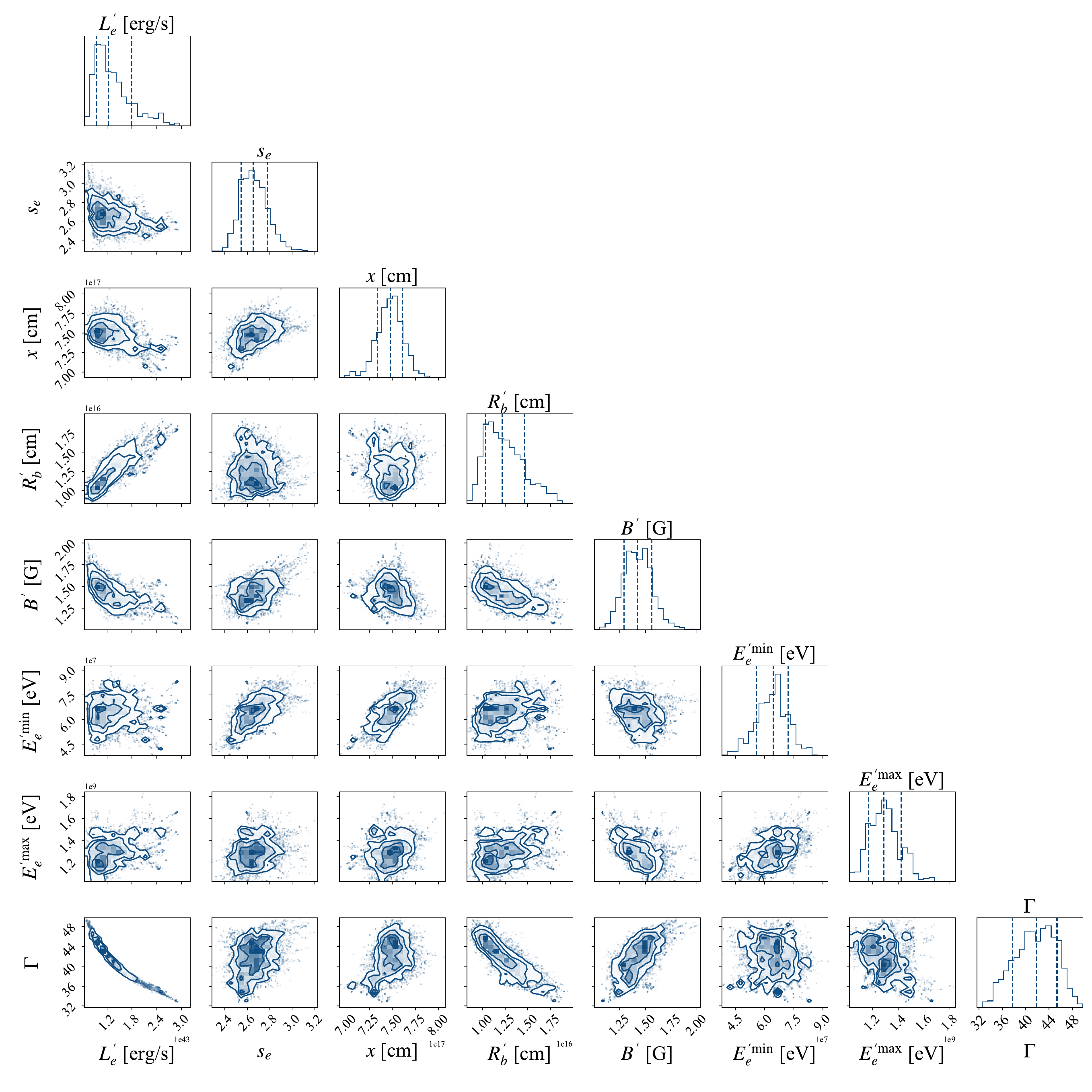}
    \caption{The same as Fig. \ref{fig:corner_plot} but for MJD 55519--55520.}
    \label{fig:corner_plot_55519_55520}
\end{figure*}

\begin{figure*}
    \includegraphics[width=2\columnwidth]{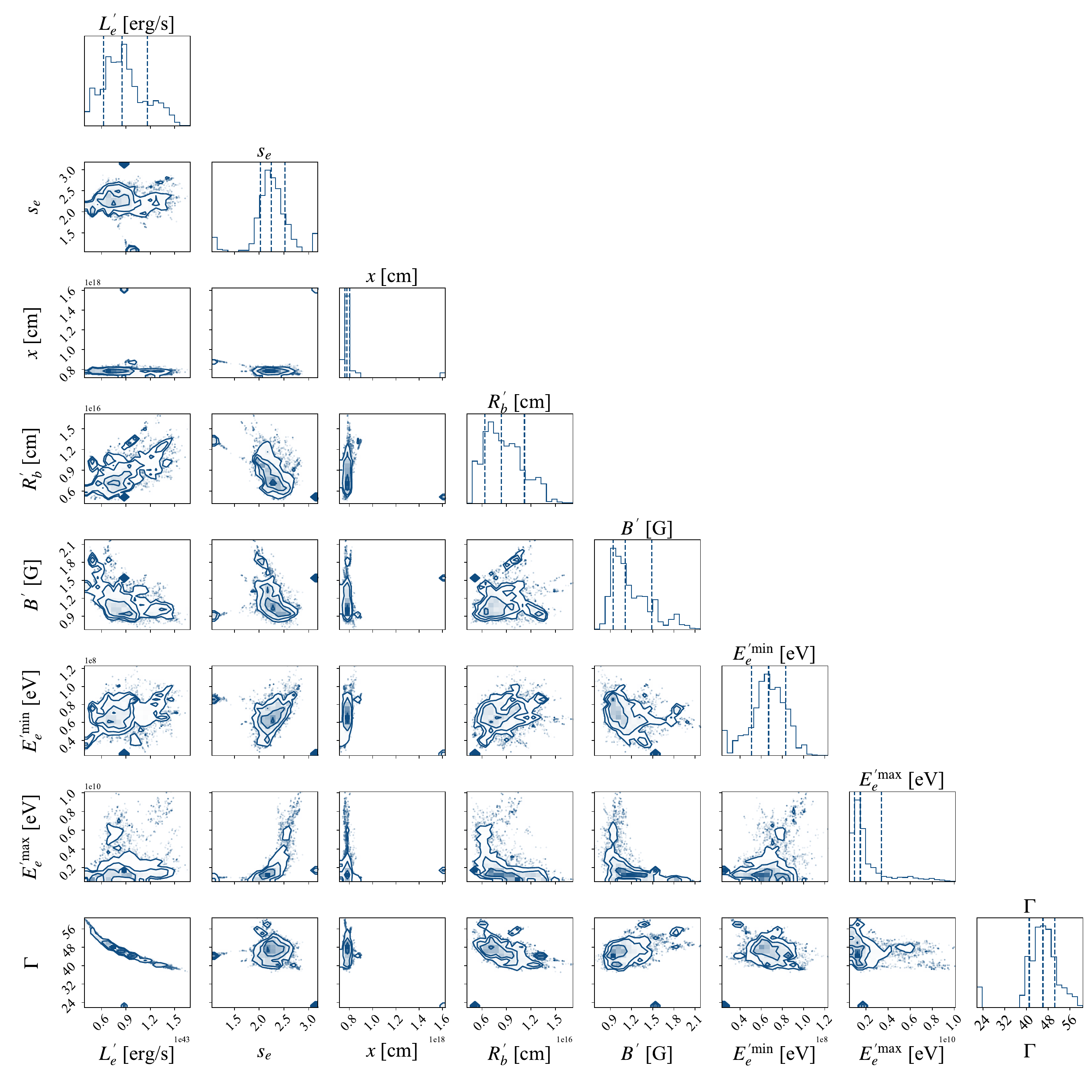}
    \caption{The same as Fig. \ref{fig:corner_plot} but for MJD 55520--55521.}
    \label{fig:corner_plot_55520_55521}
\end{figure*}

\begin{figure*}
    \includegraphics[width=2\columnwidth]{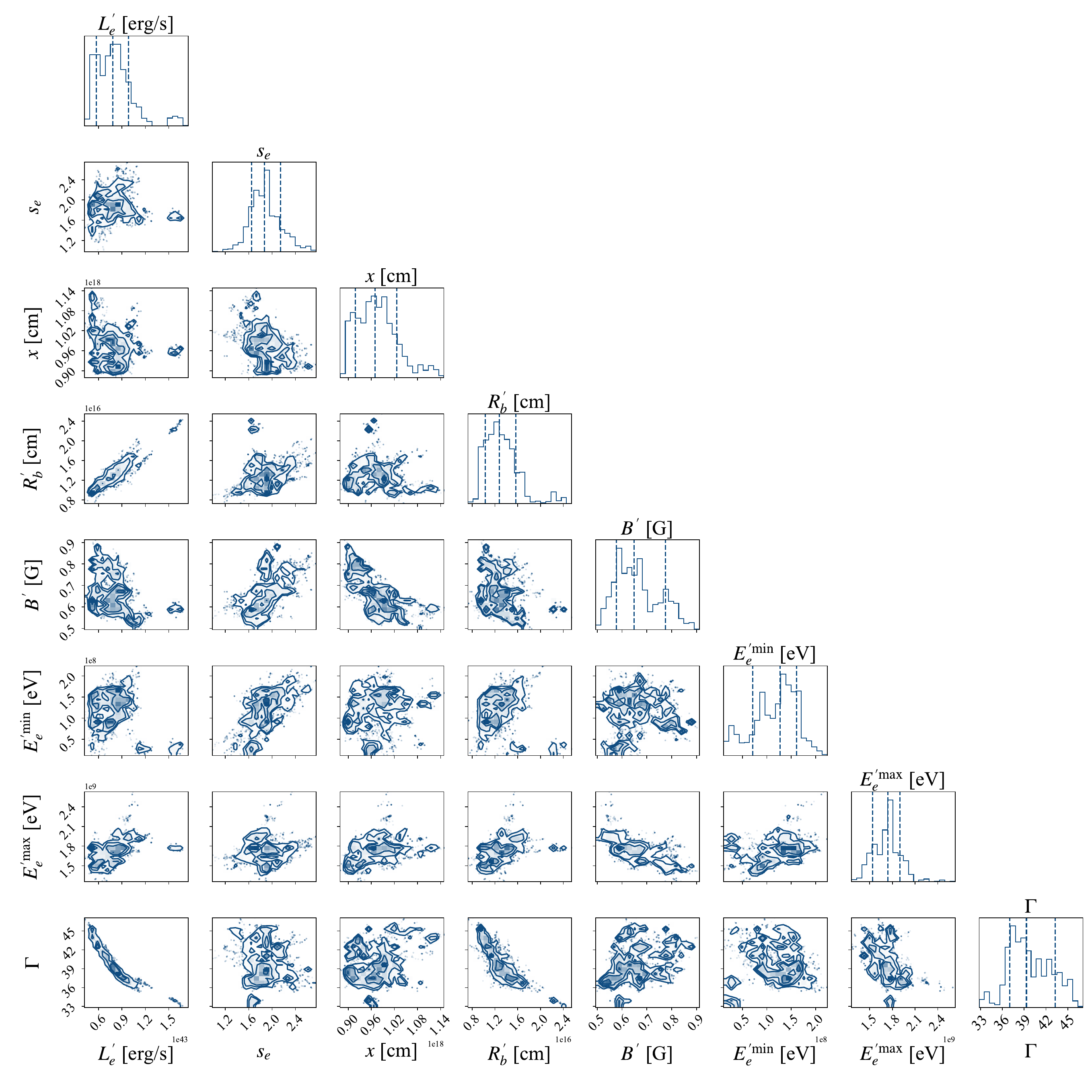}
    \caption{The same as Fig. \ref{fig:corner_plot} but for MJD 55521--55522 and, instead, 2, 500 initial steps of each walker are discarded as a burn-in and no thinning is applied due to slower convergence of the fit.}
    \label{fig:corner_plot_55521_55522}
\end{figure*}

\begin{figure*}
    \includegraphics[width=2\columnwidth]{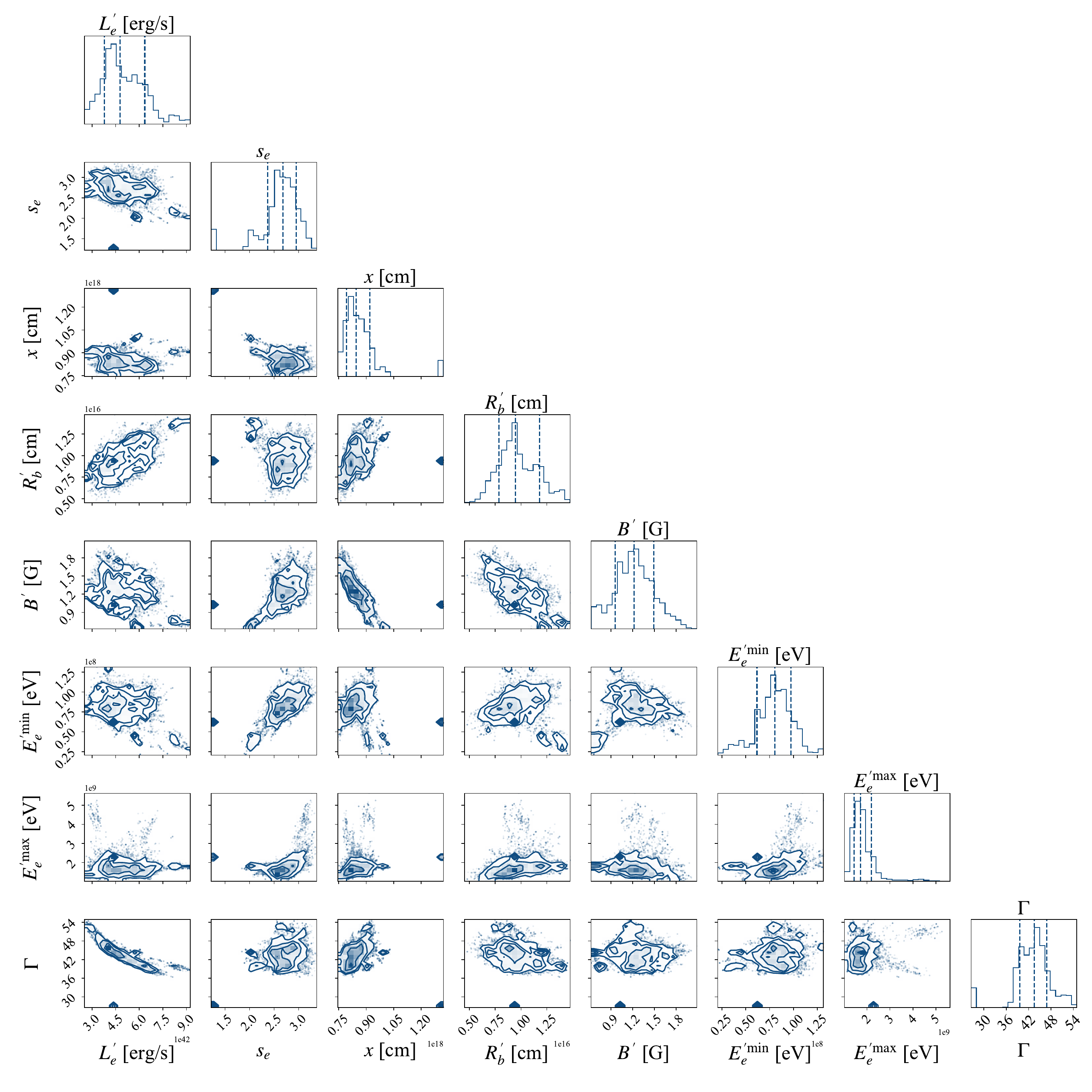}
    \caption{The same as Fig. \ref{fig:corner_plot} but for MJD 55522--55523.}
    \label{fig:corner_plot_55522_55523}
\end{figure*}

\begin{figure*}
    \includegraphics[width=2\columnwidth]{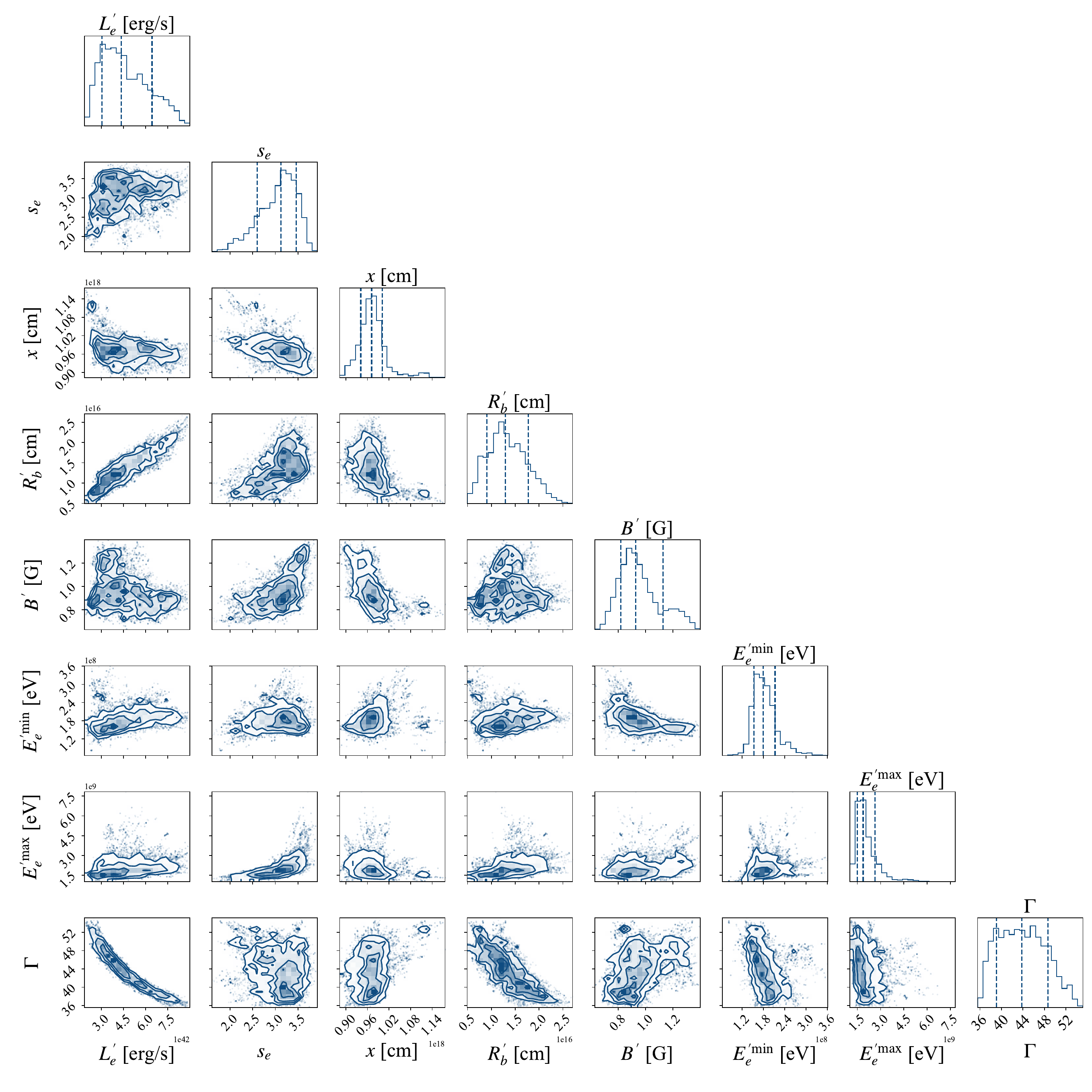}
    \caption{The same as Fig. \ref{fig:corner_plot} but for MJD 55523--55524 and, instead, $200$ initial steps of each walker are discarded as a burn-in and thinning of $5$ is applied due to a better value of $\chi^{2}_{\mathrm{red}}$ obtained in this case.}
    \label{fig:corner_plot_55523_55524}
\end{figure*}


\bsp	
\label{lastpage}
\end{document}